\documentclass[aps,prd,reprint,superscriptaddress,nofootinbib]{revtex4-1}
\usepackage{amsmath}
\usepackage{amssymb}
\usepackage{graphicx}
\usepackage[caption=false]{subfig}
\usepackage{enumitem}
\usepackage{color}
\usepackage{mathtools}
\usepackage[percent]{overpic}
\usepackage{bm}
\usepackage{rotating}
\usepackage{cancel}
\usepackage{hyperref}

\newcommand{\ket}[1]{{\left| {#1} \right>}}
\newcommand{\bra}[1]{{\left< {#1} \right|}}

\newcommand{\ii}{\mathrm{i}}
\newcommand{\trf}{\text{Tr}_{\text{f}}}

\newcommand{\tj}[6]{ \begin{pmatrix}
  #1 & #2 & #3 \\
  #4 & #5 & #6 
\end{pmatrix}}
\newcommand{\ac}{e}
\newcommand{\proj}[2]{{\left| {#1} \right>\left< {#2} \right|}}
\begin{document}

\title{Entanglement harvesting from the electromagnetic vacuum with hydrogenlike atoms}

\author{Alejandro Pozas-Kerstjens}
\affiliation{ICFO-Institut de Ciencies Fotoniques, The Barcelona Institute of Science and Technology, Av. Carl Friedrich Gauss 3, 08860 Castelldefels (Barcelona), Spain}
\author{Eduardo Mart\'{i}n-Mart\'{i}nez}
\affiliation{Institute for Quantum Computing, University of Waterloo, Waterloo, Ontario, N2L 3G1, Canada}
\affiliation{Department of Applied Mathematics, University of Waterloo, Waterloo, Ontario, N2L 3G1, Canada}
\affiliation{Perimeter Institute for Theoretical Physics, 31 Caroline St. N., Waterloo, Ontario, N2L 2Y5, Canada}


\begin{abstract}
We study how two fully-featured hydrogenlike atoms harvest entanglement from the electromagnetic field vacuum, even when the atoms are spacelike separated. We compare the electromagnetic case ---qualitatively and quantitatively--- with previous results that used  scalar fields and featureless, idealized atomic models. Our study reveals the new traits that emerge when we relax these idealizations, such as anisotropies in entanglement harvesting and the effect of exchange of angular momentum. We show that, under certain circumstances, relaxing previous idealizations makes vacuum entanglement harvesting more efficient.
\end{abstract}

\maketitle


\section{Introduction}

One of the most fundamental differences between classical and quantum theories of fields is the completely different nature of their respective ground states. While in classical theories the ground state is just a zero energy density state everywhere, the vacuum state of a quantum field, even a noninteracting one, contains classical and quantum correlations between field observables defined in different regions of spacetime, even when those regions are spacelike separated \cite{Summers1985,Summers1987}. This nonclassical behavior of the vacuum lies deep in the core of quantum theory, and is a key ingredient in phenomena such as, e.g., \textit{quantum energy teleportation} \cite{Hotta2008,Hotta2009} (with implications, among other areas, in quantum thermodynamics \cite{Frey2014}), \textit{quantum collect calling} \cite{Jonsson2015} (with implications in cosmology \cite{Blasco:2015eya,Blasco2015,Blasco2016}), or even in the black hole information loss problem \cite{Hawking1975,Hawking1976,Hawking2016,Almheiri2013,Braunstein2013}.

These correlations are, in principle, physically accessible: both classical and quantum vacuum correlations can be extracted from the field to quantum particle detectors that couple to it locally \cite{Reznik2003,Reznik2005}. This allows two parties ---Alice and Bob--- to entangle the particle detectors that each one operates, even if they remain spacelike separated during their whole existence (see e.g. \cite{Reznik2005}). The extraction of nonclassical correlations from the quantum vacuum to particle detectors was first pointed out by Valentini \cite{Valentini1991}, and later on studied by Reznik \textit{et al.} \cite{Reznik2003,Reznik2005}. This phenomenon has become known as \textit{entanglement harvesting}.

Entanglement harvesting from scalar fields is now well understood \cite{Pozas-Kerstjens2015}. In fact, the ability to harvest entanglement from quantum fields has motivated numerous works aiming to harness this phenomenon in a variety of scenarios. These range from the extraction of resources and \textit{entanglement farming} \cite{Martin-Martinez2013a} or metrology \cite{Salton2015}, to its study in cosmology, where it has been shown that entanglement harvesting is very sensitive to the geometry of the underlying spacetime \cite{Steeg2009,Martin-Martinez2012} or even its topology \cite{Martin-Martinez2016a}.

All the works mentioned above, and furthermore, all works in entanglement harvesting known to the authors, model the field-probe interaction by the simplified Unruh-DeWitt particle detector model \cite{DeWittBook} (with two exceptions, the original work by Valentini \cite{Valentini1991}---which we will address below---and Braun's work on heath-bath mediated entanglement creation \cite{Braun2002,Braun2005}---which studied energy-degenerate pointlike light-atom interactions in setups where, unlike in \cite{Valentini1991}, spacelike entanglement harvesting is not present---). The Unruh-DeWitt model consists of a linear coupling of a spherically symmetric (often pointlike) two-level quantum system and a massless, scalar quantum field (sometimes with slight variations such as a spatial smearing function that endows the two-level system with a finite size  \cite{Pozas-Kerstjens2015}). Although in principle simple, this model was shown to capture the fundamental features of the light-matter interaction in scenarios where exchange of angular momentum does not play a role \cite{Martin-Martinez2013,Alhambra2014}. The fact that this simple toy model of a particle detector can reproduce the light-matter interaction to a great extent is one of the reasons behind the strong  interest that this model (in entanglement harvesting in particular) has attracted in more applied scenarios. The success of the Unruh-DeWitt model is such that there have been works based on this model that study the feasibility of experimental implementations of entanglement harvesting in atomic systems and superconducting circuits \cite{Olson2011,Olson2012,Sabin2012}, and how clock synchronization may affect these experimental realizations of vacuum entanglement harvesting \cite{Martin-Martinez2016}.

Nevertheless, and despite its great success, the Unruh-DeWitt model fails to capture the complete interaction between fully-featured atoms and the electromagnetic field vacuum. The electromagnetic field is not a scalar but a vector field, which carries angular momentum. Moreover, realistic atomic orbitals are, in general, nonisotropic. These two features cannot be captured in the scalar Unruh-DeWitt model or any of its variants. This means that any study based on the Unruh-DeWitt model will not be able to see the inherent anisotropies and orientation dependence that entanglement harvesting has, and will not predict any effects related with the fact that the atoms can exchange angular momentum with the field (which, as we will see, may help or hinder their ability to harvest entanglement).

Furthermore, previous work on heat-bath mediated entanglement creation   \cite{Braun2002,Braun2005}, and even the original pioneering work by Valentini \cite{Valentini1991} (which is the only previous work known to the authors that discusses spacelike entanglement harvesting with an electromagnetic dipole coupling) consider neither the fine details of the atoms' exchange of angular momentum with the field nor the full features of the atomic orbital structure. As we will discuss, these aspects cannot be ignored and  yield nontrivial effects. In fact, the results in \cite{Valentini1991} are no different from those that would be obtained with a pointlike detector which is derivatively coupled to a scalar field \cite{Schlicht2004,Juarez-Aubry2014}. Also, \cite{Valentini1991} considered excited states of the detectors as opposed to the, perhaps more interesting, ground state entanglement harvesting. This fact makes it difficult to compare with other later results.

In this paper we go beyond the Unruh-DeWitt model and other alike scalar simplifications of the light-matter interaction. For that, we feature a dipole coupling between the electromagnetic field and hydrogenoid atomic probes, being very careful with the way in which atoms exchange angular momentum with the field, and with the full atomic orbital structure of the probes.

Through this analysis we will show that the harvesting of timelike and spacelike entanglement from the electromagnetic vacuum is possible using fully-featured atomic qubits, initially in their ground states. Furthermore, we will compare entanglement harvesting predictions of previous scalar models (both, Unruh-DeWitt and derivative couplings) with the electromagnetic case, and characterize their fundamental differences in detail.

We will confirm that, indeed, the scalar Unruh-DeWitt model, with either linear or derivative coupling, would yield qualitatively the same features as the electromagnetic dipole coupling if we could restrict the interactions to not exchanging angular momentum. Nevertheless, we show that in the full electromagnetic coupling both the anisotropy of the interaction and the exchange of angular momentum have a nontrivial influence on the ability to harvest entanglement from the field. 

We present a full study of how the harvesting power of transitions of fully-featured hydrogenoid atoms depends on the the anisotropic character of the atoms' excited states (through the atoms' relative orientation), and how the exchange of angular momentum impacts the harvesting of entanglement. We show that the electromagnetic field allows for the harvesting of more entanglement but with a more limited range than predicted by previous scalar or simplified derivative-coupling models.

This paper is structured as follows: In section \ref{sec:formalism} we present the different models of light-matter interaction that we will use and compare throughout the paper. Namely, the simplified Unruh-DeWitt  and derivative scalar couplings, and the electromagnetic dipole interaction. In section \ref{sec:setup} we analyze in detail the time evolution of the atoms-field system and the fine details of the hydrogenoid atom dipole coupling model. In section \ref{sec:results} we give the core of our results: In subsections \ref{sec:results} A-C we discuss the main features introduced by the vector nature of the electromagnetic coupling and how they impact entanglement harvesting. In subsection \ref{sec:comparison} we compare the entangling power of the dipole electromagnetic coupling with the models previously studied in the literature (Unruh-DeWitt and derivative). Finally we summarize our conclusions in section \ref{sec:conclusions}, and give the complete derivations of the relevant expressions used throughout the paper in the appendices.

\section{Models of light-matter interaction}\label{sec:formalism}

As discussed above, our goal is to study entanglement harvesting taking into account the electromagnetic interaction of hydrogenlike atoms with the electromagnetic field. This will allow us to go beyond previous studies that used approximated scalar-coupling models, which ignore transfer of angular momentum and possible effects of field polarization in entanglement harvesting.

The first step is to select a realistic model of light-matter interaction. We will first review two scalar models that have been used in the past to approximate the coupling of atomic electrons with the electromagnetic field. Then we will consider and thoroughly discuss the validity of an interaction model where the electron of a hydrogen atom couples dipolarly to the full electromagnetic field. This is the model that we will employ to derive our main results.

\subsection{Scalar coupling: Unruh-DeWitt Hamiltonian}

The interaction between a nonrelativistic electron of momentum $\bm p$ and an electromagnetic field defined by a vector potential $\bm A(\bm x,t)$ and a scalar potential $V(\bm x,t)$ is given by the standard classical minimal coupling Hamiltonian
\begin{equation}
H=\frac{\bm p^2}{2m}-\frac{e}{m}\bm p\cdot\bm A(\bm x,t)+\frac{e^2}{2m}\left[\bm A(\bm x,t)\right]^2+V(\bm x,t)\label{minimalcoupling}.
\end{equation}

This Hamiltonian is behind the so-called light-matter interaction when the electron is bound to an atom.  Instead of using the full Hamiltonian \eqref{minimalcoupling}, it is commonplace in the literature on entanglement harvesting to simplify the interaction and replace the electromagnetic coupling by a linear coupling between the monopole moment $\hat\mu$ of a pointlike two-level system (often referred to as \textit{detector}) and a quantum scalar field $\hat \phi$. This coupling mimicks the $\bm p \cdot \bm A$ term in \eqref{minimalcoupling}. When this model is used to capture the features of the light-matter interaction, it is very common to argue that the scalar potential term in \eqref{minimalcoupling} can be ignored (typically working in the Coulomb gauge) and that the quadratic term $\propto \bm A^2$ can be neglected since it is of higher order in $e$ (for an exception to this see, e.g., \cite{Alhambra2014}). The monopole moment in the interaction picture is given by
\begin{equation}
\hat\mu(t)=\hat\sigma^+e^{\ii \Omega t}+\hat\sigma^-e^{-\ii \Omega t},
\label{monopolemoment}
\end{equation}
where $\Omega$ is the gap between the two detector states. This monopole moment of the detector is then coupled to a scalar field  $\hat\phi$ at the position where the detector is, denoted by $\bm x_d$. The specific form of the interaction Hamiltonian is
\begin{equation}
H_\textsc{udw}=\ac\,\chi(t)\hat\mu(t)\,\hat\phi(\bm x_d,t),
\label{UdW}
\end{equation}
where $\ac$ is the coupling constant and $\chi(t)$ is a switching function that controls the time dependence of the interaction strength. This model of interaction is known as the Unruh-DeWitt particle detector model \cite{DeWittBook}, which has been extensively used in fundamental studies in quantum field theory \cite{Birrell1984}. 

It has been discussed that under the assumption of interactions without exchange of angular momentum the Unruh-DeWitt Hamiltonian captures the main features of the light-matter interaction \cite{Martin-Martinez2013,Alhambra2014}.

Oftentimes, the Unruh-DeWitt particle detector model is upgraded with a detector spatial smearing. The smearing of particle detectors may respond to the need to regularize divergences of the pointlike model \cite{Schlicht2004} or, as for example in quantum optics, to improve on the accuracy of the models of light-matter interaction considering that the atoms are not really pointlike objects, and instead they are localized in the full extension of their atomic wave functions \cite{Martin-Martinez2016,Alhambra2014}. Furthermore, as discussed in \cite{Schlicht2004} spatial smearings are sometimes introduced implicitly in some form of soft UV regularization (see, e.g., \cite{Takagi1986}).

To take into account corrections coming from the finite size of atoms, we can introduce an \textit{ad hoc} spatial profile or \textit{smearing function}, typically strongly supported on a finite spatial region, that controls how much each point of the detector in that region interacts with the field, leading to
\begin{equation}
    H_\textsc{udw}=\ac\chi(t)\int\text{d}^3\bm x\,F(\bm x-\bm x_d)\hat\mu(t)\,\hat\phi(\bm x,t).
\end{equation}

One could argue that to include the atomic orbital wave function geometry, it is natural to think that the spatial support of the atom could be associated with the spatial probability profile of the atomic wave functions \cite{Alhambra2014}. In previous works in entanglement harvesting, different spatial smearings of strong support on a compact region have been studied \cite{Pozas-Kerstjens2015}. Nevertheless, this is a feature that has to be added \textit{ad hoc} to the spatial smearing profile in \eqref{UdW}, since the atomic wave function association with the smearing function does not naturally arise in this simplified model.

\subsection{Derivative coupling}

In the same fashion that the minimal coupling Hamiltonian $\bm p \cdot \bm A$ is simplified into the Unruh-DeWitt Hamiltonian in \eqref{UdW}, the dipole coupling Hamiltonian $\bm d \cdot \bm E$ could be again simplified to a scalar coupling when the interactions involve no exchange of angular momentum. Intuitively, given that the electric field is defined from the electromagnetic vector potential as $\bm E=-\partial_t\bm A$ (in the Coulomb gauge), an interaction Hamiltonian that couples the atomic monopole moment to the time derivative of a scalar field should capture some of the features of the dipole coupling. Concretely, we can think of the following Hamiltonian:
\begin{equation}\label{derivativecoup}
H_{\textsc{udw}_\text{d}}=\ac\chi(t)\int\text{d}^3\bm x\,F(\bm x-\bm x_d)\hat\mu(t)\,\partial_t\hat\phi(\bm x,t).
\end{equation}

This model has also been employed in the analysis of entanglement harvesting \cite{Lin2016}. Additionally, it has been particularly useful in (1+1)-dimensional analyses, where the use of a derivative coupling alleviates IR divergences in the behavior of the Unruh-DeWitt model \cite{Schlicht2004,Juarez-Aubry2014,Martin-Martinez2014}. We see in \eqref{derivativecoup} that, as  in the case of Unruh-DeWitt detectors, a spatial profile can also be introduced \textit{ad hoc} in this case to account for the finite size of the atomic probes \cite{Thinh2016}.

\subsection{Dipole coupling of an atom to the electromagnetic field}

Let us now consider a model for the complete interaction of an atom with a vector electromagnetic field. We begin with the local \textit{dipole coupling} between an electric dipole and an electric field,
\begin{equation} \label{dipolecoupling}
    \hat{\bm d}\cdot\hat{\bm E}=\ac\hat{\bm x}\cdot\hat{\bm E},
\end{equation}
$\ac$ being the dipole's charge, $\hat{\bm x} $ its position operator, and $\hat{\bm E}$ the electric field operator.

This coupling is extensively used in quantum optics to describe the light-matter interaction \cite{ScullyBook}. It is well known that the leading-order contribution to atomic transitions is of a dipole nature and is governed by a term of the form \eqref{dipolecoupling}. The intensity of higher multipole transitions is strongly suppressed and only becomes relevant for transitions forbidden by the dipole selection rules (see for instance Ref. \cite{BransdenBook}). 

Indeed, the dipole coupling is only an approximation for the full electromagnetic interaction of an atomic electron with the electromagnetic field. However, it is discussed in \cite{Lamb1987,ScullyBook} that for realistically small atoms, an approximate gauge transformation yields the dipole coupling out of the full atomic-field coupling [atomic electron minimally coupled to the electromagnetic field vector potential Eq. \eqref{minimalcoupling}]. This approximation may break when the initial state of the field and the atoms is not excited and only for interaction times that are comparable or smaller than the length scale of the atoms \cite{Loukoprep}. However, for interaction times much larger than the light crossing time of the atomic radius the dipole coupling should yield a good approximation even for ground state dynamics. The coupling \eqref{dipolecoupling} is extensively used in atomic physics and quantum optics to successfully reproduce experiments \cite{Sague2007,Kramida2010} and in theoretical proposals \cite{Takagi1986,ScullyBook,Leon2009,Intravaia2015}.

The dipole coupling is also convenient since the atom couples explicitly to a gauge-invariant field observable, and because, when the approximation holds, the gauge choice made for the field degrees of freedom corresponds to the choice made in the conventional textbook solutions of the Schr\"odinger equation for an electron trapped in a Coulomb potential. That is, $\bm A=0$ in the absence of currents (for further discussion see \cite{ScullyBook,Lamb1987} and the multipolar gauge in e.g. \cite{Jackson2002}). Moreover, the fact that the commutator of $\bm E$ satisfies microcausality (vanishes for spacelike separated events) means that the results of \cite{Martin-Martinez2015} for a scalar field can be quickly reproduced here for the electromagnetic interaction and thus the interaction between two atoms through the field as given in \eqref{dipolecoupling} is fully causal. This kind of coupling has also been used in other contexts outside quantum optics, as for instance to analyze the Unruh effect with an electromagnetic field \cite{Takagi1986}, or in quantum friction \cite{Intravaia2015}. 

Let us, for simplicity and for the sake of comparison with the scalar models presented above, assume that only two levels of the atomic structure are relevant in our setup. In that case, the dipole operator enables  transitions between the atomic ground state and one single relevant excited state. When we restrict it to only two levels, the operator $\hat{\bm x}\cdot\hat{\bm E}$ in the interaction picture reads
\begin{equation}
    \hat{\bm x}\cdot\hat{\bm E}=\bra{e}\hat{\bm x}\cdot\hat{\bm E}\ket{g}e^{\ii \Omega t}\ket{e}\bra{g}+\bra{g}\hat{\bm x}\cdot\hat{\bm E}\ket{e}e^{-\ii \Omega t}\ket{g}\bra{e}.
\end{equation}

Inserting resolutions of the identity in the position eigenbasis and noting that $\psi_g(\bm x)=\bra{\bm x}\left.\!g\right>$ and \mbox{$\psi_e(\bm x)=\bra{\bm x}\left.\!e\right>$} are the position representation of the ground and excited level wave functions, respectively, the operator \eqref{dipolecoupling} can be recast as
\begin{align}
    \hat{\bm x}\cdot\hat{\bm E}(\bm x,t)=&\int\text{d}^3\bm x\left[\bm F(\bm x)\cdot\hat{\bm E}(\bm x,t)e^{\ii \Omega t}\ket{e}\bra{g}\right.\notag\\
    &\left.\qquad\,\,+\bm F^*(\bm x)\cdot\hat{\bm E}(\bm x,t)e^{-\ii \Omega t}\ket{g}\bra{e}\right],
\end{align}
where $\hat{\bm E}(\bm x,t)$ is an operator which acts on the field Hilbert space, and the spatial \textit{smearing vector} $\bm{F}(\bm x)$ is defined as
\begin{equation}
\bm F(\bm x)=\psi^*_e(\bm x)\bm x\,\psi_g(\bm x).
\label{smearing}
\end{equation}
Note that in this case, the specific form of the spatial smearing arises naturally from the coupling, and does not have to be inserted \textit{ad hoc}.

By direct comparison with Eq. \eqref{dipolecoupling}, the position-space representation of the dipole moment in the interaction picture can be written as
\begin{equation}
\hat{\bm d}(\bm x,t)=\ac\left[\bm F(\bm x)e^{\ii \Omega t}\hat{\sigma}^++\bm F^*(\bm x)e^{-\ii \Omega t}\hat{\sigma}^-\right],
\label{dipolemoment}
\end{equation}
where we have adopted the usual notation for the SU(2) ladder operators $\hat{\sigma}^+=\ket{e}\bra{g}$ and $\hat{\sigma}^-=\ket{g}\bra{e}$. With this expression for the dipole moment, the Hamiltonian for the dipole interaction reads
\begin{equation}
H_\textsc{em}=\chi(t)\int\text{d}^3\bm x\,\hat{\bm d}(\bm x-\bm x_d,t)\cdot\hat{\bm E}(\bm x,t).
\label{EMHamil}
\end{equation}

We are going to use \eqref{EMHamil} as the model of light-matter interaction with which we will analyze entanglement harvesting. In doing so, we will be able to qualitatively and quantitatively compare the results with previous models that neglected the vector nature of the field.

\section{Setup}\label{sec:setup}

\subsection{Full system dynamics}

Let us consider two hydrogenlike atoms (A and B) that interact locally with the electromagnetic vacuum via the dipole coupling Hamiltonian \eqref{EMHamil},
\begin{equation}
H_\textsc{em}=\sum_{\nu}\chi_\nu(t)\int\text{d}^3\bm x\,\hat{\bm d}_\nu(\bm x-\bm x_\nu,t)\cdot\hat{\bm E}(\bm x,t)
 \label{EMHamiltonian}
\end{equation}
in the interaction picture, where $\nu=\text{A,\,B}$ labels each of the two atoms. Each atom $\nu$ is located in position $\bm x_\nu$. We recall that the dipole moment operator is given by \eqref{dipolemoment}, that is
\begin{equation}
\hat{\bm d}_\nu(\bm x,t)=\ac_\nu\left[\bm F_\nu(\bm x)e^{\ii \Omega_\nu t}\hat{\sigma}_\nu^++\bm F_\nu^*(\bm x)e^{-\ii \Omega_\nu t}\hat{\sigma}_\nu^-\right],
\end{equation}
and
\begin{equation}
\bm F_\nu(\bm x)={\psi^*_e}_\nu(\bm x)\bm x\,{\psi_g}_\nu(\bm x).
\label{smearingvec}
\end{equation}

In addition to the study of the electromagnetic coupling, we will compare the analysis with the other scalar approximations used in past studies of entanglement harvesting, and discussed in the previous section. This way we can draw a fair comparison of the three interaction models (Unruh-DeWitt, derivative scalar coupling and dipole electromagnetic coupling). Again in the interaction picture, the scalar-field interaction Hamiltonians for two detectors coupled to the field are
\begin{equation}
    H_\textsc{udw}=\sum_{\nu}\ac_\nu\chi_\nu(t)\int\text{d}^3\bm x\,F_\nu(\bm x-\bm x_\nu)\hat{\mu}_\nu(t)\hat{\phi}(\bm x,t)
    \label{UdWHamiltonian}
\end{equation}
for the Unruh-DeWitt coupling and    
\begin{equation}
    H_{\textsc{udw}_\text{d}}=\sum_{\nu}\ac_\nu\chi_\nu(t)\int\text{d}^3\bm x F_\nu(\bm x-\bm x_\nu)\hat{\mu}_\nu(t)\partial_t\hat{\phi}(\bm x,t)
    \label{DerivativeHamiltonian}
\end{equation}
for the derivative coupling.

In the three cases in Eqs. \eqref{EMHamiltonian}, \eqref{UdWHamiltonian} and \eqref{DerivativeHamiltonian} we can express the fields in terms of plane-wave mode expansions in the standard way \cite{Birrell1984,ScullyBook}, i.e.,
\begin{align}
\hat{\phi}(\bm{x},t)=&\int\frac{\text{d}^3\bm{k}}{\sqrt{(2\pi)^32|\bm k|}}
\left[\hat{a}_{\bm{k}}e^{\ii k\cdot x}+\text{H.c.}\right],\label{scalar}\\
\partial_t\hat{\phi}(\bm{x},t)=&\int\frac{\text{d}^3\bm{k}}{\sqrt{(2\pi)^3}}\sqrt{\frac{|\bm{k}|}{2}}
\left[-\ii \hat{a}_{\bm{k}}e^{\ii k\cdot x}+\text{H.c.}\right],\label{derivative}\\
\hat{\bm{E}}(\bm{x},t)=&\!\sum_s\!\int\!\!\frac{\text{d}^3\bm{k}}{\sqrt{(2\pi)^3}}\sqrt{\frac{|\bm{k}|}{2}}
\left[-\ii \hat{a}_{\bm{k},s}\boldsymbol\epsilon(\bm{k},s)e^{\ii k\cdot x}\!+\!\text{H.c.}\right],\label{electric}
\end{align}
where $k\cdot x\coloneqq\bm k\cdot\bm x-|\bm k| t$.  Note that each $\hat{a}$ (resp. $\hat{a}^\dagger$) represents an annihilation (resp. creation) operator for a field mode of momentum $\bm k$ and (in the electromagnetic case) polarization $s$. These operators satisfy the following canonical commutation relations:
\begin{align}
[\hat{a}_{\bm{k}},\hat{a}^\dagger_{\bm{k}'}]=&\delta^{(3)}(\bm k - \bm k'),\label{commutationsc}\\
[\hat{a}_{\bm{k},s},\hat{a}^\dagger_{\bm{k}',s'}]=&\delta^{(3)}(\bm k - \bm k')\delta_{s,s'}.\label{commutationem}
\end{align}

In the case of the electromagnetic field, its vector nature is encoded in the set $\left\{\boldsymbol\epsilon(\bm k,s)\right\}_{s=1}^3$ of orthonormal (in general complex) \textit{polarization vectors}. Maxwell's equations force the electric field to satisfy $\nabla\cdot\bm E=0$, which yields the \textit{transversality condition} $\bm k\cdot\boldsymbol\epsilon(\bm k,s)=0$. This constraint restricts the sum over polarizations in Eq. \eqref{electric} to $s=1,\,2$, representing the two physical polarizations of the electromagnetic field, which are typically denoted as $\{\perp_1,\,\perp_2\}$ and satisfy the \textit{completeness relation}
\begin{equation}
\sum_{s=\perp_1,\perp_2}\boldsymbol\epsilon(\bm k,s)\otimes\boldsymbol\epsilon(\bm k,s)=\openone-\frac{\bm k\otimes\bm k}{|\bm k|^2}.\label{completeness}
\end{equation}

For completeness, an explicit derivation of this (otherwise well-known) expression can be seen in appendix \ref{app:completeness}.

The time evolution generated by each respective interaction Hamiltonian \eqref{EMHamiltonian}, \eqref{UdWHamiltonian} and \eqref{DerivativeHamiltonian} can be obtained by a perturbative Dyson expansion of the time-evolution operator
\begin{align}
U=\underbrace{\vphantom{-\ii\int_{-\infty}^{\infty}\!\!\!\text{d}t\,H(t)}\openone}_{U^{(0)}}\underbrace{-\ii\int_{-\infty}^{\infty}\!\!\!\text{d}t\,H(t)}_{U^{(1)}}\underbrace{-\!\!\int_{-\infty}^{\infty}\!\!\!\!\text{d}t\int_{-\infty}^{t}\!\!\!\!\!\!\text{d}t^{\prime}\,H(t)H(t^{\prime})}_{U^{(2)}}+\dots\label{eq:evo}
\end{align}

Therefore, the time evolution of an initial state ${\hat{\rho}}_0$, given by ${\hat{\rho}}=U{\hat{\rho}}_0U^\dagger$, can be written as a perturbative expansion in the overall coupling strength $\ac^{i+j}=\ac_\textsc{a}^i\ac_\textsc{b}^j$,
\begin{equation}
{\hat{\rho}}={\hat{\rho}}_0+{\hat{\rho}}^{(1,0)}+{\hat{\rho}}^{(0,1)}+{\hat{\rho}}^{(2,0)}+{\hat{\rho}}^{(0,2)}+{\hat{\rho}}^{(1,1)}+\mathcal{O}(\ac^3).
\end{equation}

Here the notation ${\hat{\rho}}^{(i,j)}$ represents the correction to the initial state ${\hat{\rho}}_0$ obtained by acting on it with $U^{(i)}$ from the left and with  ${U^{(j)}}^\dagger$ from the right, namely
\begin{equation}
    {\hat{\rho}}^{(i,j)}=U^{(i)}{\hat{\rho}}_0{U^{(j)}}^\dagger.
\end{equation}

We are interested in the harvesting of entanglement from the vacuum state of the field. Therefore, we consider that the initial state of the atoms-field system is
\begin{equation}
{\hat{\rho}}_0=\ket{0}\bra{0}\otimes{\hat{\rho}}_{\textsc{ab},0},
\end{equation}
where $\ket{0}$ is the vacuum state of the field and ${\hat{\rho}}_{\textsc{ab},0}$ is the joint initial state of the atoms. We will be interested in the partial state of the atoms after their interaction with the fields, which is given by
\begin{equation}
{\hat{\rho}}_\textsc{ab}=\trf({\hat{\rho}}),\label{fieldtrace}
\end{equation}
where $\trf$ represents the partial trace over the field degrees of freedom.

This means that the nondiagonal terms in the field produced by the time evolution will not be relevant to our study. In particular, any contribution ${\hat{\rho}}^{(i,j)}$ for which the parities of $i$ and $j$ are different (i.e., any correction with overall coupling constant $\ac^{i+j}$ with $i+j$ odd) will not contribute to the detectors' final state \eqref{fieldtrace}, as long as the initial state of the field  is diagonal in the Fock basis (as it is for the case of the vacuum, or any incoherent superposition of Fock states such as a thermal state).

Furthermore, instead of choosing an odd-parity initial state as in \cite{Valentini1991}, we assume that the two detectors are initially both on their respective ground states as in previous studies of entanglement harvesting from scalar fields \cite{Pozas-Kerstjens2015}. In this fashion, the initial joint state has even parity and reads
\begin{equation}
{\hat{\rho}}_{\textsc{ab},0}=\ket{g_\textsc{a}}\bra{g_\textsc{a}}\otimes\ket{g_\textsc{b}}\bra{g_\textsc{b}}.
\end{equation}

Although they have a completely different spatial structure, the monopole moment operator \eqref{monopolemoment} and the dipole moment \eqref{dipolemoment} have the same internal Hilbert space structure, namely, they are Hermitian linear combinations of the SU(2) ladder operators. This is due to the fact that, for now, we are focusing our study on a particular transition involving only two atomic levels. This in turn means that the time-evolved detectors' density matrix in the three cases discussed will have the same structure (although with notably different coefficients) when written in the basis
\begin{equation}
\{\ket{g_\textsc{a}}\otimes\ket{g_\textsc{b}},\ket{e_\textsc{a}}\otimes\ket{g_\textsc{b}},\ket{g_\textsc{a}}\otimes\ket{e_\textsc{b}},\ket{e_\textsc{a}}\otimes\ket{e_\textsc{b}}\}.
\end{equation}

Namely, the time-evolved density matrix is
\begin{align}
{\hat{\rho}}_\textsc{ab}=&\begin{pmatrix}
1-\mathcal{L}_\textsc{aa}-\mathcal{L}_\textsc{bb} & 0 & 0 & \mathcal{M}^* \\
0 & \mathcal{L}_\textsc{aa} & \mathcal{L}_\textsc{ab} & 0 \\
0 & \mathcal{L}_\textsc{ba} & \mathcal{L}_\textsc{bb} & 0 \\
\mathcal{M} & 0 & 0 & 0
\end{pmatrix}+\mathcal{O}(\ac^4).
\label{state}
\end{align}

At a first sight, one can think that this matrix fails to be positive given the conditions in Ref. \cite{Martin-Martinez2016a}. We show in appendix \ref{app:positivity} that this matrix is indeed positive at leading order, $\mathcal{O}(e^2)$, in perturbation theory.

We take the following conventions: the matrix elements in \eqref{state} will be notated as $\mathcal{L}_{\mu\nu}^\textsc{em}$ and $\mathcal{M}^\textsc{em}$ for the dipole coupling to the electromagnetic field, $\mathcal{L}_{\mu\nu}^\textsc{udw}$ and $\mathcal{M}^\textsc{udw}$ for the scalar Unruh-DeWitt coupling, and $\mathcal{L}_{\mu\nu}^{\textsc{udw}_\text{d}}$ and $\mathcal{M}^{\textsc{udw}_\text{d}}$ for the derivative coupling, respectively. In particular, the functions $\mathcal{L}_{\mu\nu}$ and $\mathcal{M}$ in the electromagnetic case can be evaluated as
\begin{align}
\mathcal{L}^\textsc{em}_{\mu\nu}=&\ac_\mu\ac_\nu\int_{-\infty}^{\infty}\text{d}t_1\int_{-\infty}^{\infty}\text{d}t_2\int\text{d}^3\bm x_1\int\text{d}^3\bm x_2\notag\\
&\times e^{\ii(\Omega_\mu t_1-\Omega_\nu t_2)}\chi_\mu(t_1)\chi_\nu(t_2)\notag\\
&\times{\bm{F}_\nu^*}^{\text{\textbf{t}}}(\bm x_2-\bm x_\nu)\text{\bf W}(\bm x_2,\bm x_1,t_2,t_1)\bm F_\mu(\bm x_1-\bm x_\mu),  \label{Lmununotint}
\end{align}
\begin{align}
\mathcal{M}&^\textsc{em}=-\ac_\textsc{a}\ac_\textsc{b}\int_{-\infty}^{\infty}\text{d}t_1\int_{-\infty}^{t_1}\text{d}t_2\int\text{d}^3\bm x_1\int\text{d}^3\bm x_2\notag\\
&\times\Big[e^{\ii(\Omega_\textsc{a} t_1+\Omega_\textsc{b} t_2)}\chi_\textsc{a}(t_1)\chi_\textsc{b}(t_2)\notag\\
&\quad\times{\bm{F}_\textsc{a}}^{\!\!\text{\textbf{t}}}(\bm x_1-\bm x_\textsc{a})\text{\bf W}(\bm x_1,\bm x_2,t_1,t_2)\bm F_\textsc{b}(\bm x_2-\bm x_\textsc{b})\notag\\
&+e^{\ii(\Omega_\textsc{b} t_1+\Omega_\textsc{a} t_2)}\chi_\textsc{b}(t_1)\chi_\textsc{a}(t_2)\notag\\
&\quad\times{\bm{F}_\textsc{b}}^{\!\!\text{\textbf{t}}}(\bm x_1-\bm x_\textsc{b})\text{\bf W}(\bm x_1,\bm x_2,t_1,t_2)\bm F_\textsc{a}(\bm x_2-\bm x_\textsc{a})\Big],\label{Mnotint}
\end{align}
where ${\bm{F}_\nu}^{\!\!\text{\textbf{t}}}$ and ${\bm{F}_\nu^*}^{\text{\textbf{t}}}$ are respectively the transpose and Hermitian conjugate of the vector ${\bm{F}_\nu}$, the spatial smearing vectors are given by Eq. \eqref{smearing} and the matrix \textbf{W} is the Wightman function 2-tensor of the field, whose components are given by
\begin{equation}
\left[\text{\bf W}\right]_{ij}=W^\textsc{em}_{ij}(\bm x,\bm x',t,t')=\bra{0}E_i(\bm x,t)E_j(\bm x',t')\ket{0}.
\label{Wightman}
\end{equation}

In the scalar cases (see, e.g., \cite{Pozas-Kerstjens2015}), Eqs. \eqref{Lmununotint} and \eqref{Mnotint} have a similar form, but the  corresponding scalar analogues of the vector quantities have to be used. I.e., the smearing vectors $\bm F(\bm x)$ are replaced by the scalar smearing functions $F(\bm x)$, and instead of the Wightman tensor one has to use the scalar Wightman function of the corresponding field,
\begin{align}
   W^\textsc{udw}(\bm x,\bm x',t,t')&=\bra{0}\hat{\phi}(\bm x,t)\hat{\phi}(\bm x',t')\ket{0},\\
   W^{\textsc{udw}_\text{d}}(\bm x,\bm x',t,t')&=\bra{0}\partial_t\hat{\phi}(\bm x,t)\partial_{t'}\hat{\phi}(\bm x',t')\ket{0}\notag\\
   &=\partial_t\partial_{t'}W^\textsc{udw}(\bm x,\bm x',t,t').
\end{align}

\subsection{Quantifying the harvested entanglement}

As stated before, we are interested in analyzing the entanglement acquired between the two atoms after the interaction with the field vacuum. To quantify the entanglement between them we will use the negativity. The negativity of a state represented by a density matrix ${\hat{\rho}}_{\textsc{ab}}$ is defined as the magnitude of the sum of negative eigenvalues of the partially transposed density matrix ${\hat{\rho}}_{\textsc{ab}}^{\text{\textbf{t}}_\nu}$ \cite{Vidal2002}. It is an entanglement monotone that for two-qubit settings only vanishes for separable states. For a state of the form \eqref{state}, the negativity (computed in full detail, for instance, in \cite{Pozas-Kerstjens2015}) is given by \mbox{$\mathcal{N}=\text{max}\left(0,\mathcal{N}^{(2)}\right)+\mathcal{O}(\ac^4)$}, where 
\begin{equation}
    \mathcal{N}^{(2)}=\!-\frac{1}{2} \left(\mathcal{L}_\textsc{aa}+\mathcal{L}_\textsc{bb}-\sqrt{\left(\mathcal{L}_\textsc{aa}-\mathcal{L}_\textsc{bb}\right)^2+4 |\mathcal{M}|^2}\right).
    \label{generalnegativity}
\end{equation}
Note that a naive inspection of the partial transpose of Eq. \eqref{state} as it stands produces the seemingly always-negative eigenvalue $E_2=-|\mathcal{L}_\textsc{ab}|^2$, potentially leading to having the atoms always entangled, regardless of the specific configuration. However, as discussed in \cite{Pozas-Kerstjens2015}, this term is $\mathcal{O}(\ac^4)$; thus the second-order expansion in \eqref{generalnegativity} is not enough to study this subdominant eigenvalue. It was shown in \cite{Pozas-Kerstjens2015} that (for the Unruh-DeWitt model) no relevant changes to the second-order result were obtained when subleading effects were considered. 

We will consider the simple case in which both atoms are identical, which translates into \mbox{$\Omega_\textsc{a}=\Omega_\textsc{b}\equiv\Omega$}, \mbox{$\ac_\textsc{a}=\ac_\textsc{b}\equiv\ac$} and \mbox{$\bm F_\textsc{a}=\bm F_\textsc{b}$}. In this case, \mbox{$\mathcal{L}_\textsc{aa}=\mathcal{L}_\textsc{bb}\equiv\mathcal{L}_{\mu\mu}$}, and Eq. \eqref{generalnegativity} simplifies to
\begin{equation}
\mathcal{N}^{(2)}=\left|\mathcal{M}\right|-\mathcal{L}_{\mu\mu}.\label{negativity}
\end{equation}

This expression allows for a very intuitive interpretation. When $\mu=\nu$, Eq. \eqref{Lmununotint} is a local term: it only depends on the properties of just one of the atoms. On the other hand, Eq. \eqref{Mnotint} depends on the properties of both atoms and therefore is a nonlocal term. This means that in Eq. \eqref{negativity} there is a direct competition between nonlocal, entangling excitations and local noise, leading to the intuitive notion that in order to have entanglement between the atoms, the nonlocal term must overcome the local, single-atom ``noisy'' excitations \cite{Reznik2005,Pozas-Kerstjens2015,Martin-Martinez2016}. It is worth noticing that, as shown in \cite{Martin-Martinez2016a}, for this case of two identical atoms, the second-order correction to the negativity is related to another conventional entanglement measure, the concurrence \cite{Wooters1998}, by the simple relation      $\mathcal{C}^{(2)}=2\mathcal{N}^{(2)}$.

\subsection{Characterization of the atomic model}

We will consider two identical, static, hydrogenlike atoms located at positions $\bm x_\textsc{a}$ and $\bm x_\textsc{b}$, which interact with the field for a finite time scale $T$ implemented through strongly-supported smooth Gaussian switching functions\footnote{Although these switching functions are not strictly compactly supported, whenever we study harvesting in ``spacelike'' regimes (at more than 7--8 sigma away from maximal light contact) we also double-test the results by performing numerical studies where we substitute the Gaussian switchings with a compactified version for which we enforce that $\chi(t)=0$ when $t> 8T/\sqrt{2}$ (i.e., 8-sigma away from the Gaussian peak) and we make sure that the difference is below the numerical double-precision threshold for zero. This was done to guarantee that the results are not an artifact of the switching function Gaussian tails.} 
\begin{equation}
\chi_\nu(t)=e^{-\frac{(t-t_\nu)^2}{T^2}},
\label{switching}
\end{equation}
where $t_{\nu}$ is the center of the Gaussian of atom $\nu$. This choice of switching function arises naturally when one thinks, for instance, of atoms flying transversely through a long optical cavity: the ground transversal mode has precisely a Gaussian profile \cite{SveltoBook}.  

We will consider only two atomic levels, the ground state being the hydrogenoid-$1s$ state \cite{GalindoBook}
\begin{equation}
\psi^\textsc{em}_g(\bm x)=\psi_{100}(\bm x)=R_{10}(|\bm x|)Y_{00}(\theta,\phi)=\frac{1}{\sqrt{\pi a_0^3}}e^{-\frac{|\bm x|}{a_0}},
\label{groundwf}
\end{equation}
where $a_0$ is the generalized Bohr radius.

In the electromagnetic case, under the realistic consideration of hydrogenlike atoms, the smearing functions (smearing vectors) are no longer rotationally invariant. Indeed, the excited state has to be at least a $2p$ level ---which is no longer spherically symmetric--- due to the angular momentum selection rules of the dipole interaction.

This shows already one of the main differences between previous scalar models and the electromagnetic interaction: new features will appear as we consider transitions with exchange of angular momentum that could not possibly be  captured by any scalar model. Furthermore, the lack of rotational invariance of the smearing vectors $\bm F_\nu$ translates into a dependence of entanglement harvesting on the relative orientation of the two atoms. For illustration, we will consider that the excited level is the nonisotropic, $2p_z$ excited state\footnote{Identical results would be obtained for the $2p_x$ and $2p_y$ levels.}:
\begin{equation}
\psi^\textsc{em}_e(\bm x)=R_{21}(|\bm x|)Y_{10}(\theta,\phi)=\frac{1}{\sqrt{32\pi a_0^5}}e^{-\frac{|\bm x|}{2a_0}}|\bm x|\cos\theta.
\label{excitedwf}
\end{equation}

\begin{figure}[h]
    \centering
    \includegraphics[width=0.30\textwidth]{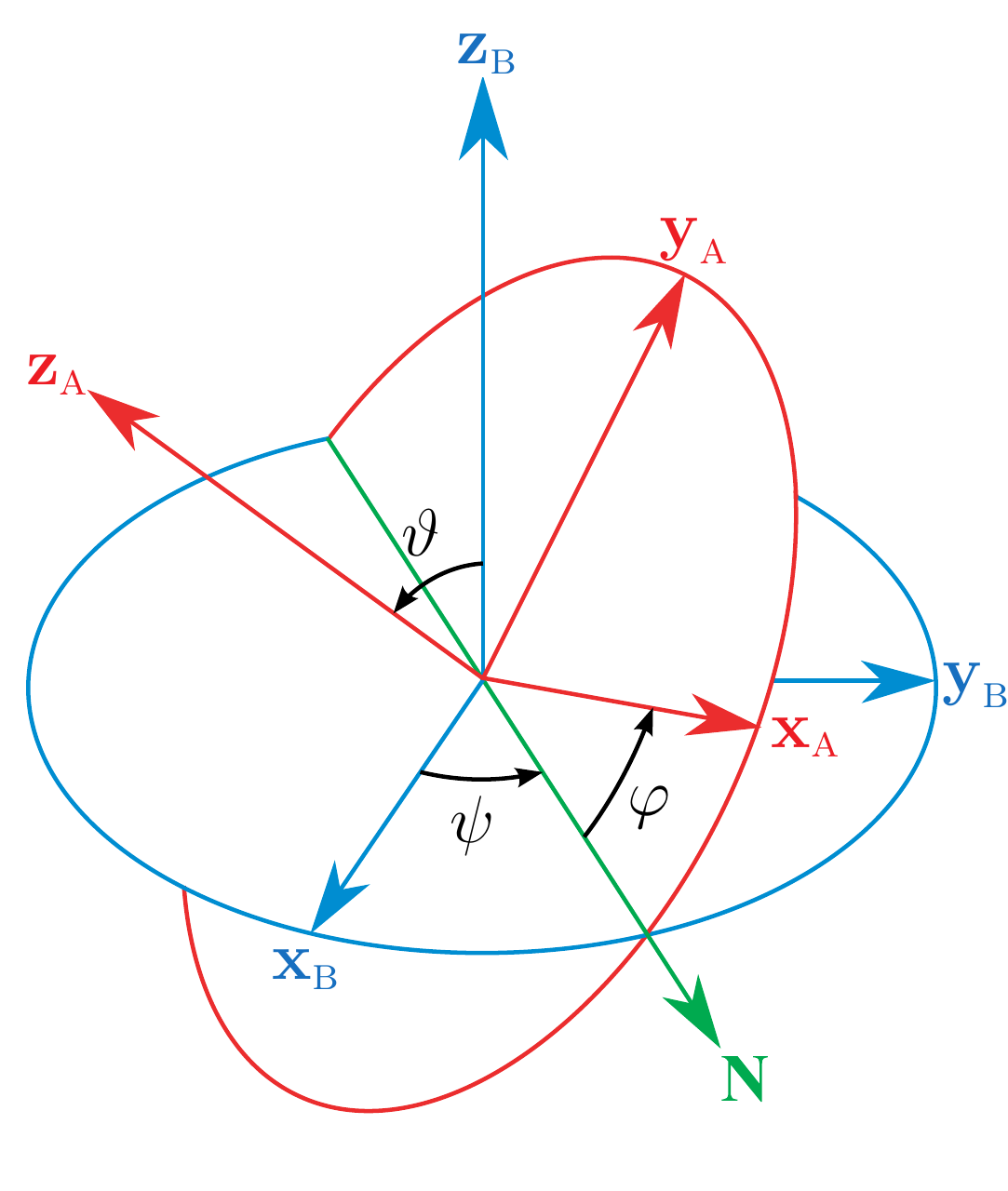}
    \caption{Euler angles characterizing the relative orientation of atom B (blue) with respect to atom A's reference frame (red).}
    \label{fig:eulerangles}
\end{figure}

Equations \eqref{groundwf} and \eqref{excitedwf} completely determine the spatial smearing vector Eq. \eqref{smearing} for each atom:
\begin{align}
   \bm F_\textsc{a}(\bm x)=&\frac{\cos\theta}{4\pi a_0^4\sqrt{2}}e^{-\frac{3|\bm x|}{2a_0}}|\bm x|^2\begin{pmatrix}
   \sin\theta\cos\phi\\
   \sin\theta\sin\phi\\
   \cos\theta
   \end{pmatrix},\\
    \bm F_\textsc{b}(\bm x)=&\frac{\cos \theta \cos \vartheta -\sin \theta \sin \vartheta  \cos (\psi +\phi )}{4\pi a_0^4\sqrt{2}}e^{-\frac{3|\bm x|}{2a_0}}
  \notag\\
  &\times|\bm x|^2\begin{pmatrix}
   \sin\theta\cos\phi\\
   \sin\theta\sin\phi\\
   \cos\theta
   \end{pmatrix},
\end{align}
where the triplet $(\psi,\vartheta,\varphi)$ denotes the three Euler angles defining the relative orientation of the reference frame of atom B with respect to atom A's reference frame (see Figure \ref{fig:eulerangles}). The different angular dependences of the smearing vectors $\bm F_\textsc{a}$ and $\bm F_\textsc{b}$ are due to the fact that we have expressed the reference frame of atom B in terms of the reference frame of atom A. In so doing, the spherical harmonics of atom B (which take a canonical form in its own reference frame) transform to A's frame linearly,
\begin{equation}
   Y^\textsc{b}_{lm}(\theta,\phi)=\sum_{\mu=-l}^l Y^\textsc{a}_{l\mu}(\theta,\phi)\mathcal{D}^l_{\mu,m}(\psi,\vartheta,\varphi),
   \label{changeofharmonics}
\end{equation}
where the different $\mathcal{D}^l_{m_1,m_2}(\psi,\vartheta,\varphi)$ are the Wigner D-functions that characterize the rotation of the angular momentum operators under changes of reference frame. For a more detailed description and the explicit definition of the Wigner D-functions, see appendix \ref{app:longcalcs}.

\section{Results}\label{sec:results}

\subsection{Local noise and correlation term for the electromagnetic field}

Under these assumptions, the calculations are (very) lengthy but straightforward. In appendix \ref{app:longcalcs} we show in very great detail how to compute the local term \eqref{Lmununotint} and the correlation term \eqref{Mnotint} for general transitions between any two arbitrary states of the two hydrogenoid atoms. In particular, for a transition between the $1s$ and the $2p_z$ orbitals, as we also show in the appendix, the two terms involved in Eq. \eqref{negativity} take the following form:
\begin{align}
\mathcal{L}^\textsc{em}_{\mu\mu}=&\ac^2\frac{49152}{\pi} a_0^2T^2\int_0^\infty\text{d}|\bm k|\frac{|\bm k|^3e^{-\frac{1}{2}T^2(\Omega+|\bm k|)^2}}{\left(4 a_0^2 |\bm k|^2+9\right)^6}, \label{LmumuEM}\\
\left|\mathcal{M}\right|^\textsc{em}=&\ac^2\frac{24576\left|\cos\vartheta\right|}{\pi}a_0^2T^2\notag\\
   &\times\bigg|\int_0^\infty\text{d}|\bm k|\,|\bm k|^3e^{-\frac{1}{2} T^2\left(\Omega^2+|\bm k|^2\right)}\notag\\
   &\times\frac{j_{0}(|\bm k||\bm x_\textsc{a}-\bm x_\textsc{b}|)+j_{2}(|\bm k||\bm x_\textsc{a}-\bm x_\textsc{b}|)}{\left(4 a_0^2 |\bm k|^2+9\right)^6}\notag\\
   &\times\left[ E(|\bm k|,t_\textsc{ba})+E(|\bm k|,-t_\textsc{ba})\right]\bigg|, \label{MEM}
\end{align}
where
\begin{equation}
E(|\bm k|,t_\textsc{ba})=e^{\ii|\bm k|t_\textsc{ba}}\text{erfc}\left(\frac{\ii T^2 |\bm k|+t_\textsc{ba}}{\sqrt{2} T}\right),
\end{equation}
$j_l(x)$ are the spherical Bessel functions of the first kind, $\text{erfc}(x)=1-\text{erf}(x)$ is the complementary error function, $t_\textsc{ba}=t_\textsc{b}-t_\textsc{a}$ represents the time delay between the switchings,  and we recall that the Euler angle $\vartheta$ is the relative angle between the axes of symmetry of the $2p_z$ orbitals of atoms A and B.

We see explicitly in  Eq. \eqref{LmumuEM} that the local noise term $\mathcal{L}_{\mu\mu}$ depends only on the properties of only one of the atoms. As expected, the dependences on the relative distance $|\bm x_\textsc{a}-\bm x_\textsc{b}|$ and relative angle between the $2p_z$ orbitals' axes of symmetry $\vartheta$ appear only in the nonlocal correlations term \eqref{MEM}.

\subsection{Orientation dependence of entanglement harvesting}

\begin{figure}[h!]
    \centering
    \includegraphics[width=0.45\textwidth]{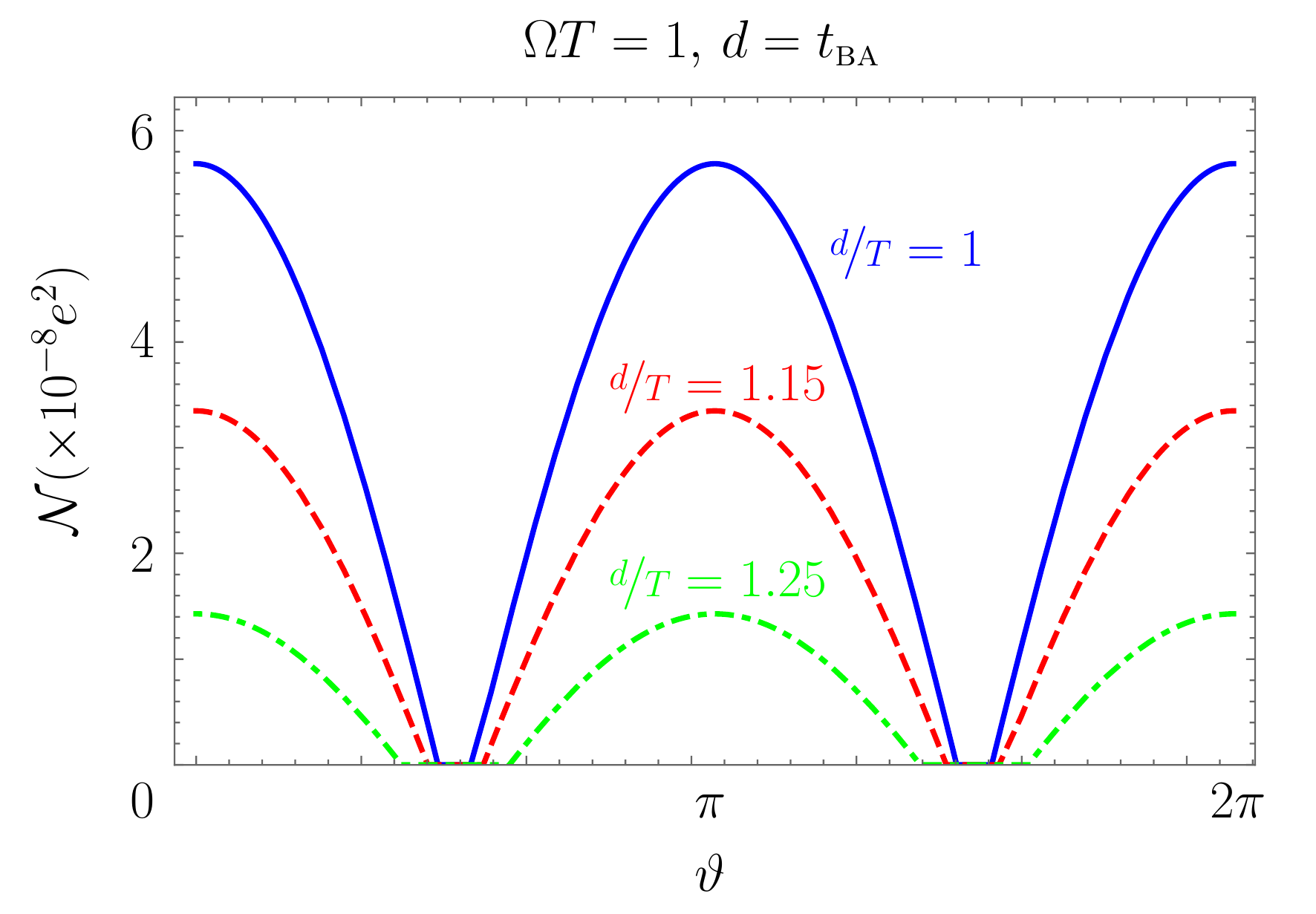}
    \caption{Negativity (leading order) when the two atoms are in full light-contact $d=t_\textsc{ba}$ for two hydrogenoid atoms with energy gap $\Omega T=1$, atomic radius $a_0\Omega=0.001$ and spatial separations (solid blue) $d/T=1$, (dashed red) $d/T=1.15$ and (dashed-dotted green) $d/T=1.25$, as a function of the relative orientation between the atoms.}
    \label{fig:orientation}
\end{figure}

Given that, as discussed above, the orbitals of the excited states are not isotropic, the relative orientation of the atoms is a new, unexplored degree of freedom that has a nontrivial influence on the ability of the atomic probes to harvest entanglement from the field. In Figure \ref{fig:orientation} we show the amount of entanglement that the two atoms harvest when placed in light contact ($d=t_\textsc{ba}$) as the relative angle $\vartheta$ between them varies. As it can be seen from \eqref{generalnegativity} and \eqref{MEM}, the same would apply for atoms in spacelike separation.

In particular, the specific dependence of the nonlocal term \eqref{MEM} on $\vartheta$ implies that atoms oriented along perpendicular axes will not be able to harvest entanglement from the field. Nevertheless, this claim needs to be qualified: what the $\vartheta$ dependence of \eqref{MEM} actually shows is that the atoms cannot harvest entanglement using the transition $1s\rightarrow 2p_z$ when the two $2p_z$ orbitals are perpendicular. Note however that, in principle, entanglement harvesting would be possible through the participation of the two atoms' $2p_x$ and $2p_y$ orbitals, which in general will not be perpendicular even when the two atoms' $2p_z$ orbitals are. Although, rigorously speaking, to assess the entanglement harvested by all the possible transitions to the $2p$ orbitals requires a higher dimensional study, the symmetry of the problem and the perturbative nature of the interaction (which makes multipartite entanglement appear only at higher order in perturbation theory, which is beyond the scope of this paper) allows us to already conclude that there is an optimal orientation of the two atoms that maximizes entanglement harvesting: when the sum of the absolute values of the projections of (the director vectors of) all the axes of B's frame on axes of A's frame is maximal. This optimal case is obtained when the $2p$ atomic orbitals of B are oriented such that their axes of symmetry maximize their angular separation with respect to the $2p$ orbitals of A. Namely, the Euler angles for all 96 maximum harvesting configurations are
\begin{align}
    (\psi,\vartheta,\varphi)&=\left(\frac{\pi}{4}+n\frac{\pi}{2},\vartheta_1,\frac{\pi}{4}+m\frac{\pi}{2}\right),\notag\\
   (\psi,\vartheta,\varphi)&=\left(\frac{\pi}{4}+n\frac{\pi}{2},\pi-\vartheta_1,\frac{\pi}{4}+m\frac{\pi}{2}\right),\notag\\
   (\psi,\vartheta,\varphi)&=\left(\psi_1+n\frac{\pi}{2},\vartheta_2,l\frac{\pi}{2}-\psi_1\right),\notag\\
   (\psi,\vartheta,\varphi)&=\left(\psi_2+n\frac{\pi}{2},\vartheta_2,l\frac{\pi}{2}-\psi_2\right),
\end{align}
for $n,\,m=0\dots3$, $l=1\dots8$, $\vartheta_1\approx1.2310$, $\vartheta_2\approx2.3005$, $\psi_1\approx0.4636$ and $\psi_2\approx1.1071$. Conversely, entanglement harvesting would be minimum when every axis of one atom's reference frame is parallel to an axis of the other atom's frame. In that configuration each axis of B's frame has zero projection onto two of the axes of A's frame.

\begin{figure}[h]
\begin{tabular}{cc}
\includegraphics[width=0.22\textwidth]{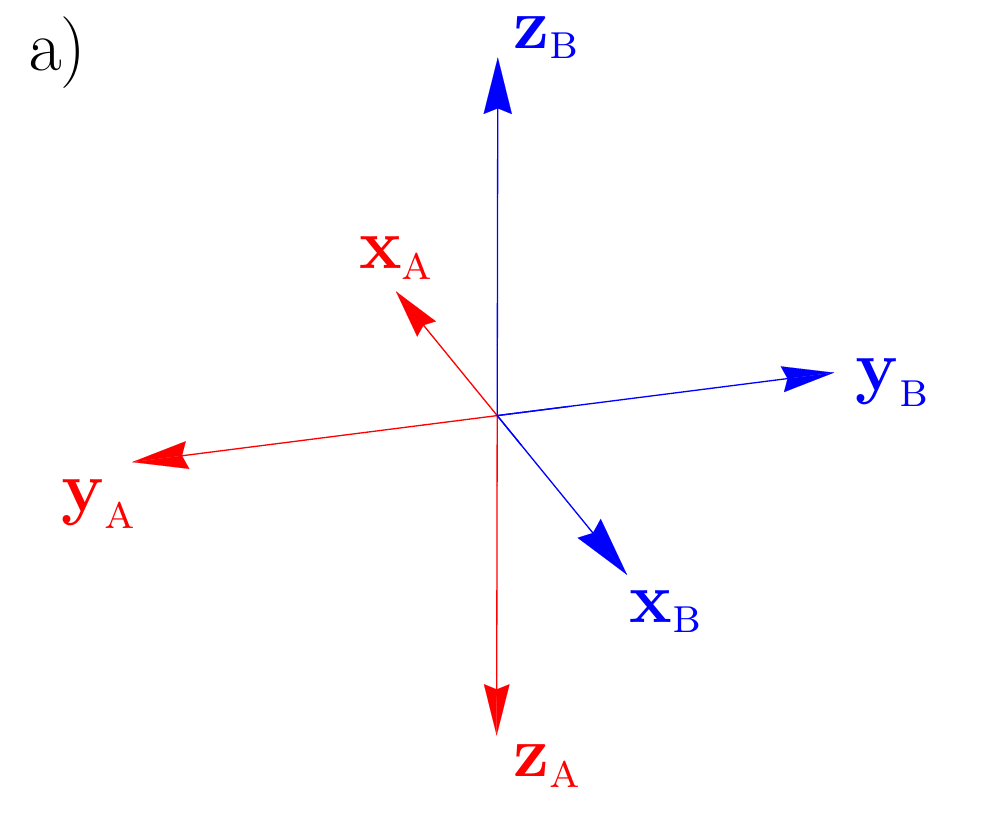} & \includegraphics[width=0.22\textwidth]{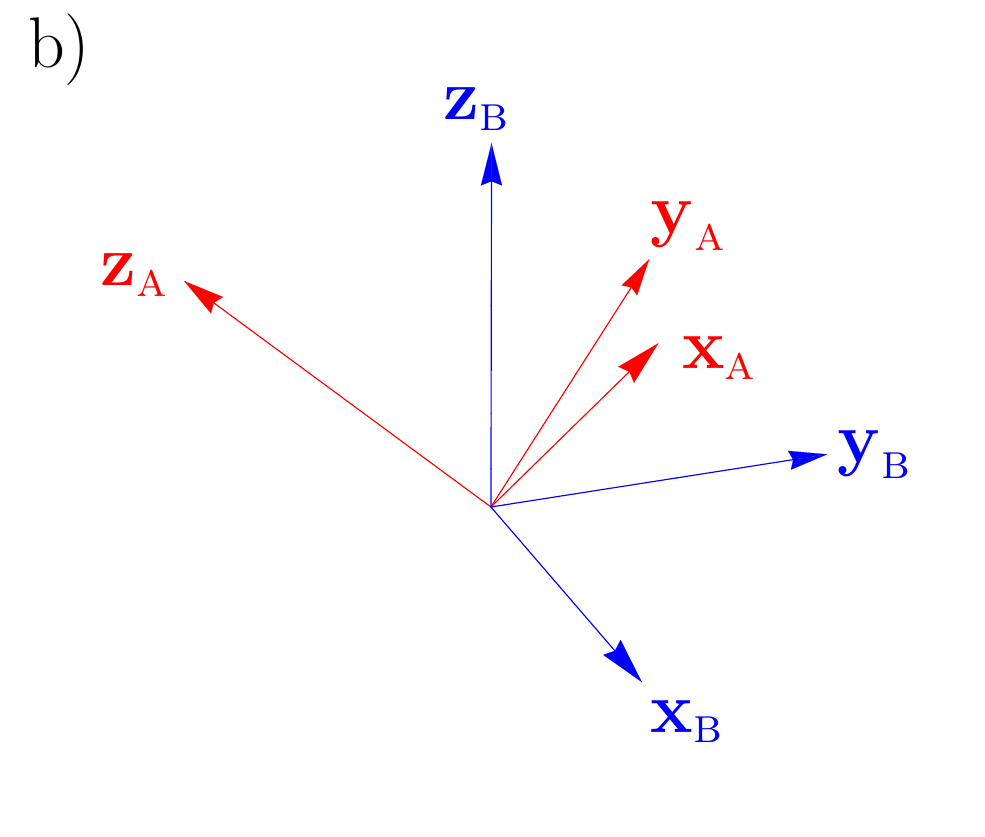}
\end{tabular}
\caption{Examples of relative orientation of the atoms that (a) minimize and (b) maximize the entanglement harvested from the field when all the $1s\rightarrow2p$ transitions are considered. The Euler angles for the configuration of maximal harvesting shown are $(\pi/4,1.231,\pi/4)$.}
\label{fig:maxminharvest}
\end{figure}

In Figure \ref{fig:maxminharvest} we show examples of configurations which minimize and maximize the entanglement harvested.

\subsection{Quantitative analysis}

We show in this section  the most relevant features of entanglement harvesting from an  electromagnetic field to two identical, hydrogenlike atoms.

We set realistic values for the parameters of the model. For the hydrogen atom's $1s\rightarrow 2p$ transition (of wavelength $\Lambda\sim 100$ nm and for $a_0\sim0.1$ nm), in natural units $a_0\Omega=a_0\Lambda^{-1}=0.001$. With these parameters we can compute the negativity from Eqs. \eqref{generalnegativity}, \eqref{LmumuEM} and \eqref{MEM}. 

We are first going to analyze how often it is possible to harvest entanglement from the field to the two atoms. For comparison, we recall  \cite{Pozas-Kerstjens2015}, where it was shown that, in the scalar case, it is possible to harvest entanglement with arbitrarily distant detectors using a smooth switching and scaling up the energy gap consistently.

\begin{figure}[h!]
    \includegraphics[width=0.42\textwidth]{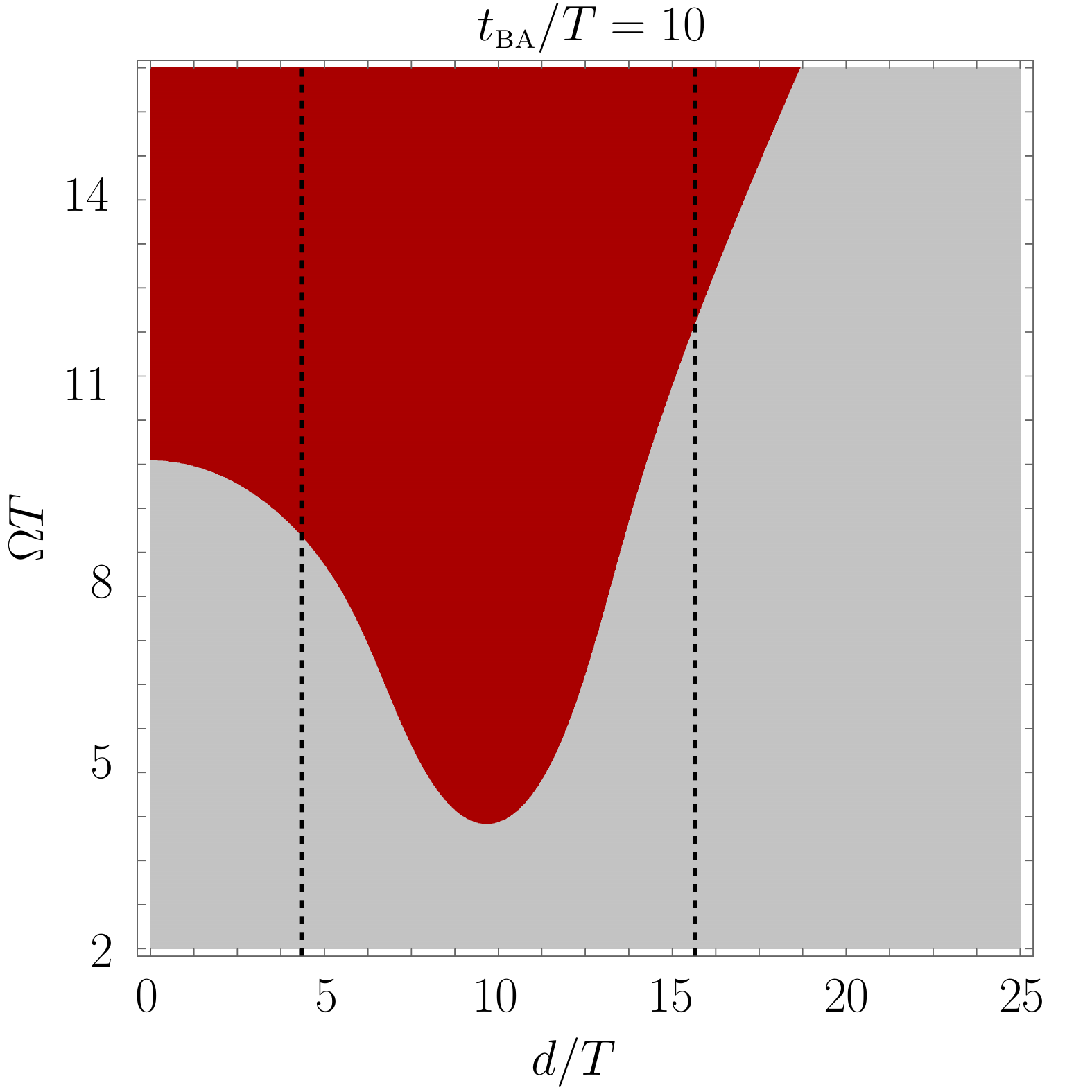}
    \caption{Binary plot showing in (dark) red the values of $d/T$ and $\Omega T$ for which entanglement harvesting is possible, and in (light) grey the values for which there is zero entanglement harvesting. We consider $a_0\Omega=0.001$ (hydrogen proportions) and that the atomic interactions with the field are delayed by a time delay  $t_\textsc{ba}/T=10$. The dashed black lines represent the boundaries of the regions of the plot where  atom B is in the lightcone of atom A. We have located these boundaries  at $8$ standard deviations of the switching function \eqref{switching}, that is $d=t_\textsc{ba} \pm 8T/\sqrt{2}$. Notice that we can harvest entanglement from arbitrarily far away distances by increasing the transition energy gap.}
    \label{fig:rngEM}
\end{figure}

In Figure \ref{fig:rngEM} we show in dark red the values of energy gap $\Omega T$ of each atom, and the spatial distance $d/T$ between them, for which entanglement can be harvested from the electromagnetic vacuum. In the figure, the delay between the switchings is set to $t_\textsc{ba}/T=10$.

Figure \ref{fig:rngEM} shows that harvesting of entanglement is possible even from regions that are arbitrarily spacelike separated, provided that the atoms have an energy gap that is large enough: given an arbitrarily large distance between the atoms, one can consider a transition energy large enough so that the atoms can harvest entanglement. This is similar to the case discussed in \cite{Pozas-Kerstjens2015} for the scalar field Unruh-DeWitt entanglement harvesting: higher energy gaps reduce the impact of the local noise, making it possible to harvest entanglement for arbitrarily separated atoms. To increase the gap of realistic atomic species one could consider higher energetic transitions $1 s\rightarrow n\,p$, for which the product  $a_0 \Omega=0.001$ does not scale with $n$.

However, increasing the gap to compensate for an increase of spatial separation has a downside: higher energy gaps also suppress the nonlocal correlation term $\mathcal{M}$ (albeit more slowly than $\mathcal{L}_{\mu\mu}$), thereby decreasing the total amount of entanglement harvested, leading to the same  ``damping-and-leakage'' effect described for the Unruh-DeWitt case \cite{Pozas-Kerstjens2015}.

\begin{figure*}[t]
\begin{tabular}{cc}
    \includegraphics[width=0.45\textwidth]{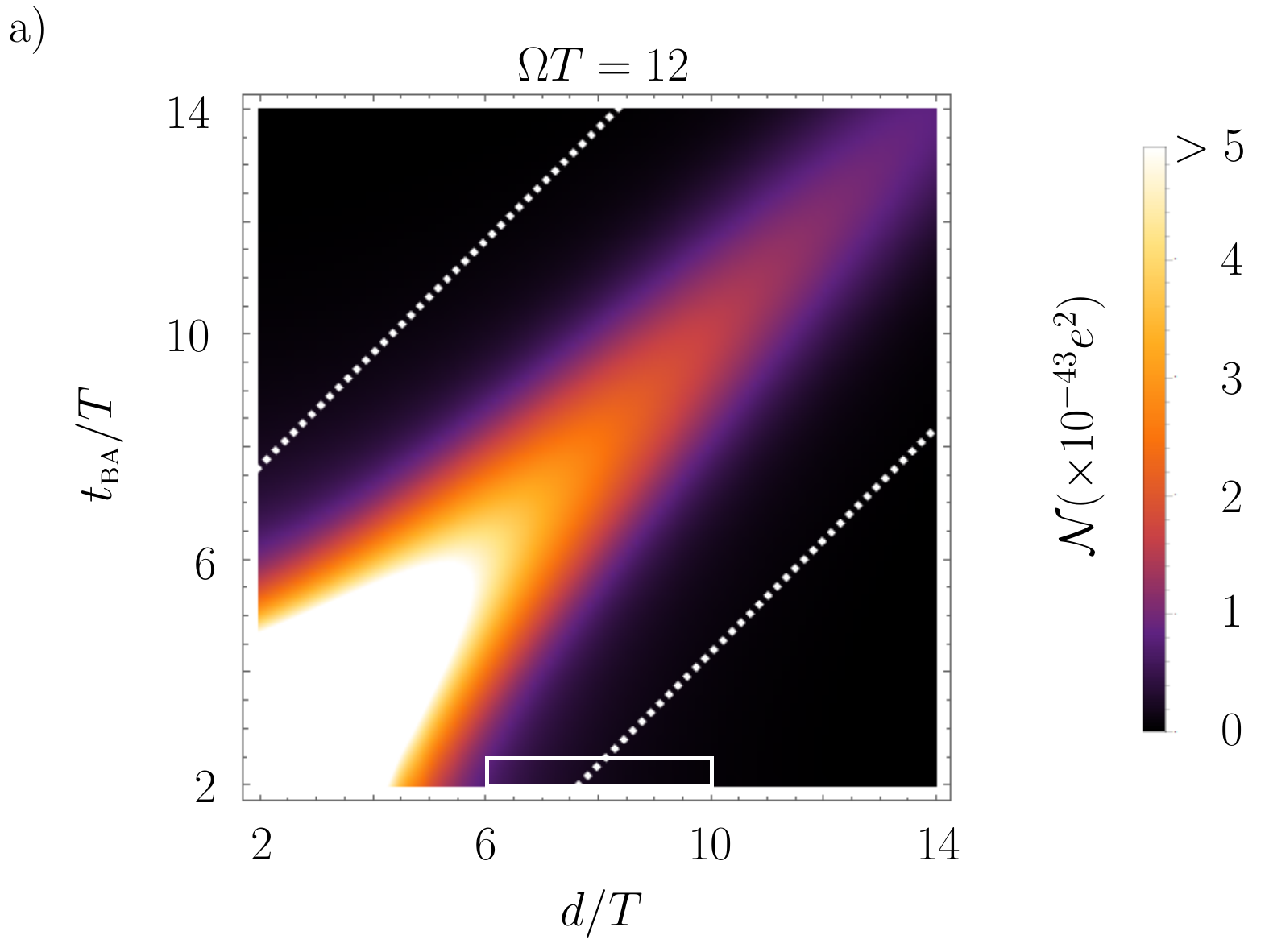} &  \includegraphics[width=0.45\textwidth]{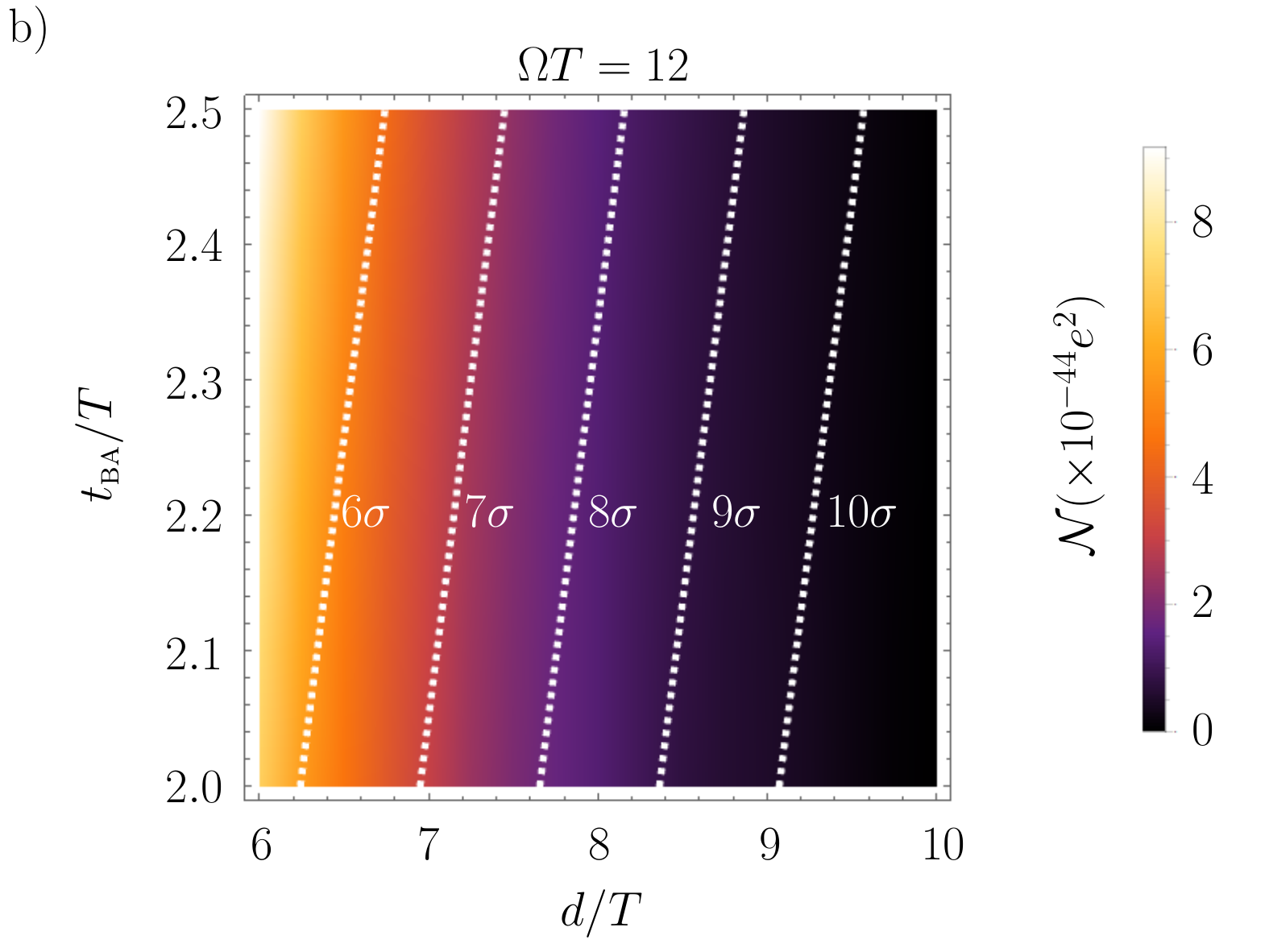}
\end{tabular}
\caption{(a) Negativity to leading order as a function of spatial distance $d$ and time delay between the atoms' interactions $t_{\textsc{ba}}$ of two hydrogenoid atoms satisfying $a_0\Omega=0.001$ for a fixed energy gap of $\Omega T=12$. The white, dashed lines represent the boundaries of the lightcone, located at $d=t_\textsc{ab}\pm 8T/\sqrt{2}$. (b) Zoom-in on the area marked with a white rectangle in (a), to demonstrate spacelike entanglement harvesting. The white dashed lines represent the distance to the point of light contact $d=t_\textsc{ab}$ in number of standard deviations $\sigma=T/\sqrt{2}$. Both for the Gaussian switching and the compactly supported cropped version of the Gaussian switching at 8$\sigma$ we observe spacelike entanglement harvesting.}
    \label{fig:spacetimeEM}
\end{figure*}

In Figure \ref{fig:spacetimeEM} we show the amount of entanglement that can be harvested from the field for a relatively large value of $\Omega T=12$ as a function of the distance between the atomic centers of mass $d/T$ and the time delay of their (recall local) interactions with the field $t_\textsc{ba}/T$. 

As expected given the results on harvesting from scalar fields \cite{Pozas-Kerstjens2015}, the entanglement is maximum inside the lightcone, where direct communication between the atoms can be achieved via exchange of real photons. Moving away from the lightcone the entanglement decreases with spatiotemporal distance, eventually leaking into the region of spacelike separation [shown in Figure \ref{fig:spacetimeEM}(b)].  Entanglement harvesting also decays very quickly with the spatial separation of the atoms, as it can be seen in Figures \ref{fig:orientation} and \ref{fig:spacetimeEM}(a).

One may perhaps be tempted to ascribe the harvested entanglement in the region deemed spacelike separated in Figure \ref{fig:spacetimeEM}(b) to the overlap of the atomic wave functions. After all, atomic orbitals are not compactly supported, and as such, the atoms are `never' completely spacelike separated. In Figure \ref{fig:spacetimeEM}(b), we see that there is entanglement harvesting when the atoms are separated by $9\sigma$ [where we recall $\sigma = T/\sqrt{2}$ where $T$ is the interaction time scale as given in \eqref{switching}]. At these points, the spatial separation of the atoms is $d\approx 10^4 a_0$ (we recall that $a_0=0.001\Omega^{-1}$). With these numbers, the atomic wave function of atom A in the region of the center of mass of atom B is suppressed by an exponential factor $e^{-10^4}$, and indeed the overlap of the two wave functions yields $\int\text{d}|\bm x| |\bm x|^2 \psi^\textsc{a}(|\bm x|)\psi^\textsc{b}(|\bm x|)\approx 10^{-4343}$. Although the entanglement harvested $\mathcal{N}$ at this distance is very small, it is still several thousands of orders of magnitude too large to be explained by the overlap of the atomic wave functions. Note that these results were obtained both for a Gaussian switching function and for a compactly supported Gaussian-like switching function that is made zero at $t-t_\nu=\pm8\sigma=\pm8T/\sqrt{2}$ (that is, at $8$ standard deviations of the center of the Gaussian), so there is no appreciable light-like contact due to the switching either. Due to the absence of direct contact between the atoms when they are spacelike separated, the entanglement is really harvested from that existing previously in the electromagnetic field. The reason why this entanglement is small is because entanglement harvesting will only be relevant in atomic physics when the atoms are separated by several times the atomic sizes (as was already discussed in \cite{Martin-Martinez2016}).

\subsection{Comparison with the scalar models}\label{sec:comparison}

Finally, let us compare the electromagnetic coupling results with  entanglement harvesting from scalar fields using the Unruh-DeWitt and derivative coupling models. Recall that in these scalar cases the smearing function needs to be introduced \textit{ad hoc} in the model. In order to compare the electromagnetic coupling with the two scalar cases \eqref{UdWHamiltonian} and \eqref{DerivativeHamiltonian} we need to be mindful of the scalar nature of the coupling that prevents transfer of angular momentum through the field. As a consequence, in the cases of harvesting from scalar fields, the transitions \mbox{$1s\rightarrow2p$} are not allowed (in a similar way that a $1s\rightarrow 2s$ transition is not allowed in the dipole electromagnetic case). Thus, in order to keep the comparison fair, we consider the most similar transition allowed by the scalar selection rules, which in this case would be between the two spherically symmetric levels of the hydrogenoid atom with lowest energy, namely the $1s$ and $2s$ states, whose position representation wave functions are given by 
\begin{align}
\psi_g(\bm x)&=\frac{1}{\sqrt{\pi a_0^3}}e^{-\frac{|\bm x|}{a_0}},\\
\psi_e(\bm x)&=\frac{1}{4\sqrt{2\pi a_0^3}}e^{-\frac{|\bm x|}{2a_0}}\left(2-\frac{|\bm x|}{a_0}\right).
\end{align}

Although obvious, one way of seeing directly from these expressions that a transition between these two levels is forbidden in the electromagnetic dipole coupling is to realize that these two levels have equal (even) parity, so the spatial smearing vector Eq. \eqref{smearing} (which has odd overall parity) vanishes when integrated over all space. 

As the smearing function in the scalar case we propose to use the scalar analogue to the smearing vector \eqref{smearing}, which associates the smearing with the two-level wave functions. Such an appropriate definition for the scalar cases with  overall even parity is 
\begin{equation}
F(\bm x)=\psi_e(\bm x)\psi_g(\bm x)=\frac{1}{4\pi a_0^3\sqrt{2}}e^{-\frac{3|\bm x|}{2a_0}}\left(2-\frac{|\bm x|}{a_0}\right).
\label{scalarsmearing}
\end{equation}

Considering this analysis, the two terms involved in Eq. \eqref{negativity} take the following form:

\begin{align}
\mathcal{L}^\textsc{udw}_{\mu\mu}\!\!=&\ac^2 \frac{32768}{\pi} a_0^4 T^2\int_0^\infty\text{d}|\bm k|\frac{|\bm k|^5e^{-\frac{1}{2}T^2(\Omega+|\bm k|)^2}}{\left(4 a_0^2 |\bm k|^2+9\right)^6}, \label{LmumuSC}\\
\left|\mathcal{M}\right|^\textsc{udw}\!\!=&\ac^2 \frac{16384}{\pi|\bm x_\textsc{a}-\bm x_\textsc{b}|} a_0^4 T^2\bigg|\int_0^\infty\!\!\!\text{d}|\bm k|\,|\bm k|^4e^{-\frac{1}{2} T^2 \left(\Omega^2+|\bm k|^2\right)}\notag\\
   &\times\!\frac{\sin(|\bm k||\bm x_\textsc{a}\!-\!\bm x_\textsc{b}|)}{\left(4 a_0^2 |\bm k|^2+9\right)^6}\left[ E(|\bm k|,t_\textsc{ba})\!+\!E(|\bm k|,-t_\textsc{ba})\right]\!\bigg|, \label{MSC}
\end{align}
\begin{align}
\mathcal{L}^{\textsc{udw}_\text{d}}_{\mu\mu}=&\ac^2 \frac{32768}{\pi} a_0^4 T^2 \int_0^\infty\text{d}|\bm k|\frac{|\bm k|^7e^{-\frac{1}{2}T^2(\Omega+|\bm k|)^2}}{\left(4 a_0^2 |\bm k|^2+9\right)^6}, \label{LmumuSCD}\\
\left|\mathcal{M}\right|^{\textsc{udw}_\text{d}}\!=&\ac^2 \frac{16384}{\pi|\bm x_\textsc{a}-\bm x_\textsc{b}|} a_0^4 T^2\bigg|\int_0^\infty\!\!\text{d}|\bm k|\,|\bm k|^6e^{-\frac{1}{2} T^2\left(\Omega^2+|\bm k|^2\right)}\notag\\
   &\times\!\frac{\sin(|\bm k||\bm x_\textsc{a}\!-\!\bm x_\textsc{b}|)}{\left(4 a_0^2 |\bm k|^2+9\right)^6}\!\left[ E(|\bm k|,t_\textsc{ba})\!+\!E(|\bm k|,-t_\textsc{ba})\right]\bigg|. \label{MSCD}
\end{align}

We see that these two pairs of expressions are very similar. Comparing on one hand Eqs. \eqref{LmumuSC} and \eqref{LmumuSCD}; and on the other hand  Eqs. \eqref{MSC} and \eqref{MSCD}, we see that the expressions in each pair only differ in an extra factor of $|\bm k|^2$ on the integrals in momentum for the derivative coupling case. This was easy to anticipate given the derivative nature of the coupling. Other than that, the expressions of the matrix elements of ${\hat{\rho}}_{\textsc{ab}}$ are identical in the Unruh-DeWitt model and in the derivative coupling model.

It is much more interesting to compare Eqs. \eqref{LmumuSC}, \eqref{LmumuSCD}, \eqref{MSC} and \eqref{MSCD} with their electromagnetic counterparts \eqref{LmumuEM} and \eqref{MEM}. There are not many significant differences in the local terms $\mathcal{L}_{\mu\mu}$: the only changes stem from the different dimensions of the couplings.

The case of the nonlocal term $\mathcal{M}$ is much more interesting, since it is here where the anisotropy of the excited levels and the vector nature of the electromagnetic field become apparent. The anisotropy of the excited levels can be seen through the factor $\cos\vartheta$ in Eq. \eqref{MEM}, where, as discussed above, $\vartheta$ is the relative angle between the $2p_z$ orbitals of the two atoms [see Figure \ref{fig:eulerangles}, Eq. \eqref{changeofharmonics} and appendix \ref{app:longcalcs}].

In addition to the anisotropy of entanglement harvesting, the vector nature of the interaction manifests itself in the appearance of the spherical Bessel functions $j_0(|\bm k| d)$ and $j_2(|\bm k| d)$ in \eqref{MEM}, substituting the scalar cases' factor  $\sin(|\bm k| d)/(|\bm k| d)$ (with $d=|\bm x_\textsc{a}-\bm x_\textsc{b}|$). Notice that $\mathcal{M}$ is the $\proj{g_\textsc{a}}{e_\textsc{a}}\otimes\proj{g_\textsc{b}}{e_\textsc{b}}$ matrix element of the density matrix ${\hat{\rho}}_{\textsc{ab}}$, so the subindices $l=0,2$ of the spherical Bessel functions in the electromagnetic case correspond exactly with the even values of possible total angular momentum that one can obtain by combining the angular momentum of the dipole coupling term $\bm d\cdot \bm E$, $l_d=1$, with the excited and ground atomic states' angular momenta $l_g=0$, $l_e=1$.

Furthermore, another way to see that the appearance of higher-order spherical Bessel functions is precisely the difference introduced by the exchange of angular momentum between the two atoms is the following: the $j_0$ summand in the integral \eqref{MEM} yields the same qualitative features of the scalar models [since  \mbox{$j_0(|\bm k| d)=\sin(|\bm k| d)/(|\bm k| d)$}]. We can wave our hands and roughly say that if we were to restrict the two atomic levels to be spherically symmetric ($l_g=l_e=0$) and thus naively replace the coupling term by the scalar case (with a zero angular momentum $l_s=0$) the only possible value of total angular momentum would be, of course, $l_e\,\oplus\, l_g\, \oplus\, l_s=0$. Under this light, it is not surprising that we recover the same behavior as in the scalar cases just by simply neglecting the contribution of the higher-order Bessel functions in the electromagnetic case.

\begin{figure}[h!]
\includegraphics[width=0.45\textwidth]{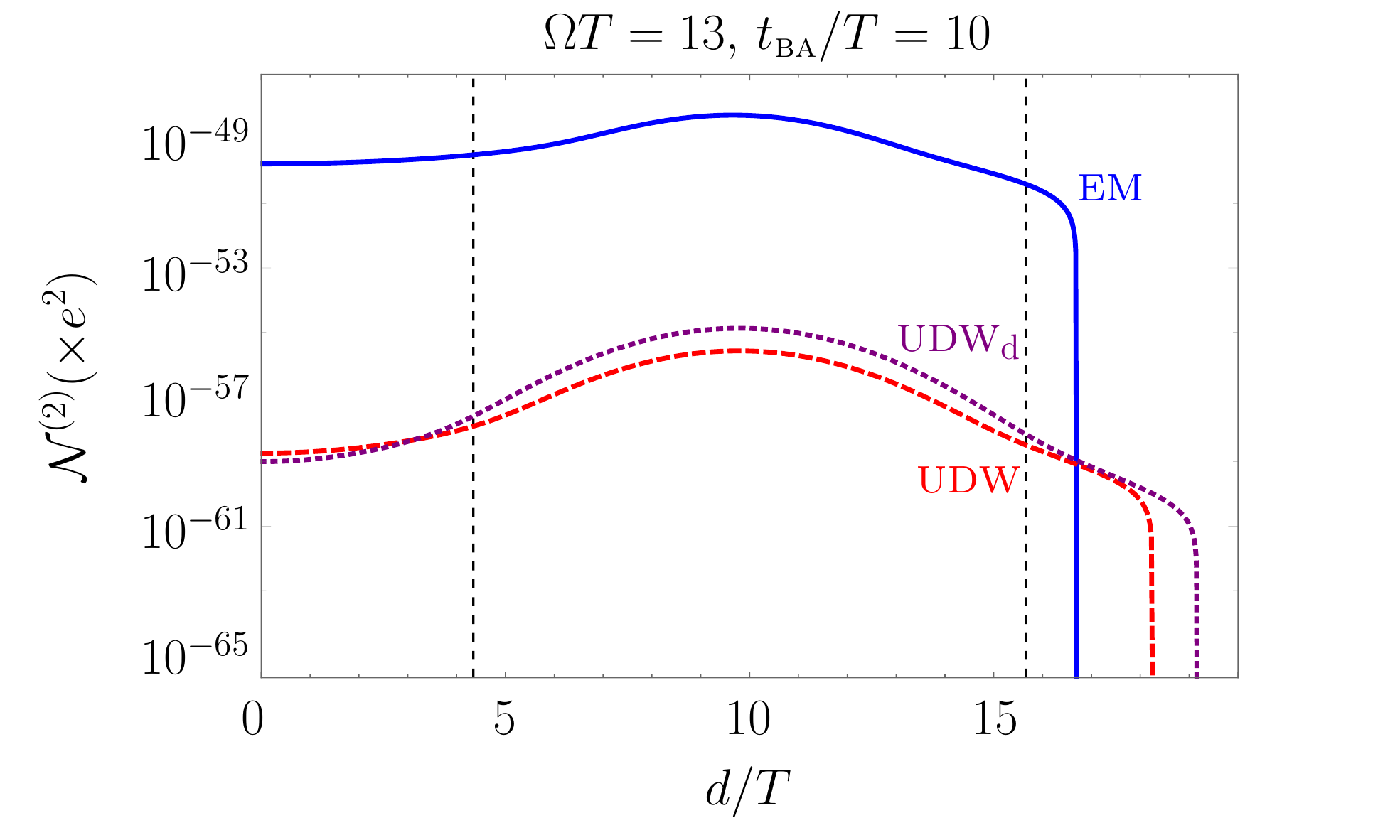}
\caption{Negativity as a function of the spatial separation between the detectors in the different models studied: (blue, solid) hydrogen atom dipolarly coupled to the electromagnetic field, (red, dashed) monopole spherically symmetric detector coupled to a scalar field and (purple, dotted) scalar derivative coupling. In the electromagnetic coupling case, the atomic $2p$ orbitals are parallel. For all models, the energy gap is chosen to be $\Omega T=13$, the detectors' size is given by $a_0\Omega=0.001$ and the time delay between the switchings is $t_\textsc{ba}/T=10$. The black dashed lines represent the boundaries of the lightcone, located at $d=t_\textsc{ba}\pm 8T/\sqrt{2}$. Electromagnetic harvesting is stronger, but has a shorter reach, than the scalar cases. Note that the sharp drop in all cases is due to the logarithmic scale.}
\label{fig:compare}
\end{figure}

Indeed, in a rough qualitative analysis, comparing the results reported in \cite{Pozas-Kerstjens2015} with Figures \ref{fig:rngEM} and \ref{fig:spacetimeEM} we can conclude that the scalar models capture the essence of the phenomenon of harvesting if we do not pay attention to the effects introduced due to the exchange of angular momentum. However, under closer scrutiny, both the amount of entanglement harvested and the distance at which the atoms can still harvest entanglement are significantly different in the different models. As an illustration of this, in Figure \ref{fig:compare} we show the values of the negativity in the three models as a function of spatial separation between the detectors, for the same values of the energy gap $\Omega T=13$ and time delay between the switchings $t_\textsc{ba}/T=10$.

Figure \ref{fig:compare} shows (1) that in the dipole model the amount of entanglement harvested from the field when the atoms' symmetry axes are parallel is much higher than in the scalar models, and (2) in the scalar models it is possible to harvest entanglement when the atoms are further away than in the dipole, electromagnetic model. Regarding the two scalar models, there exist no major differences between them, which again points to the fact that the main differences appear when we incorporate the exchange of angular momentum to the model rather than when we change from an amplitude coupling to a derivative coupling.

\section{Conclusions}\label{sec:conclusions}

We have analyzed the harvesting of entanglement from the electromagnetic vacuum using two fully-featured hydrogenlike atoms. We have done so in regimes when the two atoms are time-, light- and spacelike separated during their interaction with the field.

In stark contrast to previous work ---where the light-matter interaction was modeled by spherically symmetric Unruh-DeWitt-like detectors coupled to a scalar field--- we have analyzed the unique features emerging when we consider the vector nature of the electromagnetic interaction in the atom-field dynamics, also taking into account the atomic orbital geometry. 

In particular, there are two features that could not be studied with previously-employed idealized models which relied on scalar simplifications of the interaction: (1) the effect of the relative orientation of the atoms on their ability to harvest entanglement, and (2) the effect of the exchange of angular momentum between the atoms and the electromagnetic field.

Regarding the relative orientation between the atoms, we showed that considering only the entanglement harvested through the $1s\rightarrow 2p_z$ transition, the amount of entanglement harvested depends on the mutual projection of the axes of symmetry of the $2p_z$ orbitals of both atoms. As a consequence, atoms oriented along perpendicular axes are not capable of harvesting any entanglement with that transition at all. Extending the analysis to consider all equally possible $1s\rightarrow 2p$ transitions, we discussed the relative spatial configurations of the two atoms that maximize and minimize the entanglement harvested through these transitions, which correspond to ground-to-first excited state transitions of hydrogenoid atoms.

As for the exchange of angular momentum, we showed how it manifests itself through the appearance of higher-order spherical Bessel functions in the two-atom correlation terms. We discussed that it is precisely this exchange of angular momentum the responsible for the difference between previous idealized scalar models and the full electromagnetic model. This exchange of angular momentum affects entanglement harvesting in two different ways: (1) it permits the harvesting of more entanglement than the analogue scalar models and (2) at the same time, it constrains the spatial range in which entanglement can be harvested. In any case, we show that harvesting of entanglement with atoms placed in arbitrarily far away points is possible for large enough values of the atomic energy gaps, paying the price of a reduction in the total amount of harvested entanglement as the distance increases. Note that, since this is a fundamental study, in this manuscript we have focused on large time delays and large energy gaps for the scenarios analyzed in the figures, and thus we obtained low values of harvested entanglement. It has been studied elsewhere \cite{Retzker2005,Olson2011,Olson2012,Sabin2012,Martin-Martinez2016} that the value of harvested entanglement (even with the less powerful scalar cases) can be made large enough to be detectable under realistic parameters for shorter delays and lower gaps.

Furthermore, we have compared the dipole coupling model with previous scalar models consistently used in the literature on entanglement harvesting. On the one hand, we have shown  that these models do describe the phenomenon qualitatively when we discard the contributions coming from the exchange of angular momentum in the full electromagnetic coupling. On the other hand we have shown quantitatively how those toy models fail to capture some of the features of more realistic field-atom interactions with anisotropic atomic transitions and nonzero angular momentum exchange.

Finally, the formalism developed in this article and its appendices allows to readily generalize the model to include any arbitrary transition in the study of atomic entanglement harvesting from the electromagnetic vacuum.
\vspace*{0.3cm}

\begin{acknowledgments}
The authors gratefully acknowledge Luis J. Garay for
his comments on the final version of this paper. The authors also thank Pablo Rodriguez-Lopez for helpful discussions. The work of E. M.-M. is supported by the National Sciences and Engineering Research Council of Canada through the Discovery program. \mbox{A. P.-K.} acknowledges financial support from Fundaci\'on Obra \mbox{Social} \mbox{``la Caixa''}, Fundaci\'o Privada Cellex and the Spanish MINECO Severo Ochoa program (SEV-2015-0522). Both authors acknowledge Lionel Brits for the base image used for Figure \ref{fig:eulerangles}, distributed under the Creative Commons BY 3.0 license.
\end{acknowledgments}

\appendix

\begin{widetext}
\section{Completeness relation of the polarization vectors} \label{app:completeness}

In this appendix we derive explicitly the completeness relation \eqref{completeness}, following \cite{Cohen-TannoudjiQEDBook}. Recall that the polarization basis $\left\{\boldsymbol\epsilon(\bm k,s)\right\}_{s=1}^3$ is an orthonormal basis of $\mathbb{R}^3$. We can choose $\boldsymbol\epsilon(\bm k,3)\propto\hat{\bm k}$, the other two unit vectors being mutually orthogonal and perpendicular to this, which we denote $\boldsymbol\epsilon(\bm k,\perp_1)$ and $\boldsymbol\epsilon(\bm k,\perp_2)$. With this choice the completeness relation Eq. \eqref{completeness}, when written in components, reads
\begin{align}
\sum_{s=\perp_1,\perp_2}&\epsilon_j(\bm k,s)\epsilon_l(\bm k,s)=\epsilon_j(\bm k,\perp_1)\epsilon_l(\bm k,\perp_1)+\epsilon_j(\bm k,\perp_2)\epsilon_l(\bm k,\perp_2).
\end{align}

Now, using an auxiliary orthonormal basis of $\mathbb{R}^3$ $\left\{\mathbf{e}_i\right\}_{i=1}^3$, the sum can be recast as
\begin{align}
\sum_{s=\perp_1,\perp_2}\epsilon_j(\bm k,s)\epsilon_l(\bm k,s)=&\Big(\bm{e}_j\cdot\boldsymbol\epsilon(\bm k,\perp_1)\Big)\Big(\bm{e}_l\cdot\boldsymbol\epsilon(\bm k,\perp_1)\Big)+\Big(\bm{e}_j\cdot\boldsymbol\epsilon(\bm k,\perp_2)\Big)\Big(\bm{e}_l\cdot\boldsymbol\epsilon(\bm k,\perp_2)\Big)\notag\\
&+\left(\bm{e}_j\cdot\bm{\hat{k}}\right)\left(\bm{e}_l\cdot\bm{\hat{k}}\right)-\left(\bm{e}_j\cdot\bm{\hat{k}}\right)\left(\bm{e}_l\cdot\bm{\hat{k}}\right),
\label{completenessderivation}
\end{align}
where in the last line we have added and subtracted the same quantity. Written in this form, the first three terms of \eqref{completenessderivation} represent the coefficients of the scalar product $\bm e_j\cdot\bm e_l$ expressed in the $\{\boldsymbol\epsilon(\bm k,\perp_1),\boldsymbol\epsilon(\bm k,\perp_2),\hat{\bm k}\}$ basis, while the last term represents the product of the components of $\hat{\bm k}$ in the directions of $\bm e_j$ and $\bm e_l$. Therefore we can write
\begin{align}
\sum_{s=\perp_1,\perp_2}\epsilon_j(\bm k,s)\epsilon_l(\bm k,s)&=\bm{e}_j\cdot\bm{e}_l-\frac{k_jk_l}{|\bm{k}|^2}=\delta_{jl}-\frac{k_jk_l}{|\bm{k}|^2},
\end{align}
which is just the expression in components of Eq. \eqref{completeness}.

\section{Positivity of the density matrix} \label{app:positivity}

In this appendix we explicitly show how the time-evolved density matrix Eq. \eqref{state} satisfies the positivity conditions for an X-state given in Ref \cite{Martin-Martinez2016a}. Following \cite{Martin-Martinez2016a}, a generic density matrix for an X-state can be written as follows
\begin{equation}
{\hat{\rho}}_\textsc{x}=\begin{pmatrix}
r_{11} & 0 & 0 & r_{14}e^{-\ii\xi} \\
0 & r_{22} & r_{23}e^{-\ii\varsigma} & 0 \\
0 & r_{23}e^{\ii\varsigma} & r_{33} & 0 \\
r_{14}e^{\ii\xi} & 0 & 0 & r_{44}
\end{pmatrix},
\end{equation}
where $\{r_{ij}\}, \xi,\varsigma\in\mathbb{R}$.

A necessary condition for this matrix to represent a quantum state is that it is positive-semidefinite, which amounts to its eigenvalues being nonnegative. This restriction (as noted in \cite{Martin-Martinez2016a}) imposes the following constraints on the elements of ${\hat{\rho}}_\textsc{x}$:
\begin{align}
r_{11}&r_{44}\geq r_{14}^2, \label{positivity1}\\
r_{22}&r_{33}\geq r_{23}^2. \label{positivity2}
\end{align}

In the following, we will show explicitly how the matrix Eq. \eqref{state} is positive-semidefinite (to leading order) by analyzing its eigenvalues, and we will relate the results to the above constraints Eqs. \eqref{positivity1}, \eqref{positivity2}.

The eigenvalues of Eq. \eqref{state} are

\begin{align}
   E_1&=\frac{1}{2} \left(1-(L_\textsc{aa}+L_\textsc{bb})\ac^2+\sqrt{\left[1-\ac^2(L_\textsc{aa}+L_\textsc{bb})^2\right]^2+4 \ac^4|M|^2}\right)+\mathcal{O}(\ac^4),\\
   E_2&=\frac{1}{2} \left(1-(L_\textsc{aa}+L_\textsc{bb})\ac^2-\sqrt{\left[1-\ac^2(L_\textsc{aa}+L_\textsc{bb})^2\right]^2+4 \ac^4|M|^2}\right)+\mathcal{O}(\ac^4), \label{E2}\\
   E_3&=\frac{1}{2} \ac ^2 \left(L_\textsc{aa}+L_\textsc{bb}+\sqrt{(L_\textsc{aa}-L_\textsc{bb})^2+4 |L_\textsc{ab}|^2}\right)+\mathcal{O}(\ac^4),\\
   E_4&=\frac{1}{2} \ac ^2 \left(L_\textsc{aa}+L_\textsc{bb}-\sqrt{(L_\textsc{aa}-L_\textsc{bb})^2+4 |L_\textsc{ab}|^2}\right)+\mathcal{O}(\ac^4),
\end{align}
where $\mathcal{L}_{\mu\nu}=\ac^2 L_{\mu\nu}$ and $\mathcal{M}=\ac^2 M$, and we have assumed $\ac_\textsc{a}=\ac_\textsc{b}=\ac$ for simplicity.

It is already clear that $E_1>0$ and $E_3>0$. Expanding $E_2$ and $E_4$ in powers of the coupling strength we obtain

\begin{align}
   E_2&=-\ac^4|M|^2+\mathcal{O}(\ac^6),\\
   E_4&=\frac{1}{2} \ac ^2 \left(L_\textsc{aa}+L_\textsc{bb}-\sqrt{(L_\textsc{aa}-L_\textsc{bb})^2+4 |L_\textsc{ab}|^2}\right)+\mathcal{O}(\ac^4).
\end{align}

The eigenvalue $E_2$ seems to always be less than zero. Nevertheless, it must be noted that this contribution appears at fourth order, and therefore, to second order we have $E_2=0+\mathcal{O}(\ac^4)$, which is compatible with the matrix being positive-semidefinite at this order. A very similar issue has been discussed when computing the eigenvalues of the partial transpose of Eq. \eqref{state}, as well as in previous works \cite{Pozas-Kerstjens2015}. This result relates to the first condition \eqref{positivity1}. Recalling Eq. \eqref{state}, it is easy to see that
\begin{equation}
r_{11}=\mathcal{O}(1),\quad r_{14}=\mathcal{O}(\ac^2)>0,\quad r_{44}=\mathcal{O}(\ac^4).
\end{equation}
Therefore, to second order in the coupling strength, the inequality \eqref{positivity1} is satisfied trivially.

Notice that in order to take into account the $\mathcal{O}(\ac^4)$ contribution to $E_2$ with a consistent perturbative analysis, the full fourth-order correction to the state \eqref{state} should be computed. If we analyze $E_2$ consistently to fourth order in perturbation theory including the contributions missing in \eqref{E2}, it can be proved that, $E_2$ is strictly positive.

On the other hand, the requirement that $E_4\geq0$ implies

\begin{equation}
L_\textsc{aa}L_\textsc{bb}\geq|L_\textsc{ab}|^2,\label{conditiontwo}
\end{equation}
which is actually the second condition $r_{22}r_{33}\geq r_{23}^2$.

To check that this inequality is actually satisfied, recall the expression for $L_{\mu\nu}$, obtained from Eq. \eqref{Lmununotint},

\begin{equation}
L^\textsc{em}_{\mu\nu}=\int_{-\infty}^{\infty}\text{d}t_1\int_{-\infty}^{\infty}\text{d}t_2\int\text{d}^3\bm x_1\int\text{d}^3\bm x_2 e^{\ii(\Omega_\mu t_1-\Omega_\nu t_2)}\chi_\mu(t_1)\chi_\nu(t_2){\bm{F}_\nu^*}^{\text{\textbf{t}}}(\bm x_2-\bm x_\nu)\text{\bf W}(\bm x_2,\bm x_1,t_2,t_1)\bm F_\mu(\bm x_1-\bm x_\mu).
\end{equation}

When we insert the expressions for the smearing vectors Eq. \eqref{smearing} and the Wightman 2-tensor of the electric field Eq. \eqref{Wightman}, this expression reads

\begin{align}
L^\textsc{em}_{\mu\nu}=&\int_{-\infty}^{\infty}\text{d}t_1\int_{-\infty}^{\infty}\text{d}t_2\notag e^{\ii(\Omega_\mu t_1-\Omega_\nu t_2)}\chi_\mu(t_1)\chi_\nu(t_2)\notag\\
&\times\int\text{d}^3\bm x_1\int\text{d}^3\bm x_2\,\psi_e(\bm x_2-\bm x_\mu)\psi^*_g(\bm x_2-\bm x_\mu)\psi^*_e(\bm x_1-\bm x_\nu)\psi_g(\bm x_1-\bm x_\nu)\notag\\
&\times\int\frac{\text{d}^3\bm k}{(2\pi)^3}\frac{|\bm k|}{2}e^{-\ii|\bm k|(t_2-t_1)}e^{\ii\bm k\cdot\bm x_2}e^{-\ii\bm k\cdot\bm x_1}\,(\bm x_2-\bm x_\mu)^{\text{\textbf{t}}}\left(\openone-\frac{\bm k\otimes\bm k}{|\bm k|^2}\right)(\bm x_1-\bm x_\nu).
\end{align}

Finally, we perform the change of variables $\bm x=\bm x_1-\bm x_\nu$, $\bm x'=\bm x_2-\bm x_\mu$ to get the final expression

\begin{align}
L^\textsc{em}_{\mu\nu}=&\int_{-\infty}^{\infty}\text{d}t_1\int_{-\infty}^{\infty}\text{d}t_2\notag e^{\ii(\Omega_\mu t_1-\Omega_\nu t_2)}\chi_\mu(t_1)\chi_\nu(t_2)\int\text{d}^3\bm x\int\text{d}^3\bm x'\,\psi_e(\bm x')\psi^*_g(\bm x')\psi^*_e(\bm x)\psi_g(\bm x)\notag\\
&\times\int\frac{\text{d}^3\bm k}{(2\pi)^3}\frac{|\bm k|}{2}e^{-\ii|\bm k|(t_2-t_1)}e^{\ii\bm k\cdot\bm x'}e^{-\ii\bm k\cdot\bm x}\,\left({\bm x'}\right)^{\text{\textbf{t}}}\left(\openone-\frac{\bm k\otimes\bm k}{|\bm k|^2}\right)\bm x\,e^{\ii\bm k\cdot(\bm x_\mu-\bm x_\nu)}\notag\\
=&\sum_i\int\frac{\text{d}^3\bm k}{(2\pi)^3}\frac{|\bm k|}{2}\Bigg|\int_{-\infty}^{\infty}\text{d}t\,e^{\ii(\Omega+|\bm k|) t}\chi(t)\int\text{d}^3\bm y\,\psi^*_e(\bm y)\psi_g(\bm y) e^{-\ii\bm k\cdot\bm y}\,y_i\Bigg|^2e^{\ii\bm k\cdot(\bm x_\mu-\bm x_\nu)}\notag\\
&-\int\frac{\text{d}^3\bm k}{(2\pi)^3}\frac{|\bm k|}{2}\Bigg|\int_{-\infty}^{\infty}\text{d}t\,e^{\ii(\Omega+|\bm k|) t}\chi(t)\int\text{d}^3\bm y\,\psi^*_e(\bm y)\psi_g(\bm y) e^{-\ii\bm k\cdot\bm y}\,\frac{\bm y \cdot\bm k}{|\bm k|}\Bigg|^2e^{\ii\bm k\cdot(\bm x_\mu-\bm x_\nu)},
\end{align}
where we have already assumed that the two atoms are equal, and $y_i$ denote the components of the vector $\bm y$.

The only difference in this case between $L_\textsc{ab}$ and both $L_\textsc{aa}$ and $L_\textsc{bb}$ is the last phase factor $e^{\ii\bm k\cdot(\bm x_\mu-\bm x_\nu)}$, which is $e^{\ii\bm k\cdot(\bm x_\textsc{a}-\bm x_\textsc{b})}$ for $L_\textsc{ab}$ and $1$ for both $L_\textsc{aa}$ and $L_\textsc{bb}$. Now, given that for a nonnegative function $f(t)\ge 0\,\forall t$, 

\begin{equation}
\int_{-\infty}^\infty\text{d}t\,f(t)\geq\left|\int_{-\infty}^\infty\text{d}t\,e^{\ii \omega t}f(t)\right|
\end{equation}
it is easy to see that $L_\textsc{aa}\geq |L_\textsc{ab}|$ and $L_\textsc{bb}\geq |L_\textsc{ab}|$, thereby satisfying Eq. \eqref{conditiontwo}. Performing analogous calculations, the same conclusion can be reached for the scalar cases and arbitrary smearing functions.

\section{Explicit calculation of $\mathcal{L}^\text{\tiny{EM}
}_{\mu\mu}$ and $\mathcal{M}^\text{\tiny{EM}
}$}\label{app:longcalcs}

In this appendix we perform a step-by-step derivation of Eqs. \eqref{LmumuEM} and \eqref{MEM} starting from Eqs. \eqref{Lmununotint} and \eqref{Mnotint}. For generality, we will not fix the ground state or the excited state and we will perform the calculations as general as possible, beginning only with the assumption that both atoms have the same atomic structure (same ground and excited states), and particularizing to the $1s\rightarrow 2p_z$ transition only at the very end of each section of this appendix.

\subsection{Local term}
Let us begin with the complete expression of the local term

\begin{align}
\mathcal{L}^\textsc{em}_{\mu\mu}=&\ac_\mu^2\int_{-\infty}^{\infty}\text{d}t_1\int_{-\infty}^{\infty}\text{d}t_2\int\text{d}^3\bm x_1'\int\text{d}^3\bm x_2'\,e^{\ii\Omega_\mu (t_1-t_2)}\chi_\mu(t_1)\chi_\mu(t_2){\bm{F}_\mu^*}^{\text{\textbf{t}}}(\bm x_2'-\bm x_\mu)\text{\bf W}(\bm x_2',\bm x_1',t_2,t_1)\bm F_{\mu}(\bm x_1'-\bm x_\mu)\notag\\
=&\ac_\mu^2\int_{-\infty}^{\infty}\text{d}t_1\int_{-\infty}^{\infty}\text{d}t_2\,e^{\ii\Omega_\mu (t_1-t_2)}\chi_\mu(t_1)\chi_\mu(t_2)\int\text{d}^3\bm x_1\int\text{d}^3\bm x_2\,\psi_e(\bm x_2)\psi^*_g(\bm x_2)\psi^*_e(\bm x_1)\psi_g(\bm x_1)\notag\\
&\times\int\frac{\text{d}^3\bm k}{(2\pi)^3}\frac{|\bm k|}{2}e^{-\ii|\bm k|(t_2-t_1)}e^{\ii\bm k\cdot\bm x_2}e^{-\ii\bm k\cdot\bm x_1}\,\bm x_2^{\text{\textbf{t}}}\left(\openone-\frac{\bm k\otimes\bm k}{|\bm k|^2}\right)\bm x_1,
\label{Lmumucomplete}
\end{align}
where we have already written the smearing vector as $\bm F(\bm x)=\psi^*_e(\bm x)\,\bm x\,\psi_g(\bm x)$, as per Eq.  \eqref{smearing}, and the Wightman 2-tensor of the electric field Eq. \eqref{Wightman} is

\begin{equation}
{\bm{\mathrm{W}}}(\bm x_2,\bm x_1,t_2,t_1)=\bra{0}\hat{\bm E}(\bm x_2,t_2)\otimes\hat{\bm E}(\bm x_1,t_1)\ket{0}=\int\frac{\text{d}^3\bm k}{(2\pi)^3}\frac{|\bm k|}{2}e^{-\ii|\bm k|(t_2-t_1)}e^{\ii\bm k\cdot(\bm x_2-\bm x_1)}\left(\openone-\frac{\bm k\otimes\bm k}{|\bm k|^2}\right).
\end{equation}

First of all, we perform the change of variables $\bm x_1=\bm x_1'-\bm x_\mu$, $\bm x_2=\bm x_2'-\bm x_\mu$ to eliminate the explicit dependence of the smearing vector on the atomic position $\bm x_\mu$. After that, we choose spherical coordinates to perform the integrations and use the following decompositions involving spherical harmonics:
\begin{align}
\psi_{nlm}(\bm x)&=R_{nl}(|\bm x|)Y_{lm}(\bm{\hat x})\label{B3},\\
e^{\ii\bm x\cdot\bm y}&=\sum_{l=0}^\infty\sum_{m=-l}^l 4\pi\ii^l j_l(|\bm x||\bm y|)Y_{lm}(\bm{\hat x})Y^*_{lm}(\bm{\hat y})=\sum_{l=0}^\infty\sum_{m=-l}^l 4\pi\ii^l j_l(|\bm x||\bm y|)Y^*_{lm}(\bm{\hat x})Y_{lm}(\bm{\hat y}), \label{expoharmonics}\\
\bm x\cdot\bm y&=\frac{4\pi}{3}|\bm x||\bm y|\left[Y_{10}(\bm{\hat x})Y_{10}(\bm{\hat y})-Y_{11}(\bm{\hat x})Y_{1-1}(\bm{\hat y})-Y_{1-1}(\bm{\hat x})Y_{11}(\bm{\hat y})\right]\label{B5},
\end{align}
where the arguments of the spherical harmonics $\hat{\bm x}=(\theta_{\bm x},\phi_{\bm x})$ are the azimuthal and polar coordinates of the unit vector $\hat{\bm x}$ and $R_{nl}(|\bm x|)$ are the radial hydrogenoid wave functions \cite{GalindoBook}.

To make calculations less cumbersome, we separate Eq. \eqref{Lmumucomplete} into two parts, one with the identity matrix $\openone$ and the other with the momentum dyadic $\bm k\otimes\bm k$, and compute each of them separately.

Let us begin with the term containing the identity. Substituting \eqref{B3}, \eqref{expoharmonics} and \eqref{B5} we get

\begin{align}
\mathcal{L}^\textsc{em}_{\mu\mu}\Big|_{\openone}=&\ac_\mu^2\int_0^\infty\frac{\text{d}|\bm k|}{(2\pi)^3}\frac{|\bm k|^3}{2}\sum_{l=0}^\infty\sum_{m=-l}^l 4\pi\ii^l \sum_{l'=0}^\infty\sum_{m'=-l'}^{l'} 4\pi\ii^{l'} (-1)^{l'}\frac{4\pi}{3}\notag\\
&\times\int_{-\infty}^{\infty}\text{d}t_1\int_{-\infty}^{\infty}\text{d}t_2\,e^{\ii\Omega_\mu (t_1-t_2)}\chi_\mu(t_1)\chi_\mu(t_2)e^{-\ii|\bm k|(t_2-t_1)}\notag\\
&\times\int_0^\infty\text{d}|\bm x_2|\,|\bm x_2|^3R_{n_e,l_e}(|\bm x_2|)R^*_{n_g,l_g}(|\bm x_2|)j_l(|\bm k||\bm x_2|)\int_0^\infty\text{d}|\bm x_1|\,|\bm x_1|^3R^*_{n_e,l_e}(|\bm x_1|)R_{n_g,l_g}(|\bm x_1|)j_{l'}(|\bm k||\bm x_1|)\notag\\
&\times\int\text{d}\Omega_k\,Y_{lm}(\bm{\hat k})Y_{l'm'}(\bm{\hat k})\notag\\
&\times\int\text{d}\Omega_1 Y^*_{l_e,m_e}(\bm{\hat x}_1)Y_{l_g,m_g}(\bm{\hat x}_1)Y^*_{l'm'}(\bm{\hat x}_1)\int\text{d}\Omega_2 Y_{l_e,m_e}(\bm{\hat x}_2)Y^*_{l_g,m_g}(\bm{\hat x}_2)Y^*_{lm}(\bm{\hat x}_2)\notag\\
&\times\left[Y_{10}(\bm{\hat x}_1)Y_{10}(\bm{\hat x}_2)-Y_{11}(\bm{\hat x}_1)Y_{1-1}(\bm{\hat x}_2)-Y_{1-1}(\bm{\hat x}_1)Y_{11}(\bm{\hat x}_2)\right],\label{Lone}
\end{align}
where the $(-1)^{l'}$ factor comes from the identity $Y_{lm}(-\bm{\hat r})=(-1)^l Y_{lm}(\bm{\hat r})$ and $j_l(x)$ are the spherical Bessel functions.

Written in this form, almost each line of Eq. \eqref{Lone} can be computed separately. For instance, using the Gaussian switching function \eqref{switching} the time integrals in the second line yield
\begin{equation}
   \int_{-\infty}^{\infty}\text{d}t_1\int_{-\infty}^{\infty}\text{d}t_2\,e^{\ii\Omega_\mu (t_1-t_2)}\chi_\mu(t_1)\chi_\mu(t_2)e^{-\ii|\bm k|(t_2-t_1)}=\pi T^2 e^{-\frac{1}{2}T^2(\Omega_\mu+|\bm k|)^2}.
\end{equation}

The fourth line can be readily evaluated: $\int\text{d}\Omega_k\,Y_{lm}(\bm{\hat k})Y_{l'm'}(\bm{\hat k})=(-1)^{m'}\delta_{l,l'}\delta_{m,-m'}$, directly from the orthogonality relations of spherical harmonics [the $(-1)^{m'}$ factor comes from the fact that $Y^*_{lm}=(-1)^m Y_{l-m}$]. The simple form that this integral on solid angle takes allows us to easily compute the sums in $m'$ and $l'$. The two last lines can be computed using the following identity involving the integral of four spherical harmonics over the unit sphere $S^2$,

\begin{align}
   \int\text{d}&\Omega\,Y^*_{l_1m_1}(\theta,\phi)Y_{l_2m_2}(\theta,\phi)Y^*_{l_3m_3}(\theta,\phi)Y_{l_4m_4}(\theta,\phi)\notag\\
   &=\sum_{\lambda=0}^\infty\sum_{\mu=-\lambda}^\lambda\frac{2\lambda+1}{4\pi}\sqrt{(2l_1+1)(2l_2+1)(2l_3+1)(2l_4+1)}\tj{l_1}{l_3}{\lambda}{0}{0}{0}\tj{l_1}{l_3}{\lambda}{-m_1}{-m_3}{-\mu}\tj{l_2}{l_4}{\lambda}{0}{0}{0}\tj{l_2}{l_4}{\lambda}{m_2}{m_4}{\mu}\label{fourharmonics},
\end{align}
where $\text{d}\Omega=\text{d}(\cos\theta)\text{d}\phi$ and $\begin{pmatrix} l_1 & l_2 & l_3 \\ m_1 & m_2 & m_3 \end{pmatrix}$ represents the Wigner $3j$-symbols.

Using \eqref{fourharmonics}, the sums over $l'$, $m$ and $m'$ of all the integrals over solid angles ---the last three lines of Eq. \eqref{Lone}--- yield
\begin{align}
&\sum_{l'}\sum_m\sum_{m'}\ii^{l+l'}(-1)^{l'}j_{l'}(|\bm k||\bm x_1|)\int\text{d}\Omega_k\,Y_{lm}(\bm{\hat k})Y_{l'm'}(\bm{\hat k})\int\text{d}\Omega_1 Y^*_{l_e,m_e}(\bm{\hat x}_1)Y_{l_g,m_g}(\bm{\hat x}_1)Y^*_{l'm'}(\bm{\hat x}_1)\notag\\
&\quad\times\int\text{d}\Omega_2 Y_{l_e,m_e}(\bm{\hat x}_2)Y^*_{l_g,m_g}(\bm{\hat x}_2)Y^*_{lm}(\bm{\hat x}_2)\left[Y_{10}(\bm{\hat x}_1)Y_{10}(\bm{\hat x}_2)-Y_{11}(\bm{\hat x}_1)Y_{1-1}(\bm{\hat x}_2)-Y_{1-1}(\bm{\hat x}_1)Y_{11}(\bm{\hat x}_2)\right]\notag\\
&=\frac{3(-1)^{m_g-m_e}\ii^{2l}(-1)^l}{(4\pi)^2}(2l+1)(2l_e+1)(2l_g+1)\sum_{\lambda,\lambda'}(2\lambda+1)(2\lambda'+1)\tj{l}{l_g}{\lambda}{0}{0}{0}\tj{l_e}{1}{\lambda}{0}{0}{0}\tj{l}{l_e}{\lambda'}{0}{0}{0}\tj{l_g}{1}{\lambda'}{0}{0}{0}\notag\\
&\quad\times j_{l}(|\bm k||\bm x_1|)\left[\tj{l_e}{1}{\lambda}{m_e}{0}{-m_e}\tj{l}{l_g}{\lambda}{m_g-m_e}{-m_g}{m_e}\tj{l_g}{1}{\lambda'}{m_g}{0}{-m_g}\tj{l}{l_e}{\lambda'}{m_e-m_g}{-m_e}{m_g}\right.\notag\\
&\qquad+\tj{l_e}{1}{\lambda}{m_e}{-1}{1-m_e}\tj{l}{l_g}{\lambda}{1+m_g-m_e}{-m_g}{m_e-1}\tj{l_g}{1}{\lambda'}{m_g}{1}{-1-m_g}\tj{l}{l_e}{\lambda'}{m_e-m_g-1}{-m_e}{1+m_g}\notag\\
&\qquad+\left.\tj{l_e}{1}{\lambda}{m_e}{1}{-1-m_e}\tj{l}{l_g}{\lambda}{m_g-m_e-1}{-m_g}{m_e+1}\tj{l_g}{1}{\lambda'}{m_g}{-1}{1-m_g}\tj{l}{l_e}{\lambda'}{1+m_e-m_g}{-m_e}{m_g-1}\right].
\end{align}

After substituting all these in Eq. \eqref{Lone} we obtain
\begin{align}
\mathcal{L}^\textsc{em}_{\mu\mu}\Big|_{\openone}=&\ac_\mu^2\int_0^\infty\frac{\text{d}|\bm k|}{(2\pi)^3}\frac{|\bm k|^3}{2}\sum_{l=0}^\infty (4\pi)^2\ii^{2l} (-1)^{l}\frac{4\pi}{3}\pi T^2 e^{-\frac{1}{2}T^2(\Omega_\mu+|\bm k|)^2}\notag\\
&\times\int_0^\infty\text{d}|\bm x_2|\,|\bm x_2|^3R_{n_e,l_e}(|\bm x_2|)R^*_{n_g,l_g}(|\bm x_2|)j_l(|\bm k||\bm x_2|)\int_0^\infty\text{d}|\bm x_1|\,|\bm x_1|^3R^*_{n_e,l_e}(|\bm x_1|)R_{n_g,l_g}(|\bm x_1|)j_{l}(|\bm k||\bm x_1|)\notag\\
&\times\frac{3(-1)^{m_g-m_e}}{(4\pi)^2}(2l+1)(2l_e+1)(2l_g+1)\sum_{\lambda,\lambda'}(2\lambda+1)(2\lambda'+1)\tj{l}{l_g}{\lambda}{0}{0}{0}\tj{l_e}{1}{\lambda}{0}{0}{0}\tj{l}{l_e}{\lambda'}{0}{0}{0}\tj{l_g}{1}{\lambda'}{0}{0}{0}\notag\\
&\times\left[\tj{l_e}{1}{\lambda}{m_e}{0}{-m_e}\tj{l}{l_g}{\lambda}{m_g-m_e}{-m_g}{m_e}\tj{l_g}{1}{\lambda'}{m_g}{0}{-m_g}\tj{l}{l_e}{\lambda'}{m_e-m_g}{-m_e}{m_g}\right.\notag\\
   &\quad+\tj{l_e}{1}{\lambda}{m_e}{-1}{1\!-\!m_e}\tj{l}{l_g}{\lambda}{1\!+\!m_g\!-\!m_e}{-m_g}{m_e\!-\!1}\tj{l_g}{1}{\lambda'}{m_g}{1}{-1\!-\!m_g}\tj{l}{l_e}{\lambda'}{m_e\!-\!m_g\!-\!1}{-m_e}{1\!+\!m_g}\notag\\
   &\quad+\left.\tj{l_e}{1}{\lambda}{m_e}{1}{-1\!-\!m_e}\tj{l}{l_g}{\lambda}{m_g\!-\!m_e\!-\!1}{-m_g}{m_e\!+\!1}\tj{l_g}{1}{\lambda'}{m_g}{-1}{1\!-\!m_g}\tj{l}{l_e}{\lambda'}{1\!+\!m_e\!-\!m_g}{-m_e}{m_g\!-\!1}\right].
\end{align}

This expression is fully general, for any two arbitrary levels of the hydrogenoid atom coupled dipolarly to the field in the vacuum. No more integrations can be performed unless we specify which particular atomic electron states are the ground and excited states  $g,\,e$. 

Before doing that, let us compute the contribution to the local term \eqref{Lmumucomplete} that is proportional to the momentum dyadic $\bm k\otimes\bm k$. The strategy we will follow will be the same as in the previous case. After substituting in \eqref{B3}, \eqref{expoharmonics} and \eqref{B5} the contribution reads
\begin{align}
\mathcal{L}^\textsc{em}_{\mu\mu}\Big|_{\bm k\otimes\bm k}=&\ac_\mu^2\int_0^\infty\frac{\text{d}|\bm k|}{(2\pi)^3}\frac{|\bm k|^3}{2}\sum_{l=0}^\infty\sum_{m=-l}^l 4\pi\ii^l \sum_{l'=0}^\infty\sum_{m'=-l'}^{l'} 4\pi\ii^{l'} (-1)^{l'}\left(\frac{4\pi}{3}\right)^2\pi T^2 e^{-\frac{1}{2}T^2(\Omega_\mu+|\bm k|)^2}\notag\\
&\times\int_0^\infty\text{d}|\bm x_2|\,|\bm x_2|^3R_{n_e,l_e}(|\bm x_2|)R^*_{n_g,l_g}(|\bm x_2|)j_l(|\bm k||\bm x_2|)\int_0^\infty\text{d}|\bm x_1|\,|\bm x_1|^3R^*_{n_e,l_e}(|\bm x_1|)R_{n_g,l_g}(|\bm x_1|)j_{l'}(|\bm k||\bm x_1|)\notag\\
&\times\int\text{d}\Omega_k\,Y_{lm}(\bm{\hat k})Y_{l'm'}(\bm{\hat k})\int\text{d}\Omega_1 Y^*_{l_e,m_e}(\bm{\hat x}_1)Y_{l_g,m_g}(\bm{\hat x}_1)Y^*_{l'm'}(\bm{\hat x}_1)\int\text{d}\Omega_2 Y_{l_e,m_e}(\bm{\hat x}_2)Y^*_{l_g,m_g}(\bm{\hat x}_2)Y^*_{lm}(\bm{\hat x}_2)\notag\\
&\times\left[Y_{10}(\bm{\hat x}_1)Y_{10}(\bm{\hat k})\!-\!Y_{11}(\bm{\hat x}_1)Y_{1-1}(\bm{\hat k})\!-\!Y_{1-1}(\bm{\hat x}_1)Y_{11}(\bm{\hat k})\right]\!\!\left[Y_{10}(\bm{\hat k})Y_{10}(\bm{\hat x}_2)\!-\!Y_{11}(\bm{\hat k})Y_{1-1}(\bm{\hat x}_2)\!-\!Y_{1-1}(\bm{\hat k})Y_{11}(\bm{\hat x}_2)\right],
\end{align}
where we have already performed the integrals in time.

The novelty introduced by the dyadic is that the integral in the solid angle of $\bm k$ is now no longer trivial, although it can be readily performed using Eq. \eqref{fourharmonics}. Integrating over all solid angles we obtain

\begin{align}
\sum_{m,m'}&\int\text{d}\Omega_k\,Y_{lm}(\bm{\hat k})Y_{l'm'}(\bm{\hat k})\int\text{d}\Omega_1 Y^*_{l_e,m_e}(\bm{\hat x}_1)Y_{l_g,m_g}(\bm{\hat x}_1)Y^*_{l'm'}(\bm{\hat x}_1)\int\text{d}\Omega_2 Y_{l_e,m_e}(\bm{\hat x}_2)Y^*_{l_g,m_g}(\bm{\hat x}_2)Y^*_{lm}(\bm{\hat x}_2)\notag\\
&\times\left[Y_{10}(\bm{\hat x}_1)Y_{10}(\bm{\hat k})-Y_{11}(\bm{\hat x}_1)Y_{1-1}(\bm{\hat k})-Y_{1-1}(\bm{\hat x}_1)Y_{11}(\bm{\hat k})\right]\!\left[Y_{10}(\bm{\hat k})Y_{10}(\bm{\hat x}_2)-Y_{11}(\bm{\hat k})Y_{1-1}(\bm{\hat x}_2)-Y_{1-1}(\bm{\hat k})Y_{11}(\bm{\hat x}_2)\right]\notag\\
=&9(2l+1)(2l'+1)(2l_g+1)(2l_e+1)\sum_{\lambda'\lambda''}\frac{(2\lambda'+1)(2\lambda''+1)}{(4\pi)^3}
\tj{l_e}{1}{\lambda'}{0}{0}{0}
\tj{l_g}{1}{\lambda''}{0}{0}{0}
\tj{l'}{l_e}{\lambda''}{0}{0}{0}
\tj{l}{l_g}{\lambda'}{0}{0}{0}\notag\\
&\times\left(A_\mathcal{L}+B_\mathcal{L}\right),
\end{align}
where the two quantities $A_\mathcal{L}$ and $B_\mathcal{L}$ read

\begin{align}
A_\mathcal{L}=&\sum_\lambda (2\lambda+1)
\tj{1}{1}{\lambda}{0}{0}{0}^2
\tj{l}{l'}{\lambda}{0}{0}{0}\tj{l}{l_g}{\lambda'}{m_g-m_e}{-m_g}{m_e}
\tj{l_e}{1}{\lambda'}{m_e}{0}{-m_e}\notag\\
&\quad\times\tj{l}{l'}{\lambda}{m_e-m_g}{m_g-m_e}{0}
\tj{l'}{l_e}{\lambda''}{m_e-m_g}{-m_e}{m_g}
\tj{l_g}{1}{\lambda''}{m_g}{0}{-m_g}\notag\\
&+\sum_{\lambda}(2\lambda+1)
\tj{1}{1}{\lambda}{0}{0}{0}
\tj{1}{1}{\lambda}{0}{1}{-1}
\tj{l}{l'}{\lambda}{0}{0}{0}
\notag\\
&\quad\times\left[\tj{l}{l'}{\lambda}{m_e-m_g-1}{m_g-m_e}{1}
\tj{l'}{l_e}{\lambda''}{m_e-m_g}{-m_e}{m_g}
\tj{l_g}{1}{\lambda''}{m_g}{0}{-m_g}\right.\notag\\
&\qquad\times\tj{l}{l_g}{\lambda'}{m_g-m_e+1}{-m_g}{m_e-1}\tj{l_e}{1}{\lambda'}{m_e}{-1}{1-m_e}\notag\\
&\quad+\tj{l}{l'}{\lambda}{m_e-m_g}{m_g-m_e-1}{1}
\tj{l'}{l_e}{\lambda''}{m_e-m_g+1}{-m_e}{m_g-1}
\tj{l_g}{1}{\lambda''}{m_g}{-1}{1-m_g}\notag\\
&\qquad\times\left.\tj{l}{l_g}{\lambda'}{m_g-m_e}{-m_g}{m_e}\tj{l_e}{1}{\lambda'}{m_e}{0}{-m_e}\right]\notag\\
&+\sum_{\lambda}(2\lambda+1)
\tj{1}{1}{\lambda}{0}{0}{0}
\tj{1}{1}{\lambda}{0}{-1}{1}
\tj{l}{l'}{\lambda}{0}{0}{0}
\notag\\
&\quad\times\left[\tj{l}{l'}{\lambda}{m_e-m_g+1}{m_g-m_e}{-1}
\tj{l'}{l_e}{\lambda''}{m_e-m_g}{-m_e}{m_g}
\tj{l_g}{1}{\lambda''}{m_g}{0}{-m_g}\right.\notag\\
&\qquad\times\tj{l}{l_g}{\lambda'}{m_g-m_e-1}{-m_g}{m_e+1}
\tj{l_e}{1}{\lambda'}{m_e}{1}{-1-m_e}\notag\\
&\quad+\tj{l}{l'}{\lambda}{m_e-m_g}{m_g-m_e+1}{-1}
\tj{l'}{l_e}{\lambda''}{m_e-m_g-1}{-m_e}{m_g+1}
\tj{l_g}{1}{\lambda''}{m_g}{1}{-1-m_g}\notag\\
&\left.\qquad\times\tj{l}{l_g}{\lambda'}{m_g-m_e}{-m_g}{m_e}
\tj{l_e}{1}{\lambda'}{m_e}{0}{-m_e}\right]\notag\\
&+\sum_{\lambda}(2\lambda+1)
\tj{1}{1}{\lambda}{0}{0}{0}
\tj{1}{1}{\lambda}{1}{-1}{0}
\tj{l}{l'}{\lambda}{0}{0}{0}
\notag\\
&\quad\times
\left[\tj{l}{l'}{\lambda}{m_e-m_g-1}{m_g-m_e+1}{0}
\tj{l'}{l_e}{\lambda''}{m_e-m_g-1}{-m_e}{m_g+1}
\tj{l_g}{1}{\lambda''}{m_g}{1}{-1-m_g}\right.\notag\\
&\qquad\times\tj{l}{l_g}{\lambda'}{m_g-m_e+1}{-m_g}{m_e-1}
\tj{l_e}{1}{\lambda'}{m_e}{-1}{1-m_e}\notag\\
&\quad+\tj{l}{l'}{\lambda}{m_e-m_g+1}{m_g-m_e-1}{0}
\tj{l'}{l_e}{\lambda''}{m_e-m_g+1}{-m_e}{m_g-1}
\tj{l_g}{1}{\lambda''}{m_g}{-1}{1-m_g}\notag\\
&\left.\qquad\times\tj{l}{l_g}{\lambda'}{m_g-m_e-1}{-m_g}{m_e+1}
\tj{l_e}{1}{\lambda'}{m_e}{1}{-1-m_e}\right].
\end{align}

\begin{align}
B_\mathcal{L}=&\sqrt{\frac{2}{3}}
\tj{l}{l'}{2}{0}{0}{0}
\tj{l}{l'}{2}{m_e-m_g-1}{m_g-m_e-1}{2}
\tj{l'}{l_e}{\lambda''}{1+m_e-m_g}{-m_e}{m_g-1}
\notag\\
&\quad\times\tj{l_g}{1}{\lambda''}{m_g}{-1}{1-m_g}
\tj{l_e}{1}{\lambda'}{m_e}{-1}{1-m_e}
\tj{l}{l_g}{\lambda'}{1-m_e+m_g}{-m_g}{m_e-1}\notag\\
&+\sqrt{\frac{2}{3}}
\tj{l}{l'}{2}{0}{0}{0}
\tj{l}{l'}{2}{m_e-m_g+1}{m_g-m_e+1}{-2}
\tj{l'}{l_e}{\lambda''}{m_e-m_g-1}{-m_e}{m_g+1}\notag\\
&\quad\times\tj{l_g}{1}{\lambda''}{m_g}{1}{-1-m_g}
\tj{l}{l_g}{\lambda'}{m_g-m_e-1}{-m_g}{m_e+1}
\tj{l_e}{1}{\lambda'}{m_e}{1}{-1-m_e},
\end{align}

Now we particularize the local term $\mathcal{L}_{\mu\mu}$ to the atomic transition studied in the main text. Namely, we consider the ground state of both detectors to be the hydrogenoid-$1s$ orbital, and the excited state a hydrogenoid-$2p_z$ orbital. Therefore, we have $l_e=1$, $l_g=0$, $m_e=0$ and $m_g=0$. In this scenario, the identity term \eqref{Lone} reads

\begin{align}
\mathcal{L}^\textsc{em}_{\mu\mu}\Big|_{\openone}=&\ac_\mu^2\int_0^\infty\frac{\text{d}|\bm k|}{(2\pi)^3}\frac{|\bm k|^3}{2}\sum_{l=0}^\infty (4\pi)^2\frac{4\pi}{3}\pi T^2 e^{-\frac{1}{2}T^2(\Omega_\mu+|\bm k|)^2}\notag\\
&\times\int_0^\infty\text{d}|\bm x_2|\,|\bm x_2|^3R_{2,1}(|\bm x_2|)R^*_{1,0}(|\bm x_2|)j_l(|\bm k||\bm x_2|)\int_0^\infty\text{d}|\bm x_1|\,|\bm x_1|^3R^*_{2,1}(|\bm x_1|)R_{1,0}(|\bm x_1|)j_{l}(|\bm k||\bm x_1|)\notag\\
&\times\frac{3}{(4\pi)^2}(2l+1)(2+1)(0+1)\sum_{\lambda,\lambda'}(2\lambda+1)(2\lambda'+1)
\tj{l}{0}{\lambda}{0}{0}{0}
\tj{1}{1}{\lambda}{0}{0}{0}
\tj{l}{1}{\lambda'}{0}{0}{0}
\tj{0}{1}{\lambda'}{0}{0}{0}\notag\\
&\times\left[\tj{1}{1}{\lambda}{0}{0}{0}
   \tj{l}{0}{\lambda}{0}{0}{0}
   \tj{0}{1}{\lambda'}{0}{0}{0}
   \tj{l}{1}{\lambda'}{0}{0}{0}
   +\tj{1}{1}{\lambda}{0}{-1}{1}
   \tj{l}{0}{\lambda}{1}{0}{-1}
   \tj{0}{1}{\lambda'}{0}{1}{-1}
   \tj{l}{1}{\lambda'}{-1}{0}{1}\right.\notag\\
   &\quad+\left.\tj{1}{1}{\lambda}{0}{1}{-1}
   \tj{l}{0}{\lambda}{-1}{0}{1}
   \tj{0}{1}{\lambda'}{0}{-1}{1}
   \tj{l}{1}{\lambda'}{1}{0}{-1}\right].
\end{align}

Using the properties of the Wigner $3j$-symbols it is clear that the only nonzero contributions to the sum are $\lambda=0,1,2$ and $\lambda'=1$. Additionally, the first $3j$-symbol enforces that the nonzero terms in the sum are only those satisfying $l=\lambda$. Performing the sums, the expression is simplified to

\begin{align}
    \mathcal{L}^\textsc{em}_{\mu\mu}\Big|_{\openone}=&\frac{\ac_\mu^2}{12\pi}T^2\int_0^\infty\text{d}|\bm k|\,|\bm k|^3
e^{-\frac{1}{2}T^2(\Omega_\mu+|\bm k|)^2}\int_0^\infty\text{d}|\bm x_2|\,|\bm x_2|^3R_{2,1}(|\bm x_2|)R^*_{1,0}(|\bm x_2|)\int_0^\infty\text{d}|\bm x_1|\,|\bm x_1|^3R^*_{2,1}(|\bm x_1|)R_{1,0}(|\bm x_1|)\notag\\
&\times\left[j_0(|\bm k||\bm x_1|)j_0(|\bm k||\bm x_2|)+2 j_2(|\bm k||\bm x_1|)j_2(|\bm k||\bm x_2|)\right].
\label{Lonewithoutspace}
\end{align}

The integrals over $\text{d}|\bm x_1|$ and $\text{d}|\bm x_2|$ can now be readily evaluated using the identity
\begin{equation}
   \int_0^\infty\text{d}r\, r^3R_{2,1}(r)R_{1,0}(r)j_l(|\bm k|r)=8 \sqrt{2 \pi } 3^{-l-\frac{11}{2}} a_0^{l+1} |\bm k|^l \Gamma (l+5) \, _2\tilde{F}_1\left(\frac{l+5}{2},\frac{l+6}{2};l+\frac{3}{2};-\frac{4}{9}  a_0^2 |\bm k|^2\right),
   \label{spaceintegral}
\end{equation}
where $_2\tilde{F}_1\left(a,b;c;z\right)={}_2F_1(a,b;c;z)/\Gamma(z)$ is the regularized hypergeometric function.

After particularizing Eq. \eqref{spaceintegral} to the cases appearing in Eq. \eqref{Lonewithoutspace}, we arrive to the final expression for the identity contribution to the local term

\begin{align}
\mathcal{L}^\textsc{em}_{\mu\mu}\Big|_{\openone}=&\ac_\mu^2\frac{663552}{\pi}a_0^2T^2  \int_0^\infty\text{d}|\bm k|\,|\bm k|^3
e^{-\frac{1}{2}T^2(\Omega_\mu+|\bm k|)^2}\frac{16 a_0^4 |\bm k|^4-8 a_0^2 |\bm k|^2+9}{\left(4 a_0^2 |\bm k|^2+9\right)^8}.
\label{Loneend}
\end{align}

Next, we apply the same procedure and techniques to the term with the dyadic contribution in momenta, which yields
\begin{align}
\mathcal{L}^\textsc{em}_{\mu\mu}\Big|_{\bm k\otimes\bm k}=&\ac_\mu^2\int_0^\infty\frac{\text{d}|\bm k|}{(2\pi)^3}\frac{|\bm k|^3}{2}\sum_{l=0}^\infty 4\pi\ii^l \sum_{l'=0}^\infty 4\pi\ii^{l'} (-1)^{l'}\left(\frac{4\pi}{3}\right)^2\pi T^2 e^{-\frac{1}{2}T^2(\Omega_\mu+|\bm k|)^2}\frac{1}{(4\pi)^3}9(2l+1)(2l'+1)\cdot 1\cdot 3\notag\\
&\times\int_0^\infty\text{d}|\bm x_2|\,|\bm x_2|^3R_{2,1}(|\bm x_2|)R^*_{1,0}(|\bm x_2|)j_l(|\bm k||\bm x_2|)\int_0^\infty\text{d}|\bm x_1|\,|\bm x_1|^3R^*_{2,1}(|\bm x_1|)R_{1,0}(|\bm x_1|)j_{l'}(|\bm k||\bm x_1|)\notag\\
&\times\bigg\{(2l+1)(2+1)\sum_{\lambda}(2\lambda+1)
\tj{1}{1}{\lambda}{0}{0}{0}^2
\tj{l}{l'}{\lambda}{0}{0}{0}^2
\tj{l}{0}{l}{0}{0}{0}^2
\tj{1}{1}{l}{0}{0}{0}^2
\tj{l'}{1}{1}{0}{0}{0}^2
\tj{0}{1}{1}{0}{0}{0}^2\notag\\
&\quad+\sqrt{\frac{2}{3}}(2+1)(2l+1)
\tj{l}{l'}{2}{0}{0}{0}
\tj{l}{l'}{2}{-1}{-1}{2}
\tj{l'}{1}{1}{0}{0}{0}
\tj{l'}{1}{1}{1}{0}{-1}
\tj{0}{1}{1}{0}{0}{0}\notag\\
&\qquad\times\tj{0}{1}{1}{0}{-1}{1}
\tj{l}{0}{l}{0}{0}{0}
\tj{l}{0}{l}{1}{0}{-1}
\tj{1}{1}{l}{0}{0}{0}
\tj{1}{1}{l}{0}{-1}{1}\notag\\
&\quad+\sqrt{\frac{2}{3}}(2+1)(2l+1)
\tj{l}{l'}{2}{0}{0}{0}
\tj{l}{l'}{2}{1}{1}{-2}
\tj{l'}{1}{1}{0}{0}{0}\notag\\
&\qquad\times\tj{l'}{1}{1}{-1}{0}{1}
\tj{0}{1}{1}{0}{0}{0}
\tj{0}{1}{1}{0}{1}{-1}
\tj{l}{0}{l}{0}{0}{0}
\tj{l}{0}{l}{-1}{0}{1}
\tj{1}{1}{l}{0}{0}{0}
\tj{1}{1}{l}{0}{1}{-1}\notag\\
&\quad+\sum_{\lambda}(2\lambda+1)(2l+1)(2+1)
\tj{1}{1}{\lambda}{0}{0}{0}
\tj{1}{1}{\lambda}{0}{1}{-1}
\tj{l}{l'}{\lambda}{0}{0}{0}
\tj{l}{0}{l}{0}{0}{0}
\tj{1}{1}{l}{0}{0}{0}
\tj{0}{1}{1}{0}{0}{0}
\tj{l'}{1}{1}{0}{0}{0}
\notag\\
&\qquad\times
\left[\tj{l}{l'}{\lambda}{-1}{0}{1}
\tj{l'}{1}{1}{0}{0}{0}
\tj{0}{1}{1}{0}{0}{0}
\tj{l}{0}{l}{1}{0}{-1}
\tj{1}{1}{l}{0}{-1}{1}\right.\notag\\
&\qquad\quad\left.+\tj{l}{l'}{\lambda}{0}{-1}{1}
\tj{l'}{1}{1}{1}{0}{-1}
\tj{0}{1}{1}{0}{-1}{1}
\tj{l}{0}{l}{0}{0}{0}
\tj{1}{1}{l}{0}{0}{0}\right]\notag\\
&\quad+\sum_{\lambda}(2\lambda+1)(2l+1)(2+1)
\tj{1}{1}{\lambda}{0}{0}{0}
\tj{1}{1}{\lambda}{0}{-1}{1}
\tj{l}{l'}{\lambda}{0}{0}{0}
\tj{l}{0}{l}{0}{0}{0}
\tj{1}{1}{l}{0}{0}{0}
\tj{0}{1}{1}{0}{0}{0}
\tj{l'}{1}{1}{0}{0}{0}
\notag\\
&\qquad\times
\left[\tj{l}{l'}{\lambda}{1}{0}{-1}
\tj{l'}{1}{1}{0}{0}{0}
\tj{0}{1}{1}{0}{0}{0}
\tj{l}{0}{l}{-1}{0}{1}
\tj{1}{1}{l}{0}{1}{-1}\right.\notag\\
&\quad\qquad\left.+\tj{l}{l'}{\lambda}{0}{1}{-1}
\tj{l'}{1}{1}{-1}{0}{1}
\tj{0}{1}{1}{0}{1}{-1}
\tj{l}{0}{l}{0}{0}{0}
\tj{1}{1}{l}{0}{0}{0}\right]\notag\\
&\quad+\sum_{\lambda}(2\lambda+1)(2l+1)(2+1)
\tj{1}{1}{\lambda}{0}{0}{0}
\tj{1}{1}{\lambda}{1}{-1}{0}
\tj{l}{l'}{\lambda}{0}{0}{0}
\tj{l}{0}{l}{0}{0}{0}
\tj{1}{1}{l}{0}{0}{0}
\tj{0}{1}{1}{0}{0}{0}
\tj{l'}{1}{1}{0}{0}{0}\notag\\
&\qquad\times
\left[\tj{l}{l'}{\lambda}{-1}{1}{0}
\tj{l'}{1}{1}{-1}{0}{1}
\tj{0}{1}{1}{0}{1}{-1}
\tj{l}{0}{l}{1}{0}{-1}
\tj{1}{1}{l}{0}{-1}{1}\right.\notag\\
&\quad\qquad+\left.\tj{l}{l'}{\lambda}{1}{-1}{0}
\tj{l'}{1}{1}{1}{0}{-1}
\tj{0}{1}{1}{0}{-1}{1}
\tj{l}{0}{l}{-1}{0}{1}
\tj{1}{1}{l}{0}{1}{-1}\right]\bigg\}.
\end{align}

The properties of the Wigner $3j$-symbols can be used to cancel many of the summands. In particular, we can restrict the sums to just $\lambda,l,l'=0,1,2$, leading to the expression
\begin{align}
\mathcal{L}^\textsc{em}_{\mu\mu}\Big|_{\bm k\otimes\bm k}=&\ac_\mu^2\frac{T^2}{36\pi}\int_0^\infty\text{d}|\bm k|\,|\bm k|^3e^{-\frac{1}{2}T^2(\Omega_\mu+|\bm k|)^2}\int_0^\infty\text{d}|\bm x_1|\,|\bm x_1|^3R^*_{2,1}(|\bm x_1|)R_{1,0}(|\bm x_1|)\left[j_{0}(|\bm k||\bm x_1|)-2j_2(|\bm k||\bm x_1|)\right]
\notag\\
&\times\int_0^\infty\text{d}|\bm x_2|\,|\bm x_2|^3R_{2,1}(|\bm x_2|)R^*_{1,0}(|\bm x_2|)\left[j_0(|\bm k||\bm x_2|)-  2j_2(|\bm k||\bm x_2|)\right]\notag\\
=&\ac_\mu^2\frac{24576}{\pi}T^2 a_0^2\int_0^\infty\text{d}|\bm k|\,|\bm k|^3e^{-\frac{1}{2}T^2(\Omega_\mu+|\bm k|)^2}\frac{\left(20 a_0^2 |\bm k|^2-9\right)^2}{\left(4 a_0^2 |\bm k|^2+9\right)^8}.
\label{Lkk}
\end{align}

Finally, subtracting Eq. \eqref{Lkk} from Eq. \eqref{Loneend} ---recall Eq. \eqref{Lmumucomplete}--- one arrives at the expression of the local term Eq. \eqref{LmumuEM},

\begin{equation}
\mathcal{L}_{\mu\mu}=\ac^2\frac{49152}{\pi}T^2 a_0^2\int_0^\infty\text{d}|\bm k|\frac{|\bm k|^3e^{-\frac{1}{2}T^2(\Omega_\mu+|\bm k|)^2}}{\left(4 a_0^2 |\bm k|^2+9\right)^6}.
\end{equation}

\subsection{Nonlocal term}

The nonlocal term \eqref{MEM} contains two different summands, which differ on the order of the subindices A and B. For each of them, as in the case of the local term, one can perform a separation into one part containing the identity and another one containing the momentum dyadic. In this part of the appendix we will compute explicitly the first summand of the term, which we will call $\mathcal{M}^\textsc{ab}$, and derive the other from this one using symmetry arguments. The explicit expression for $\mathcal{M}^\textsc{ab}$ is

\begin{align}
\mathcal{M}^\textsc{ab}=&-\ac_\textsc{a}\ac_\textsc{b}\int_{-\infty}^{\infty}\text{d}t_1\int_{-\infty}^{t_1}\text{d}t_2\int\text{d}^3\bm x_1\int\text{d}^3\bm x_2\,e^{\ii(\Omega_\textsc{a} t_1+\Omega_\textsc{b} t_2)}\chi_\textsc{a}(t_1)\chi_\textsc{b}(t_2)\bm{F}_\textsc{a}^{\text{\textbf{t}}}(\bm x_1)\text{\bf W}(\bm x_1+\bm x_\textsc{a},\bm x_2+\bm x_\textsc{b},t_1,t_2)\bm F_\textsc{b}(\bm x_2)\notag\\
=&-\ac_\textsc{a}\ac_\textsc{b}\int_{-\infty}^{\infty}\text{d}t_1\int_{-\infty}^{t_1}\text{d}t_2\int\text{d}^3\bm x_1\int\text{d}^3\bm x_2\,e^{\ii(\Omega_\textsc{a} t_1+\Omega_\textsc{b} t_2)}\chi_\textsc{a}(t_1)\chi_\textsc{b}(t_2)\notag\\
&\times\int\frac{\text{d}^3\bm k}{(2\pi)^3}\frac{|\bm k|}{2}e^{-\ii|\bm k|(t_1-t_2)}e^{\ii\bm k\cdot\bm x_1}e^{-\ii\bm k\cdot\bm x_2}e^{\ii\bm k\cdot(\bm x_\textsc{a}-\bm x_\textsc{b})}\bm x_1^{\text{\textbf{t}}}\left(\openone-\frac{\bm k\otimes\bm k}{|\bm k|^2}\right)\bm x_2\psi^*_{e_\textsc{a}}(\bm x_1)\psi_{g_\textsc{a}}(\bm x_1)\psi^*_{e_\textsc{b}}(\bm x_2)\psi_{g_\textsc{b}}(\bm x_2),
\label{Mabexpanded}
\end{align}
where we have already performed the translations $\bm x_1=\bm x_1'-\bm x_\textsc{a}$, $\bm x_2=\bm x_2'-\bm x_\textsc{b}$ to eliminate the explicit dependence on $\bm x_\textsc{a}$ and $\bm x_\textsc{b}$ from the smearing vectors.

Note that the correlation term \eqref{Mabexpanded} depends on the relative spatial orientation of the two atoms. With the aim of defining a common reference frame for the two atoms, we will refer the orientation of atom B to the reference frame of atom A. This means that if the spherical harmonics used to describe atom A's orbitals are $Y^\textsc{a}_{lm}(\theta_\textsc{a},\phi_\textsc{a})=Y_{lm}(\theta_\textsc{a},\phi_\textsc{a})$, atom B's angular wave functions will be given by \cite{Morrison1987}

\begin{equation}
   Y^\textsc{b}_{lm}(\theta_\textsc{b},\phi_\textsc{b})=\sum_{\mu=-l}^l Y_{l\mu}(\theta_\textsc{b},\phi_\textsc{b})\mathcal{D}^l_{\mu,m}(\psi,\vartheta,\varphi),
\end{equation}
where $(\psi,\,\vartheta,\,\varphi)$ are the Euler angles characterizing the rotation between the reference frames of B and A (see Figure \ref{fig:eulerangles}), and $\mathcal{D}^l_{\mu,m}$ represents the Wigner D-matrix coefficients, which characterize the rotation of the angular momentum operators and are given by
\begin{equation}
   \mathcal{D}^l_{\mu,m}(\psi,\vartheta,\varphi)=e^{-\ii(\mu\psi+m\varphi)}\sqrt{\frac{(l-m)!(l+\mu)!}{(l+m)!(l-\mu)!}}\frac{\cos(\vartheta/2)^{2l+m-\mu}[-\sin(\vartheta/2)]^{\mu-m}}{(\mu-m)!} {}_2F_1\!\left(\mu\!-\!l,-m\!-\!l;\mu\!-\!m+1;\tan^2\frac{\vartheta}{2}\right).
   \end{equation}

Taking this subtlety into account, and in the same way as we did with the local term $\mathcal{L}_{\mu\mu}$, we separate the term $\mathcal{M}^\textsc{ab}$ into two parts, one containing the identity matrix and the other one containing the momentum dyadic $\bm k \otimes\bm k$. The contribution to $\mathcal{M}^\textsc{ab}$ containing the identity matrix is
\begin{align}
\mathcal{M}^\textsc{ab}\Big|_{\openone}=&-\ac_\textsc{a}\ac_\textsc{b}\int_0^\infty\frac{\text{d}|\bm k|}{(2\pi)^3}\frac{|\bm k|^3}{2}\sum_{l=0}^\infty\sum_{m=-l}^l 4\pi\ii^l \sum_{l'=0}^\infty\sum_{m'=-l'}^{l'} 4\pi\ii^{l'} (-1)^{l'}\sum_{l''=0}^\infty\sum_{m''=-l''}^{l''}4\pi\ii^{l''}j_{l''}(|\bm k||\bm x_\textsc{a}-\bm x_\textsc{b}|) \frac{4\pi}{3}\notag\\
&\times\int_{-\infty}^{\infty}\text{d}t_1\int_{-\infty}^{t_1}\text{d}t_2\,e^{\ii(\Omega_\textsc{a} t_1+\Omega_\textsc{b} t_2)}\chi_\textsc{a}(t_1)\chi_\textsc{b}(t_2)e^{-\ii|\bm k|(t_1-t_2)}\notag\\
&\times\int_0^\infty\text{d}|\bm x_1|\,|\bm x_1|^3R^*_{n_e,l_e}(|\bm x_1|)R_{n_g,l_g}(|\bm x_1|)j_{l}(|\bm k||\bm x_1|)\int_0^\infty\text{d}|\bm x_2|\,|\bm x_2|^3R^*_{n_e,l_e}(|\bm x_2|)R_{n_g,l_g}(|\bm x_2|)j_{l'}(|\bm k||\bm x_2|)\notag\\
&\times\int\text{d}\Omega_k\,Y_{lm}(\bm{\hat k})Y_{l'm'}(\bm{\hat k})Y_{l''m''}(\bm{\hat k})Y^*_{l''m''}(\widehat{\Delta\bm x})\notag\\
&\times\int\text{d}\Omega_1 (Y^\textsc{a}_{l_e,m_e})^*(\bm{\hat x}_1)Y^\textsc{a}_{l_g,m_g}(\bm{\hat x}_1)Y^*_{lm}(\bm{\hat x}_1)\int\text{d}\Omega_2 (Y^\textsc{b}_{l_e,m_e})^*(\bm{\hat x}_2)Y^\textsc{b}_{l_g,m_g}(\bm{\hat x}_2)Y^*_{l'm'}(\bm{\hat x}_2)\notag\\
&\quad\times\left[Y_{10}(\bm{\hat x}_1)Y_{10}(\bm{\hat x}_2)-Y_{11}(\bm{\hat x}_1)Y_{1-1}(\bm{\hat x}_2)-Y_{1-1}(\bm{\hat x}_1)Y_{11}(\bm{\hat x}_2)\right],
\end{align}
where $\widehat{\Delta\bm x}$ is the unit vector pointing in the direction of $\bm x_\textsc{a}-\bm x_\textsc{b}$.

This term is actually very similar to its local counterpart \eqref{Lone}. Nevertheless, now we have an extra term $e^{\ii\bm k\cdot(\bm x_\textsc{a}-\bm x_\textsc{b})}$, which needs to be decomposed into spherical harmonics as well using \eqref{expoharmonics}. Therefore there appear three spherical harmonics in the integral over $\text{d}\Omega_{\bm k}$ and five in each integral over $\text{d}\Omega_{1}$ and $\text{d}\Omega_2$. These integrals can be readily evaluated using the identities (although implicit in the expressions, all the spherical harmonics depend on the same angles $\theta$, $\phi$)
\begin{align}
\int\text{d}&\Omega\,Y_{l_1m_1}Y_{l_2m_2}Y_{l_3m_3}=\sqrt{\frac{(2l_1+1)(2l_2+1)(2l_3+1)}{4\pi}}\tj{l_1}{l_2}{l_3}{0}{0}{0}\tj{l_1}{l_2}{l_3}{m_1}{m_2}{m_3}\label{threeharmonics}\\
   \int\text{d}&\Omega\,Y_{l_1m_1}Y_{l_2m_2}Y_{l_3m_3}Y_{l_4m_4}Y_{l_5m_5}=\sum_{\lambda,\lambda'=0}^\infty\frac{(2\lambda+1)(2\lambda'+1)}{4\pi}\sqrt{\frac{(2l_1+1)(2l_2+1)(2l_3+1)(2l_4+1)(2l_5+1)}{4\pi}}\notag\\
   &\times\tj{l_1}{l_2}{\lambda}{0}{0}{0}\tj{l_1}{l_2}{\lambda}{m_1}{m_2}{-m_1-m_2}\tj{l_3}{\lambda'}{\lambda}{0}{0}{0}\tj{l_3}{\lambda'}{\lambda}{m_3}{m_4+m_5}{m_1+m_2}\tj{l_4}{l_5}{\lambda'}{0}{0}{0}\tj{l_4}{l_5}{\lambda'}{m_4}{m_5}{-m_4-m_5}.
   \label{fiveharmonics}
\end{align}

We can orient the $z$-axis in the integral over $\bm k$ along the vector $\bm x_\textsc{a}-\bm x_\textsc{b}$. In that way we can write \mbox{$Y^*_{l''m''}(\widehat{\Delta\bm x})=\delta_{m''0}\sqrt{(2l''+1)/(4\pi)}$}, and thus easily perform the sum over $m''$.

The next step is to perform the integrals over the solid angles and the sums over $m$ and $m'$:
\begin{align}
\sum_{m,m'}&\int\text{d}\Omega_k\,Y_{lm}(\bm{\hat k})Y_{l'm'}(\bm{\hat k})Y_{l''0}(\bm{\hat k})\int\text{d}\Omega_1\,Y^*_{l_e,m_e}(\bm{\hat x}_1)Y_{l_g,m_g}(\bm{\hat x}_1)Y^*_{lm}(\bm{\hat x}_1)\notag\\
&\times\int\text{d}\Omega_2\sum_{\epsilon=-l_e}^{l_e}(\mathcal{D}^{l_e}_{\epsilon,m_e})^* Y^*_{l_e,\epsilon}(\bm{\hat x}_2)\sum_{\eta=-l_g}^{l_g}\mathcal{D}^{l_g}_{\eta,m_g}Y_{l_g,\eta}(\bm{\hat x}_2)Y^*_{l'm'}(\bm{\hat x}_2)\notag\\
&\quad\times\left[Y_{10}(\bm{\hat x}_1)Y_{10}(\bm{\hat x}_2)-Y_{11}(\bm{\hat x}_1)Y_{1-1}(\bm{\hat x}_2)-Y_{1-1}(\bm{\hat x}_1)Y_{11}(\bm{\hat x}_2)\right]Y^*_{l''0}(\widehat{\Delta\bm x})\notag\\
=&\frac{3}{(4\pi)^3}(2l_e+1)(2l_g+1)(2l+1)(2l'+1)(2l''+1)\sum_\eta\mathcal{D}^{l_g}_{\eta,m_g}(\mathcal{D}^{l_e}_{\eta+m_g-m_e,m_e})^*\notag\\
&\times\sum_{\lambda,\lambda'}(2\lambda+1)(2\lambda'+1)
\tj{l}{l_e}{\lambda}{0}{0}{0}
\tj{l_g}{1}{\lambda}{0}{0}{0}
\tj{l'}{l_e}{\lambda'}{0}{0}{0}
\tj{l_g}{1}{\lambda'}{0}{0}{0}
\tj{l}{l'}{l''}{0}{0}{0}\notag\\
&\quad\times\left[\tj{l_g}{1}{\lambda}{m_g}{0}{-m_g}
\tj{l}{l_e}{\lambda}{m_e-m_g}{-m_e}{m_g}
\tj{l_g}{1}{\lambda'}{\eta}{0}{-\eta}\right.\notag\\
&\qquad\quad\times\tj{l'}{l_e}{\lambda'}{m_g-m_e}{m_e-m_g-\eta}{\eta}
\tj{l}{l'}{l''}{m_g-m_e}{m_e-m_g}{0}\notag\\
&\qquad-\tj{l_g}{1}{\lambda}{m_g}{1}{-1-m_g}
\tj{l}{l_e}{\lambda}{m_e-m_g-1}{-m_e}{m_g+1}
\tj{l_g}{1}{\lambda'}{\eta}{-1}{1-\eta}\notag\\
&\qquad\quad\times\tj{l'}{l_e}{\lambda'}{m_g-m_e+1}{m_e-m_g-\eta}{\eta-1}
\tj{l}{l'}{l''}{m_g-m_e+1}{m_e-m_g-1}{0}\notag\\
&\qquad-\tj{l_g}{1}{\lambda}{m_g}{-1}{1-m_g}
\tj{l}{l_e}{\lambda}{1+m_e-m_g}{-m_e}{m_g-1}
\tj{l_g}{1}{\lambda'}{\eta}{1}{-1-\eta}\notag\\
&\qquad\quad\times\left.\tj{l'}{l_e}{\lambda'}{m_g-m_e-1}{m_e-m_g-\eta}{\eta+1}
\tj{l}{l'}{l''}{m_g-m_e-1}{1+m_e-m_g}{0}\right].
\end{align}

As it was the case for the local term, this term cannot be further simplified without specifying the specific atomic levels of the ground and excited states.

We address now the contribution of the dyadic $\bm k\otimes\bm k$, which reads

\begin{align}
\mathcal{M}^\textsc{ab}\Big|_{\bm k\otimes\bm k}\!\!\!\!=&-\ac_\textsc{a}\ac_\textsc{b}\int_0^\infty\frac{\text{d}|\bm k|}{(2\pi)^3}\frac{|\bm k|^3}{2}\sum_{l=0}^\infty\sum_{m=-l}^l 4\pi\ii^l \sum_{l'=0}^\infty\sum_{m'=-l'}^{l'} 4\pi\ii^{l'} (-1)^{l'}\sum_{l''=0}^\infty4\pi\ii^{l''}j_{l''}(|\bm k||\bm x_\textsc{a}-\bm x_\textsc{b}|)\sqrt{\frac{2l''+1}{4\pi}}\left(\frac{4\pi}{3}\right)^2\notag\\
&\times\int_{-\infty}^{\infty}\text{d}t_1\int_{-\infty}^{t_1}\text{d}t_2\,e^{\ii(\Omega_\textsc{a} t_1+\Omega_\textsc{b} t_2)}\chi_\textsc{a}(t_1)\chi_\textsc{b}(t_2)e^{-\ii|\bm k|(t_1-t_2)}\notag\\
&\times\int_0^\infty\text{d}|\bm x_1|\,|\bm x_1|^3R^*_{n_e,l_e}(|\bm x_1|)R_{n_g,l_g}(|\bm x_1|)j_{l}(|\bm k||\bm x_1|)\int_0^\infty\text{d}|\bm x_2|\,|\bm x_2|^3R^*_{n_e,l_e}(|\bm x_2|)R_{n_g,l_g}(|\bm x_2|)j_{l'}(|\bm k||\bm x_2|)\notag\\
&\times\int\!\!\text{d}\Omega_k\,Y_{lm}(\bm{\hat k})Y_{l'm'}(\bm{\hat k})Y_{l''0}(\bm{\hat k})\!\!\int\!\!\text{d}\Omega_1 (Y^\textsc{a}_{l_e,m_e})^*(\bm{\hat x}_1)Y^\textsc{a}_{l_g,m_g}(\bm{\hat x}_1)Y^*_{lm}(\bm{\hat x}_1)\!\!\int\!\!\text{d}\Omega_2 (Y^\textsc{b}_{l_e,m_e})^*(\bm{\hat x}_2)Y^\textsc{b}_{l_g,m_g}(\bm{\hat x}_2)Y^*_{l'm'}(\bm{\hat x}_2)\notag\\
&\,\,\,\,\times\!\left[Y_{10}(\bm{\hat x}_1)Y_{10}(\bm{\hat k})\!-\!Y_{11}(\bm{\hat x}_1)Y_{1-1}(\bm{\hat k})\!-\!Y_{1-1}(\bm{\hat x}_1)Y_{11}(\bm{\hat k})\right]\!\!\left[Y_{10}(\bm{\hat k})Y_{10}(\bm{\hat x}_2)\!-\!Y_{11}(\bm{\hat k})Y_{1-1}(\bm{\hat x}_2)\!-\!Y_{1-1}(\bm{\hat k})Y_{11}(\bm{\hat x}_2)\right]\!\!.
\end{align}

In this case, the sums over $m$ and $m'$ and the integrals over solid angles in the two last lines yield

\begin{align}
   \sum_{m,m'}&\int\text{d}\Omega_k\,Y_{lm}(\bm{\hat k})Y_{l'm'}(\bm{\hat k})Y_{l''0}(\bm{\hat k})\int\text{d}\Omega_1 Y^*_{l_e,m_e}(\bm{\hat x}_1)Y_{l_g,m_g}(\bm{\hat x}_1)Y^*_{lm}(\bm{\hat x}_1)\notag\\
   &\quad\times\int\text{d}\Omega_2\sum_\epsilon (\mathcal{D}^{l_e}_{\epsilon,m_e})^*Y^*_{l_e,\epsilon}(\bm{\hat x}_2)\sum_\eta\mathcal{D}^{l_g}_{\eta,m_g}Y_{l_g,\eta}(\bm{\hat x}_2)Y^*_{l'm'}(\bm{\hat x}_2)\notag\\
&\qquad\times\left[Y_{10}(\bm{\hat x}_1)Y_{10}(\bm{\hat k})\!-\!Y_{11}(\bm{\hat x}_1)Y_{1-1}(\bm{\hat k})-Y_{1-1}(\bm{\hat x}_1)Y_{11}(\bm{\hat k})\right]\!\left[Y_{10}(\bm{\hat k})Y_{10}(\bm{\hat x}_2)-Y_{11}(\bm{\hat k})Y_{1-1}(\bm{\hat x}_2)-Y_{1-1}(\bm{\hat k})Y_{11}(\bm{\hat x}_2)\right]\notag\\
=&\sum_{\lambda'\lambda''\lambda'''}\frac{(2\lambda'+1)(2\lambda''+1)(2\lambda'''+1)}{(4\pi)^3}(2l+1)(2l'+1)(2l_g+1)(2l_e+1)9\sqrt{\frac{2l''+1}{4\pi}}\sum_{\eta=-l_g}^{l_g}(\mathcal{D}^{l_e}_{\eta+m_g-m_e,m_e})^*\mathcal{D}^{l_g}_{\eta,m_g}\notag\\
&\times\tj{l}{l'}{\lambda'}{0}{0}{0}
\tj{l_e}{l}{\lambda''}{0}{0}{0}
\tj{l_g}{1}{\lambda''}{0}{0}{0}
\tj{l_g}{1}{\lambda'''}{0}{0}{0}
\tj{l_e}{l'}{\lambda'''}{0}{0}{0}\left(A_\mathcal{M}+B_\mathcal{M}\right),
\end{align}
where now the quantities $B_\mathcal{M}$ and $A_\mathcal{M}$ are

\begin{align}
B_\mathcal{M}=&\sqrt{\frac{2}{3}}\tj{l''}{\lambda'}{2}{0}{0}{0}
\tj{l''}{\lambda'}{2}{0}{-2}{2}
\tj{l}{l'}{\lambda'}{m_g-m_e-1}{m_e-m_g-1}{2}
\tj{l_e}{l}{\lambda''}{-m_e}{1+m_e-m_g}{m_g-1}\notag\\
&\quad\times \tj{l_g}{1}{\lambda''}{m_g}{-1}{1-m_g}
\tj{l_e}{l'}{\lambda'''}{m_e-m_g-\eta}{m_g-m_e+1}{\eta-1}
\tj{l_g}{1}{\lambda'''}{\eta}{-1}{1-\eta}\notag\\
&+\sqrt{\frac{2}{3}}\tj{l''}{\lambda'}{2}{0}{0}{0}
\tj{l''}{\lambda'}{2}{0}{2}{-2}
 \tj{l}{l'}{\lambda'}{m_g-m_e+1}{1+m_e-m_g}{-2}
 \tj{l_e}{l}{\lambda''}{-m_e}{m_e-m_g-1}{m_g+1}\notag\\
&\quad\times\tj{l_g}{1}{\lambda''}{m_g}{1}{-1-m_g}
\tj{l_e}{l'}{\lambda'''}{m_e-m_g-\eta}{m_g-m_e-1}{\eta+1}
\tj{l_g}{1}{\lambda'''}{\eta}{1}{-1-\eta},
\end{align}

\begin{align}
A_\mathcal{M}=&\sum_{\lambda}(2\lambda+1)\tj{l''}{\lambda'}{\lambda}{0}{0}{0}
\tj{l''}{\lambda'}{\lambda}{0}{0}{0}
\tj{1}{1}{\lambda}{0}{0}{0}^2
 \tj{l}{l'}{\lambda'}{m_g-m_e}{m_e-m_g}{0}
 \tj{l_e}{l}{\lambda''}{-m_e}{m_e-m_g}{m_g}\notag\\
&\quad\times \tj{l_g}{1}{\lambda''}{m_g}{0}{-m_g}
\tj{l_e}{l'}{\lambda'''}{m_e-m_g-\eta}{m_g-m_e}{\eta}
\tj{l_g}{1}{\lambda'''}{\eta}{0}{-\eta}\notag\\
&-\sum_{\lambda}(2\lambda+1)\tj{1}{1}{\lambda}{0}{0}{0}
\tj{1}{1}{\lambda}{0}{1}{-1}
\tj{l''}{\lambda'}{\lambda}{0}{0}{0}
\tj{l''}{\lambda'}{\lambda}{0}{-1}{1}\notag\\
&\quad\times\left[\tj{l}{l'}{\lambda'}{m_g-m_e}{m_e-m_g-1}{1}
\tj{l_e}{l}{\lambda''}{-m_e}{m_e-m_g}{m_g}
\tj{l_g}{1}{\lambda''}{m_g}{0}{-m_g}\right.\notag\\
&\qquad\quad\times\tj{l_e}{l'}{\lambda'''}{m_e-m_g-\eta}{m_g-m_e+1}{\eta-1}
\tj{l_g}{1}{\lambda'''}{\eta}{-1}{1-\eta}\notag\\
&\qquad+\tj{l}{l'}{\lambda'}{m_g-m_e-1}{m_e-m_g}{1}
\tj{l_e}{l}{\lambda''}{-m_e}{1+m_e-m_g}{m_g-1}
\tj{l_g}{1}{\lambda''}{m_g}{-1}{1-m_g}\notag\\
&\qquad\quad\times\left.\tj{l_e}{l'}{\lambda'''}{m_e-m_g-\eta}{m_g-m_e}{\eta}
\tj{l_g}{1}{\lambda'''}{\eta}{0}{-\eta}\right]\notag\\
&-\sum_{\lambda}(2\lambda+1)\tj{1}{1}{\lambda}{0}{0}{0}
\tj{1}{1}{\lambda}{0}{-1}{1}
\tj{l''}{\lambda'}{\lambda}{0}{0}{0}
\tj{l''}{\lambda'}{\lambda}{0}{1}{-1}\notag\\
&\quad\times\left[\tj{l}{l'}{\lambda'}{m_g-m_e}{1+m_e-m_g}{-1}
\tj{l_e}{l}{\lambda''}{-m_e}{m_e-m_g}{m_g}
\tj{l_g}{1}{\lambda''}{m_g}{0}{-m_g}\right.\notag\\
&\qquad\quad\times\tj{l_e}{l'}{\lambda'''}{m_e-m_g-\eta}{m_g-m_e-1}{\eta+1}
\tj{l_g}{1}{\lambda'''}{\eta}{1}{-1-\eta}\notag\\
&\qquad+\tj{l}{l'}{\lambda'}{m_g-m_e+1}{m_e-m_g}{-1}
\tj{l_e}{l}{\lambda''}{-m_e}{m_e-m_g-1}{m_g+1}
\tj{l_g}{1}{\lambda''}{m_g}{1}{-1-m_g}\notag\\
&\qquad\quad\times\left.\tj{l_e}{l'}{\lambda'''}{m_e-m_g-\eta}{m_g-m_e}{\eta}
\tj{l_g}{1}{\lambda'''}{\eta}{0}{-\eta}\right]\notag\\
&+\sum_{\lambda}(2\lambda+1)\tj{1}{1}{\lambda}{0}{0}{0}
\tj{1}{1}{\lambda}{1}{-1}{0}
\tj{l''}{\lambda'}{\lambda}{0}{0}{0}
\tj{l''}{\lambda'}{\lambda}{0}{0}{0}\notag\\
&\quad\times\left[\tj{l}{l'}{\lambda'}{m_g-m_e+1}{m_e-m_g-1}{0}
\tj{l_e}{l}{\lambda''}{-m_e}{m_e-m_g-1}{1+m_g}
\tj{l_g}{1}{\lambda''}{m_g}{1}{-1-m_g}\right.\notag\\
&\qquad\quad\times\left.\tj{l_e}{l'}{\lambda'''}{m_e-m_g-\eta}{m_g-m_e+1}{\eta-1}
\tj{l_g}{1}{\lambda'''}{\eta}{-1}{1-\eta}\right.\notag\\
&\qquad+\tj{l}{l'}{\lambda'}{m_g-m_e-1}{1+m_e-m_g}{0}
\tj{l_e}{l}{\lambda''}{-m_e}{1+m_e-m_g}{m_g-1}
\tj{l_g}{1}{\lambda''}{m_g}{-1}{1-m_g}\notag\\
&\qquad\quad\times\left.\tj{l_e}{l'}{\lambda'''}{m_e-m_g-\eta}{m_g-m_e-1}{\eta+1}
\tj{l_g}{1}{\lambda'''}{\eta}{1}{-1-\eta}\right].
\end{align}

We particularize now to the scenario described in the paper, recall, the ground states being hydrogenoid-$1s$ states and the excited states being hydrogenoid-$2p_z$ states. In this particular situation, the contribution of the nonlocal term proportional to the identity reads

\begin{align}
\mathcal{M}^\textsc{ab}\Big|_{\openone}=&-\ac_\textsc{a}\ac_\textsc{b}\int_0^\infty\frac{\text{d}|\bm k|}{(2\pi)^3}\frac{|\bm k|^3}{2}\sum_{l=0}^\infty 4\pi\ii^l \sum_{l'=0}^\infty 4\pi\ii^{l'} (-1)^{l'}\sum_{l''=0}^\infty4\pi\ii^{l''}j_{l''}(|\bm k||\bm x_\textsc{a}-\bm x_\textsc{b}|)\frac{4\pi}{3}\notag\\
&\times\int_{-\infty}^{\infty}\text{d}t_1\int_{-\infty}^{t_1}\text{d}t_2\,e^{\ii(\Omega_\textsc{a} t_1+\Omega_\textsc{b} t_2)}\chi_\textsc{a}(t_1)\chi_\textsc{b}(t_2)e^{-\ii|\bm k|(t_1-t_2)}\notag\\
&\times\int_0^\infty\text{d}|\bm x_1|\,|\bm x_1|^3R^*_{2,1}(|\bm x_1|)R_{1,0}(|\bm x_1|)j_{l}(|\bm k||\bm x_1|)\int_0^\infty\text{d}|\bm x_2|\,|\bm x_2|^3R^*_{2,1}(|\bm x_2|)R_{1,0}(|\bm x_2|)j_{l'}(|\bm k||\bm x_2|)\notag\\
&\quad\times\frac{3}{(4\pi)^3}(2+1)(0+1)(2l+1)(2l'+1)(2l''+1)\sum_\eta\mathcal{D}^{0}_{\eta,0}(\mathcal{D}^{1}_{\eta,0})^*\notag\\
&\quad\times\sum_{\lambda,\lambda'}(2\lambda+1)(2\lambda'+1)
\tj{l}{1}{\lambda}{0}{0}{0}
\tj{0}{1}{\lambda}{0}{0}{0}
\tj{l'}{1}{\lambda'}{0}{0}{0}
\tj{0}{1}{\lambda'}{0}{0}{0}
\tj{l}{l'}{l''}{0}{0}{0}\notag\\
&\quad\times\left[\tj{0}{1}{\lambda}{0}{0}{0}
\tj{l}{1}{\lambda}{0}{0}{0}
\tj{0}{1}{\lambda'}{\eta}{0}{-\eta}
\tj{l'}{1}{\lambda'}{0}{-\eta}{\eta}
\tj{l}{l'}{l''}{0}{0}{0}\right.\notag\\
&\qquad-\tj{0}{1}{\lambda}{0}{1}{-1}
\tj{l}{1}{\lambda}{-1}{0}{1}
\tj{0}{1}{\lambda'}{\eta}{-1}{1-\eta}
\tj{l'}{1}{\lambda'}{1}{-\eta}{\eta-1}
\tj{l}{l'}{l''}{1}{-1}{0}\notag\\
&\qquad-\left.\tj{0}{1}{\lambda}{0}{-1}{1}
\tj{l}{1}{\lambda}{1}{0}{-1}
\tj{0}{1}{\lambda'}{\eta}{1}{-1-\eta}
\tj{l'}{1}{\lambda'}{-1}{-\eta}{\eta+1}
\tj{l}{l'}{l''}{-1}{1}{0}\right].
\end{align}

The Wigner D-matrix coefficients are nonzero only for $\eta=0$, for which $\mathcal{D}^0_{0,0}=1$ and $\mathcal{D}^1_{0,0}(\psi,\vartheta,\varphi)=\cos\vartheta$. Using the properties of the $3j$-symbols we see that the sums have nonzero terms only for $\lambda=1$ and $\lambda'=1$. Additionally, we also obtain the restrictions $l=0,1,2$, $l'=0,1,2$ and, as a consequence of these last two, $l''=0,1,2,3,4$. Therefore, the computation of the sums yields

\begin{align}
\mathcal{M}^\textsc{ab}\Big|_{\openone}=&-\ac_\textsc{a}\ac_\textsc{b}\frac{\cos\vartheta}{12\pi^2}\int_0^\infty\text{d}|\bm k|\,|\bm k|^3\int_{-\infty}^{\infty}\text{d}t_1\int_{-\infty}^{t_1}\text{d}t_2\,e^{\ii(\Omega_\textsc{a} t_1+\Omega_\textsc{b} t_2)}\chi_\textsc{a}(t_1)\chi_\textsc{b}(t_2)e^{-\ii|\bm k|(t_1-t_2)}\notag\\
&\times\int_0^\infty\text{d}|\bm x_1|\,|\bm x_1|^3R^*_{2,1}(|\bm x_1|)R_{1,0}(|\bm x_1|)\int_0^\infty\text{d}|\bm x_2|\,|\bm x_2|^3R^*_{2,1}(|\bm x_2|)R_{1,0}(|\bm x_2|)\notag\\
&\quad\times\Big\{ j_{0}(|\bm k||\bm x_\textsc{a}-\bm x_\textsc{b}|) \left[ j_{0}(|\bm k||\bm x_1|) j_{0}(|\bm k||\bm x_2|)+2 j_{2}(|\bm k||\bm x_1|) j_{2}(|\bm k||\bm x_2|)\right]\notag\\
&\qquad+2 j_{2}(|\bm k||\bm x_\textsc{a}-\bm x_\textsc{b}|) \left[ j_{2}(|\bm k||\bm x_1|) j_{0}(|\bm k||\bm x_2|)+ j_{0}(|\bm k||\bm x_1|) j_{2}(|\bm k||\bm x_2|)- j_{2}(|\bm k||\bm x_1|) j_{2}(|\bm k||\bm x_2|)\right]\Big\}\notag\\
=&-\ac_\textsc{a}\ac_\textsc{b}\frac{\cos\vartheta}{12 \pi^2}\int_0^\infty\text{d}|\bm k|\,|\bm k|^3\int_{-\infty}^{\infty}\text{d}t_1\int_{-\infty}^{t_1}\text{d}t_2\,e^{\ii(\Omega_\textsc{a} t_1+\Omega_\textsc{b} t_2)}\chi_\textsc{a}(t_1)\chi_\textsc{b}(t_2)e^{-\ii|\bm k|(t_1-t_2)}\notag\\
&\quad\times\!\left[j_{0}(|\bm k||\bm x_\textsc{a}-\bm x_\textsc{b}|) \frac{7962624 a_0^2 \left(16 a_0^4 |\bm k|^4-8 a_0^2 |\bm k|^2+9\right)}{\left(4 a_0^2 |\bm k|^2+9\right)^8}-j_{2}(|\bm k||\bm x_\textsc{a}-\bm x_\textsc{b}|)\frac{28311552 a_0^4 |\bm k|^2 \left(8 a_0^2 |\bm k|^2-9\right)}{\left(4 a_0^2 |\bm k|^2+9\right)^8}\right].
\end{align}

Analogously, the contribution containing the $\bm k\otimes\bm k$ dyadic reads
\begin{align}
\mathcal{M}^\textsc{ab}\Big|_{\bm k\otimes\bm k}=&-\ac_\textsc{a}\ac_\textsc{b}(\mathcal{D}^{1}_{0,0})^*\mathcal{D}^{0}_{0,0}\int_0^\infty\frac{\text{d}|\bm k|}{(2\pi)^3}\frac{|\bm k|^3}{2}\sum_{l=0}^\infty 4\pi\ii^l \sum_{l'=0}^\infty 4\pi\ii^{l'} (-1)^{l'}\sum_{l''=0}^\infty4\pi\ii^{l''}j_{l''}(|\bm k||\bm x_\textsc{a}-\bm x_\textsc{b}|) \left(\frac{4\pi}{3}\right)^2\notag\\
&\times\int_{-\infty}^{\infty}\text{d}t_1\int_{-\infty}^{t_1}\text{d}t_2\,e^{\ii(\Omega_\textsc{a} t_1+\Omega_\textsc{b} t_2)}\chi_\textsc{a}(t_1)\chi_\textsc{b}(t_2)e^{-\ii|\bm k|(t_1-t_2)}\notag\\
&\times\int_0^\infty\text{d}|\bm x_1|\,|\bm x_1|^3R^*_{2,1}(|\bm x_1|)R_{1,0}(|\bm x_1|)j_{l}(|\bm k||\bm x_1|)\int_0^\infty\text{d}|\bm x_2|\,|\bm x_2|^3R^*_{2,1}(|\bm x_2|)R_{1,0}(|\bm x_2|)j_{l'}(|\bm k||\bm x_2|)\notag\\
&\quad\times\sum_{\lambda'}\frac{(2\lambda'+1)(2+1)(2+1)}{(4\pi)^3}(2l+1)(2l'+1)(0+1)(2+1)9\frac{2l''+1}{4\pi}\notag\\
&\quad\times\tj{l}{l'}{\lambda'}{0}{0}{0}
\tj{1}{l}{1}{0}{0}{0}
\tj{0}{1}{1}{0}{0}{0}
\tj{1}{l'}{1}{0}{0}{0}
\tj{0}{1}{1}{0}{0}{0}\notag\\
&\qquad\times\Bigg\{\sqrt{\frac{2}{3}}\tj{l''}{\lambda'}{2}{0}{0}{0}
\tj{l''}{\lambda'}{2}{0}{-2}{2}
\tj{l}{l'}{\lambda'}{-1}{-1}{2}
\tj{1}{l}{1}{0}{1}{-1}\tj{0}{1}{1}{0}{-1}{1}
\tj{1}{l'}{1}{0}{1}{-1}
\tj{0}{1}{1}{0}{-1}{1}\notag\\
&\qquad\quad+\sqrt{\frac{2}{3}}\tj{l''}{\lambda'}{2}{0}{0}{0}
\tj{l''}{\lambda'}{2}{0}{2}{-2}
\tj{l}{l'}{\lambda'}{1}{1}{-2}
\tj{1}{l}{1}{0}{-1}{1}\tj{0}{1}{1}{0}{1}{-1}
\tj{1}{l'}{1}{0}{-1}{1}
\tj{0}{1}{1}{0}{1}{-1}\notag\\
&\qquad\quad+\sum_\lambda(2\lambda+1)\tj{l''}{\lambda'}{\lambda}{0}{0}{0}^2
\tj{1}{1}{\lambda}{0}{0}{0}^2
\tj{l}{l'}{\lambda'}{0}{0}{0}
\tj{1}{l}{1}{0}{0}{0}
\tj{0}{1}{1}{0}{0}{0}
\tj{1}{l'}{1}{0}{0}{0}
\tj{0}{1}{1}{0}{0}{0}\notag\\
&\qquad\quad-\sum_{\lambda}(2\lambda+1)\tj{1}{1}{\lambda}{0}{0}{0}
\tj{1}{1}{\lambda}{0}{1}{-1}
\tj{l''}{\lambda'}{\lambda}{0}{0}{0}
\tj{l''}{\lambda'}{\lambda}{0}{-1}{1}
\notag\\
&\qquad\qquad\times\left[\tj{l}{l'}{\lambda'}{0}{-1}{1}
\tj{1}{l}{1}{0}{0}{0}
\tj{0}{1}{1}{0}{0}{0}
\tj{1}{l'}{1}{0}{1}{-1}
\tj{0}{1}{1}{0}{-1}{1}\right.\notag\\
&\left.\qquad\qquad\quad+\tj{l}{l'}{\lambda'}{-1}{0}{1}
\tj{1}{l}{1}{0}{1}{-1}
\tj{0}{1}{1}{0}{-1}{1}
\tj{1}{l'}{1}{0}{0}{0}
\tj{0}{1}{1}{0}{0}{0}\right]\notag\\
&\qquad\quad-\sum_{\lambda}(2\lambda+1)\tj{1}{1}{\lambda}{0}{0}{0}
\tj{1}{1}{\lambda}{0}{-1}{1}
\tj{l''}{\lambda'}{\lambda}{0}{0}{0}
\tj{l''}{\lambda'}{\lambda}{0}{1}{-1}
\notag\\
&\qquad\qquad\times\left[\tj{l}{l'}{\lambda'}{0}{1}{-1}
\tj{1}{l}{1}{0}{0}{0}
\tj{0}{1}{1}{0}{0}{0}
\tj{1}{l'}{1}{0}{-1}{1}
\tj{0}{1}{1}{0}{1}{-1}\right.\notag\\
&\left.\qquad\qquad\quad+\tj{l}{l'}{\lambda'}{1}{0}{-1}
\tj{1}{l}{1}{0}{-1}{1}
\tj{0}{1}{1}{0}{1}{-1}
\tj{1}{l'}{1}{0}{0}{0}
\tj{0}{1}{1}{0}{0}{0}\right]\notag\\
&\qquad\quad+\sum_{\lambda}(2\lambda+1)\tj{1}{1}{\lambda}{0}{0}{0}
\tj{1}{1}{\lambda}{1}{-1}{0}
\tj{l''}{\lambda'}{\lambda}{0}{0}{0}
\tj{l''}{\lambda'}{\lambda}{0}{0}{0}\notag\\
&\qquad\qquad\times\left[\tj{l}{l'}{\lambda'}{1}{-1}{0}
\tj{1}{l}{1}{0}{-1}{1}
\tj{0}{1}{1}{0}{1}{-1}
\tj{1}{l'}{1}{0}{1}{-1}
\tj{0}{1}{1}{0}{-1}{1}\right.\notag\\
&\left.\qquad\qquad\quad+\tj{l}{l'}{\lambda'}{-1}{1}{0}
\tj{1}{l}{1}{0}{1}{-1}
\tj{0}{1}{1}{0}{-1}{1}
\tj{1}{l'}{1}{0}{-1}{1}
\tj{0}{1}{1}{0}{1}{-1}\right]\Bigg\}
\notag\\
=&-\ac_\textsc{a}\ac_\textsc{b}\frac{\cos\vartheta}{36\pi^2}\int\text{d}|\bm k||\bm k|^3\left[j_{0}(|\bm k||\bm x_\textsc{a}-\bm x_\textsc{b}|)-2j_{2}(|\bm k||\bm x_\textsc{a}-\bm x_\textsc{b}|)\right]\notag\\
&\times\int_{-\infty}^{\infty}\text{d}t_1\int_{-\infty}^{t_1}\text{d}t_2\,e^{\ii(\Omega_\textsc{a} t_1+\Omega_\textsc{b} t_2)}\chi_\textsc{a}(t_1)\chi_\textsc{b}(t_2)e^{-\ii|\bm k|(t_1-t_2)}\notag\\
&\times\int\text{d}|\bm x_1||\bm x_1|^3R^*_{2,1}(|\bm x_1|)R_{1,0}(|\bm x_1|)\left[j_{0}(|\bm k||\bm x_1|)-2j_{2}(|\bm k||\bm x_1|)\right]\notag\\
&\times\int\text{d}|\bm x_2||\bm x_2|^3R^*_{2,1}(|\bm x_2|)R_{1,0}(|\bm x_2|)\left[j_{0}(|\bm k||\bm x_2|)-2j_{2}(|\bm k||\bm x_2|)\right].
\end{align}

Finally, performing the integrals over $\text{d}|\bm x_1|$ and $\text{d}|\bm x_2|$ yields

\begin{align}
\mathcal{M}^\textsc{ab}\Big|_{\bm k\otimes\bm k}=&-\ac_\textsc{a}\ac_\textsc{b}\frac{24576\cos\vartheta}{\pi^2}a_0^2\int_0^\infty\text{d}|\bm k|\,|\bm k|^3\left[j_{0}(|\bm k||\bm x_\textsc{a}-\bm x_\textsc{b}|)-2j_{2}(|\bm k||\bm x_\textsc{a}-\bm x_\textsc{b}|)\right]\notag\\
&\times\int_{-\infty}^{\infty}\text{d}t_1\int_{-\infty}^{t_1}\text{d}t_2\,e^{\ii(\Omega_\textsc{a} t_1+\Omega_\textsc{b} t_2)}\chi_\textsc{a}(t_1)\chi_\textsc{b}(t_2)e^{-\ii|\bm k|(t_1-t_2)}\frac{(9-20a_0^2|\bm k|^2)^2}{(4a_0^2|\bm k|^2+9)^8},
\end{align}
and therefore, subtracting $\left.\mathcal{M}^\textsc{ab}\right|_{\bm k\otimes\bm k}$ from $\left.\mathcal{M}^\textsc{ab}\right|_{\openone}$, the complete contribution $\mathcal{M}^\textsc{ab}$ is

\begin{align}
\mathcal{M}^\textsc{ab}=&-\ac_\textsc{a}\ac_\textsc{b}\frac{49152\cos\vartheta}{\pi^2}a_0^2\int_0^\infty\text{d}|\bm k|\,|\bm k|^3\frac{j_{0}(|\bm k||\bm x_\textsc{a}-\bm x_\textsc{b}|)+j_{2}(|\bm k||\bm x_\textsc{a}-\bm x_\textsc{b}|)}{(4a_0^2|\bm k|^2+9)^6}\notag\\
&\times\int_{-\infty}^{\infty}\text{d}t_1\int_{-\infty}^{t_1}\text{d}t_2\,e^{\ii(\Omega_\textsc{a} t_1+\Omega_\textsc{b} t_2)}\chi_\textsc{a}(t_1)\chi_\textsc{b}(t_2)e^{-\ii|\bm k|(t_1-t_2)}.
\label{MAB}
\end{align}

One could think that computing the term $\mathcal{M}^\textsc{ba}$ just amounts to switching the labels $A\leftrightarrow B$ in the expression above. Nevertheless, this is not quite the case, since when we performed the integrals over solid angles we wrote the angular wave functions of atom B with respect the reference frame of atom A. We need to implement the following additional substitutions of Euler angles to obtain $\mathcal{M}_\textsc{ba}$ from $\mathcal{M}_\textsc{ab}$:
\begin{align}
   \psi_{\textsc{b}\rightarrow\textsc{a}}=-\varphi_{\textsc{a}\rightarrow\textsc{b}},\qquad
   \vartheta_{\textsc{b}\rightarrow\textsc{a}}=-\vartheta_{\textsc{a}\rightarrow\textsc{b}},\qquad
   \varphi_{\textsc{b}\rightarrow\textsc{a}}=-\psi_{\textsc{a}\rightarrow\textsc{b}}.
\end{align}
In the case under study here, this just amounts (apart from changing the labels $A\leftrightarrow B$) to changing $\vartheta\rightarrow-\vartheta$ in Eq. \eqref{MAB}.
Therefore the complete nonlocal term $\mathcal{M}$ is

\begin{align}
\mathcal{M}=&-\ac_\textsc{a}\ac_\textsc{b}\frac{49152\cos\vartheta}{\pi^2}a_0^2\int_0^\infty\text{d}|\bm k|\,|\bm k|^3\frac{j_{0}(|\bm k||\bm x_\textsc{a}-\bm x_\textsc{b}|)+j_{2}(|\bm k||\bm x_\textsc{a}-\bm x_\textsc{b}|)}{(4a_0^2|\bm k|^2+9)^6}\notag\\
&\times\int_{-\infty}^{\infty}\text{d}t_1\int_{-\infty}^{t_1}\text{d}t_2\,\left[e^{\ii(\Omega_\textsc{a} t_1+\Omega_\textsc{b} t_2)}\chi_\textsc{a}(t_1)\chi_\textsc{b}(t_2)e^{-\ii|\bm k|(t_1-t_2)}+e^{\ii(\Omega_\textsc{b} t_1+\Omega_\textsc{a} t_2)}\chi_\textsc{b}(t_1)\chi_\textsc{a}(t_2)e^{-\ii|\bm k|(t_1-t_2)}\right].
\label{Mfinal}
\end{align}

Finally, let us perform the integrals in time for Gaussian switching functions, which admit closed-form expressions. Recall $\chi_\mu(t)=e^{-(t-t_\mu)^2/T^2}$. Therefore the time integrals take the form
\begin{align}
   &\int_{-\infty}^{\infty}\text{d}t_1\int_{-\infty}^{t_1}\text{d}t_2\,\left[e^{\ii(\Omega_\textsc{a} t_1+\Omega_\textsc{b} t_2)}e^{-\frac{(t_1-t_\textsc{a})^2}{T^2}}e^{-\frac{(t_2-t_\textsc{b})^2}{T^2}}e^{-\ii|\bm k|(t_1-t_2)}+e^{\ii(\Omega_\textsc{b} t_1+\Omega_\textsc{a} t_2)}e^{-\frac{(t_1-t_\textsc{b})^2}{T^2}}e^{-\frac{(t_2-t_\textsc{a})^2}{T^2}}e^{-\ii|\bm k|(t_1-t_2)}\right]\notag\\
   =&\sqrt{\pi}\frac{T}{2}\int_{-\infty}^{\infty}\text{d}t_1\,\left\{e^{\ii(\Omega_\textsc{a} +\Omega_\textsc{b}) t_\textsc{b}}e^{-\frac{1}{4}(|\bm k|+\Omega_\textsc{b})^2T^2}e^{\ii(\Omega_\textsc{a}-|\bm k|)t_1}e^{-\frac{(t_1+t_\textsc{ba})^2}{T^2}}\left[1+\text{erf}\left(\frac{t_1}{T}-\ii \frac{T}{2}(|\bm k|+\Omega_\textsc{b})\right)\right]\right.\notag\\
   &\left.+e^{\ii(\Omega_\textsc{a} +\Omega_\textsc{b}) t_\textsc{a}}e^{-\frac{1}{4}(|\bm k|+\Omega_\textsc{a})^2T^2}e^{\ii(\Omega_\textsc{b}-|\bm k|)t_1}e^{-\frac{(t_1-t_\textsc{ba})^2}{T^2}}\left[1+\text{erf}\left(\frac{t_1}{T}-\ii \frac{T}{2}(|\bm k|+\Omega_\textsc{a})\right)\right]\right\}\notag\\
   =&\sqrt{\pi} \frac{T^2}{2} e^{\ii(\Omega_\textsc{a}+\Omega_\textsc{b}) t_\textsc{b}} \left[\sqrt{\pi}e^{-\frac{1}{4}(\Omega_\textsc{a}^2+\Omega_\textsc{b}^2+2|\bm k|^2)T^2}e^{\frac{1}{2}|\bm k|(\Omega_\textsc{a}-\Omega_\textsc{b})T^2}e^{\ii(|\bm k|-\Omega_\textsc{a})t_\textsc{ba}}+e^{-\frac{t_\textsc{ba}^2}{T^2}}I\left(\frac{T}{2}(|\bm k|+\Omega_\textsc{b}),T(|\bm k|-\Omega_\textsc{a})+2\ii\frac{ t_\textsc{ba}}{T}\right)\right]\notag\\
   &+\sqrt{\pi} \frac{T^2}{2} \left[\sqrt{\pi}e^{\ii(\Omega_\textsc{a}+\Omega_\textsc{b})t_\textsc{b}}e^{-\frac{1}{4}(\Omega_\textsc{a}^2+\Omega_\textsc{b}^2+2|\bm k|^2)T^2}e^{\frac{1}{2}|\bm k|(\Omega_\textsc{b}-\Omega_\textsc{a})T^2}e^{-\ii(\Omega_\textsc{a}+|\bm k|)t_\textsc{ba}}\right.\notag\\
   &\qquad\left.+e^{\ii(\Omega_\textsc{a}+\Omega_\textsc{b})t_\textsc{a}}e^{-\frac{t_\textsc{ba}^2}{T^2}}I\left(\frac{T}{2}(|\bm k|+\Omega_\textsc{a}),T(|\bm k|+\Omega_\textsc{a})+\ii\frac{ t_\textsc{ba}}{T}\right)\right]\notag\\
   =&\frac{1}{2} \pi  T^2 \left[\text{erfc}\left(\frac{2 t_\textsc{ba}+\ii T^2 (2 |\bm k|-\Omega_\textsc{a}+\Omega_\textsc{b})}{2 \sqrt{2} T}\right)+e^{-|\bm k| \left(T^2 (\Omega_\textsc{a}-\Omega_\textsc{b})+2 \ii t_\textsc{ba}\right)} \text{erfc}\left(\frac{-2 t_\textsc{ba}+\ii T^2 (2 |\bm k|+\Omega_\textsc{a}-\Omega_\textsc{b})}{2 \sqrt{2} T}\right)\right]\notag\\
   &\times e^{\frac{1}{4} \left(-2 |\bm k|^2 T^2+2 |\bm k| \left[T^2 (\Omega_\textsc{a}-\Omega_\textsc{b})+2 \ii t_\textsc{ba}\right]-T^2 \left(\Omega_\textsc{a}^2+\Omega_\textsc{b}^2\right)+4 \ii t_\textsc{b} (\Omega_\textsc{a}+\Omega_\textsc{b})-4 \ii t_\textsc{ba} \Omega_\textsc{a}\right)},
\end{align}
where

\begin{equation}
I(a,b)=\int_{-\infty}^\infty\text{d}x\,e^{-a^2-\ii b x-x^2}\text{erf}(x-\ii a) =-\ii\sqrt{\pi}e^{-a^2-\frac{b^2}{4}}\text{erf}\left(\frac{a+\frac{b}{2}}{\sqrt{2}}\right)
\end{equation}
is explicitly computed in \cite{Pozas-Kerstjens2015}.
\end{widetext}

\bibliography{bibliography}

\begin{thebibliography}{56}%
\makeatletter
\providecommand \@ifxundefined [1]{%
 \@ifx{#1\undefined}
}%
\providecommand \@ifnum [1]{%
 \ifnum #1\expandafter \@firstoftwo
 \else \expandafter \@secondoftwo
 \fi
}%
\providecommand \@ifx [1]{%
 \ifx #1\expandafter \@firstoftwo
 \else \expandafter \@secondoftwo
 \fi
}%
\providecommand \natexlab [1]{#1}%
\providecommand \enquote  [1]{``#1''}%
\providecommand \bibnamefont  [1]{#1}%
\providecommand \bibfnamefont [1]{#1}%
\providecommand \citenamefont [1]{#1}%
\providecommand \href@noop [0]{\@secondoftwo}%
\providecommand \href [0]{\begingroup \@sanitize@url \@href}%
\providecommand \@href[1]{\@@startlink{#1}\@@href}%
\providecommand \@@href[1]{\endgroup#1\@@endlink}%
\providecommand \@sanitize@url [0]{\catcode `\\12\catcode `\$12\catcode
  `\&12\catcode `\#12\catcode `\^12\catcode `\_12\catcode `\%12\relax}%
\providecommand \@@startlink[1]{}%
\providecommand \@@endlink[0]{}%
\providecommand \url  [0]{\begingroup\@sanitize@url \@url }%
\providecommand \@url [1]{\endgroup\@href {#1}{\urlprefix }}%
\providecommand \urlprefix  [0]{URL }%
\providecommand \Eprint [0]{\href }%
\providecommand \doibase [0]{http://dx.doi.org/}%
\providecommand \selectlanguage [0]{\@gobble}%
\providecommand \bibinfo  [0]{\@secondoftwo}%
\providecommand \bibfield  [0]{\@secondoftwo}%
\providecommand \translation [1]{[#1]}%
\providecommand \BibitemOpen [0]{}%
\providecommand \bibitemStop [0]{}%
\providecommand \bibitemNoStop [0]{.\EOS\space}%
\providecommand \EOS [0]{\spacefactor3000\relax}%
\providecommand \BibitemShut  [1]{\csname bibitem#1\endcsname}%
\let\auto@bib@innerbib\@empty
\bibitem [{\citenamefont {Summers}\ and\ \citenamefont
  {Werner}(1985)}]{Summers1985}%
  \BibitemOpen
  \bibfield  {author} {\bibinfo {author} {\bibfnamefont {S.~J.}\ \bibnamefont
  {Summers}}\ and\ \bibinfo {author} {\bibfnamefont {R.}~\bibnamefont
  {Werner}},\ }\href {\doibase 10.1016/0375-9601(85)90093-3} {\bibfield
  {journal} {\bibinfo  {journal} {Phys. Lett. A}\ }\textbf {\bibinfo {volume}
  {110}},\ \bibinfo {pages} {257} (\bibinfo {year} {1985})}\BibitemShut
  {NoStop}%
\bibitem [{\citenamefont {Summers}\ and\ \citenamefont
  {Werner}(1987)}]{Summers1987}%
  \BibitemOpen
  \bibfield  {author} {\bibinfo {author} {\bibfnamefont {S.~J.}\ \bibnamefont
  {Summers}}\ and\ \bibinfo {author} {\bibfnamefont {R.}~\bibnamefont
  {Werner}},\ }\href {\doibase 10.1063/1.527733} {\bibfield  {journal}
  {\bibinfo  {journal} {J. Math. Phys.}\ }\textbf {\bibinfo {volume} {28}},\
  \bibinfo {pages} {2440} (\bibinfo {year} {1987})}\BibitemShut {NoStop}%
\bibitem [{\citenamefont {Hotta}(2008)}]{Hotta2008}%
  \BibitemOpen
  \bibfield  {author} {\bibinfo {author} {\bibfnamefont {M.}~\bibnamefont
  {Hotta}},\ }\href {\doibase 10.1103/PhysRevD.78.045006} {\bibfield  {journal}
  {\bibinfo  {journal} {Phys. Rev. D}\ }\textbf {\bibinfo {volume} {78}},\
  \bibinfo {pages} {045006} (\bibinfo {year} {2008})}\BibitemShut {NoStop}%
\bibitem [{\citenamefont {Hotta}(2009)}]{Hotta2009}%
  \BibitemOpen
  \bibfield  {author} {\bibinfo {author} {\bibfnamefont {M.}~\bibnamefont
  {Hotta}},\ }\href {\doibase 10.1143/JPSJ.78.034001} {\bibfield  {journal}
  {\bibinfo  {journal} {J. Phys. Soc. Jpn}\ }\textbf {\bibinfo {volume} {78}},\
  \bibinfo {pages} {034001} (\bibinfo {year} {2009})}\BibitemShut {NoStop}%
\bibitem [{\citenamefont {Frey}\ \emph {et~al.}(2014)\citenamefont {Frey},
  \citenamefont {Funo},\ and\ \citenamefont {Hotta}}]{Frey2014}%
  \BibitemOpen
  \bibfield  {author} {\bibinfo {author} {\bibfnamefont {M.}~\bibnamefont
  {Frey}}, \bibinfo {author} {\bibfnamefont {K.}~\bibnamefont {Funo}}, \ and\
  \bibinfo {author} {\bibfnamefont {M.}~\bibnamefont {Hotta}},\ }\href
  {\doibase 10.1103/PhysRevE.90.012127} {\bibfield  {journal} {\bibinfo
  {journal} {Phys. Rev. E}\ }\textbf {\bibinfo {volume} {90}},\ \bibinfo
  {pages} {012127} (\bibinfo {year} {2014})}\BibitemShut {NoStop}%
\bibitem [{\citenamefont {Jonsson}\ \emph {et~al.}(2015)\citenamefont
  {Jonsson}, \citenamefont {Mart{\'{i}}n-Mart{\'{i}}nez},\ and\ \citenamefont
  {Kempf}}]{Jonsson2015}%
  \BibitemOpen
  \bibfield  {author} {\bibinfo {author} {\bibfnamefont {R.~H.}\ \bibnamefont
  {Jonsson}}, \bibinfo {author} {\bibfnamefont {E.}~\bibnamefont
  {Mart{\'{i}}n-Mart{\'{i}}nez}}, \ and\ \bibinfo {author} {\bibfnamefont
  {A.}~\bibnamefont {Kempf}},\ }\href {\doibase 10.1103/PhysRevLett.114.110505}
  {\bibfield  {journal} {\bibinfo  {journal} {Phys. Rev. Lett.}\ }\textbf
  {\bibinfo {volume} {114}},\ \bibinfo {pages} {110505} (\bibinfo {year}
  {2015})}\BibitemShut {NoStop}%
\bibitem [{\citenamefont {Blasco}\ \emph
  {et~al.}(2015{\natexlab{a}})\citenamefont {Blasco}, \citenamefont {Garay},
  \citenamefont {Mart\'in-Benito},\ and\ \citenamefont
  {Mart\'{i}n-Mart\'{i}nez}}]{Blasco:2015eya}%
  \BibitemOpen
  \bibfield  {author} {\bibinfo {author} {\bibfnamefont {A.}~\bibnamefont
  {Blasco}}, \bibinfo {author} {\bibfnamefont {L.~J.}\ \bibnamefont {Garay}},
  \bibinfo {author} {\bibfnamefont {M.}~\bibnamefont {Mart\'in-Benito}}, \ and\
  \bibinfo {author} {\bibfnamefont {E.}~\bibnamefont
  {Mart\'{i}n-Mart\'{i}nez}},\ }\href {\doibase 10.1103/PhysRevLett.114.141103}
  {\bibfield  {journal} {\bibinfo  {journal} {Phys. Rev. Lett.}\ }\textbf
  {\bibinfo {volume} {114}},\ \bibinfo {pages} {141103} (\bibinfo {year}
  {2015}{\natexlab{a}})}\BibitemShut {NoStop}%
\bibitem [{\citenamefont {Blasco}\ \emph
  {et~al.}(2015{\natexlab{b}})\citenamefont {Blasco}, \citenamefont {Garay},
  \citenamefont {Mart{\'{i}}n-Benito},\ and\ \citenamefont
  {Mart{\'{i}}n-Mart{\'{i}}nez}}]{Blasco2015}%
  \BibitemOpen
  \bibfield  {author} {\bibinfo {author} {\bibfnamefont {A.}~\bibnamefont
  {Blasco}}, \bibinfo {author} {\bibfnamefont {L.~J.}\ \bibnamefont {Garay}},
  \bibinfo {author} {\bibfnamefont {M.}~\bibnamefont {Mart{\'{i}}n-Benito}}, \
  and\ \bibinfo {author} {\bibfnamefont {E.}~\bibnamefont
  {Mart{\'{i}}n-Mart{\'{i}}nez}},\ }\href {\doibase 10.1139/cjp-2014-0567}
  {\bibfield  {journal} {\bibinfo  {journal} {Can. J. Phys.}\ }\textbf
  {\bibinfo {volume} {93}},\ \bibinfo {pages} {968} (\bibinfo {year}
  {2015}{\natexlab{b}})}\BibitemShut {NoStop}%
\bibitem [{\citenamefont {Blasco}\ \emph {et~al.}(2016)\citenamefont {Blasco},
  \citenamefont {Garay}, \citenamefont {Mart{\'{i}}n-Benito},\ and\
  \citenamefont {Mart{\'{i}}n-Mart{\'{i}}nez}}]{Blasco2016}%
  \BibitemOpen
  \bibfield  {author} {\bibinfo {author} {\bibfnamefont {A.}~\bibnamefont
  {Blasco}}, \bibinfo {author} {\bibfnamefont {L.~J.}\ \bibnamefont {Garay}},
  \bibinfo {author} {\bibfnamefont {M.}~\bibnamefont {Mart{\'{i}}n-Benito}}, \
  and\ \bibinfo {author} {\bibfnamefont {E.}~\bibnamefont
  {Mart{\'{i}}n-Mart{\'{i}}nez}},\ }\href {\doibase 10.1103/PhysRevD.93.024055}
  {\bibfield  {journal} {\bibinfo  {journal} {Phys. Rev. D}\ }\textbf {\bibinfo
  {volume} {93}},\ \bibinfo {pages} {024055} (\bibinfo {year}
  {2016})}\BibitemShut {NoStop}%
\bibitem [{\citenamefont {Hawking}(1975)}]{Hawking1975}%
  \BibitemOpen
  \bibfield  {author} {\bibinfo {author} {\bibfnamefont {S.~W.}\ \bibnamefont
  {Hawking}},\ }\href {\doibase 10.1007/BF02345020} {\bibfield  {journal}
  {\bibinfo  {journal} {Commun. Math. Phys.}\ }\textbf {\bibinfo {volume}
  {43}},\ \bibinfo {pages} {199} (\bibinfo {year} {1975})}\BibitemShut
  {NoStop}%
\bibitem [{\citenamefont {Hawking}(1976)}]{Hawking1976}%
  \BibitemOpen
  \bibfield  {author} {\bibinfo {author} {\bibfnamefont {S.~W.}\ \bibnamefont
  {Hawking}},\ }\href {\doibase 10.1103/PhysRevD.14.2460} {\bibfield  {journal}
  {\bibinfo  {journal} {Phys. Rev. D}\ }\textbf {\bibinfo {volume} {14}},\
  \bibinfo {pages} {2460} (\bibinfo {year} {1976})}\BibitemShut {NoStop}%
\bibitem [{\citenamefont {Hawking}\ \emph {et~al.}(2016)\citenamefont
  {Hawking}, \citenamefont {Perry},\ and\ \citenamefont
  {Strominger}}]{Hawking2016}%
  \BibitemOpen
  \bibfield  {author} {\bibinfo {author} {\bibfnamefont {S.~W.}\ \bibnamefont
  {Hawking}}, \bibinfo {author} {\bibfnamefont {M.~J.}\ \bibnamefont {Perry}},
  \ and\ \bibinfo {author} {\bibfnamefont {A.}~\bibnamefont {Strominger}},\
  }\href {\doibase 10.1103/PhysRevLett.116.231301} {\bibfield  {journal}
  {\bibinfo  {journal} {Phys. Rev. Lett.}\ }\textbf {\bibinfo {volume} {116}},\
  \bibinfo {pages} {231301} (\bibinfo {year} {2016})}\BibitemShut {NoStop}%
\bibitem [{\citenamefont {Almheiri}\ \emph {et~al.}(2013)\citenamefont
  {Almheiri}, \citenamefont {Marolf}, \citenamefont {Polchinski},\ and\
  \citenamefont {Sully}}]{Almheiri2013}%
  \BibitemOpen
  \bibfield  {author} {\bibinfo {author} {\bibfnamefont {A.}~\bibnamefont
  {Almheiri}}, \bibinfo {author} {\bibfnamefont {D.}~\bibnamefont {Marolf}},
  \bibinfo {author} {\bibfnamefont {J.}~\bibnamefont {Polchinski}}, \ and\
  \bibinfo {author} {\bibfnamefont {J.}~\bibnamefont {Sully}},\ }\href
  {\doibase 10.1007/JHEP02(2013)062} {\bibfield  {journal} {\bibinfo  {journal}
  {JHEP}\ }\textbf {\bibinfo {volume} {2013}},\ \bibinfo {pages} {62} (\bibinfo
  {year} {2013})}\BibitemShut {NoStop}%
\bibitem [{\citenamefont {Braunstein}\ \emph {et~al.}(2013)\citenamefont
  {Braunstein}, \citenamefont {Pirandola},\ and\ \citenamefont
  {{\.{Z}}yczkowski}}]{Braunstein2013}%
  \BibitemOpen
  \bibfield  {author} {\bibinfo {author} {\bibfnamefont {S.~L.}\ \bibnamefont
  {Braunstein}}, \bibinfo {author} {\bibfnamefont {S.}~\bibnamefont
  {Pirandola}}, \ and\ \bibinfo {author} {\bibfnamefont {K.}~\bibnamefont
  {{\.{Z}}yczkowski}},\ }\href {\doibase 10.1103/PhysRevLett.110.101301}
  {\bibfield  {journal} {\bibinfo  {journal} {Phys. Rev. Lett.}\ }\textbf
  {\bibinfo {volume} {110}},\ \bibinfo {pages} {101301} (\bibinfo {year}
  {2013})}\BibitemShut {NoStop}%
\bibitem [{\citenamefont {Reznik}(2003)}]{Reznik2003}%
  \BibitemOpen
  \bibfield  {author} {\bibinfo {author} {\bibfnamefont {B.}~\bibnamefont
  {Reznik}},\ }\href {\doibase 10.1023/A:1022875910744} {\bibfield  {journal}
  {\bibinfo  {journal} {Found. Phys.}\ }\textbf {\bibinfo {volume} {33}},\
  \bibinfo {pages} {167} (\bibinfo {year} {2003})}\BibitemShut {NoStop}%
\bibitem [{\citenamefont {Reznik}\ \emph {et~al.}(2005)\citenamefont {Reznik},
  \citenamefont {Retzker},\ and\ \citenamefont {Silman}}]{Reznik2005}%
  \BibitemOpen
  \bibfield  {author} {\bibinfo {author} {\bibfnamefont {B.}~\bibnamefont
  {Reznik}}, \bibinfo {author} {\bibfnamefont {A.}~\bibnamefont {Retzker}}, \
  and\ \bibinfo {author} {\bibfnamefont {J.}~\bibnamefont {Silman}},\ }\href
  {\doibase 10.1103/PhysRevA.71.042104} {\bibfield  {journal} {\bibinfo
  {journal} {Phys. Rev. A}\ }\textbf {\bibinfo {volume} {71}},\ \bibinfo
  {pages} {042104} (\bibinfo {year} {2005})}\BibitemShut {NoStop}%
\bibitem [{\citenamefont {Valentini}(1991)}]{Valentini1991}%
  \BibitemOpen
  \bibfield  {author} {\bibinfo {author} {\bibfnamefont {A.}~\bibnamefont
  {Valentini}},\ }\href {\doibase 10.1016/0375-9601(91)90952-5} {\bibfield
  {journal} {\bibinfo  {journal} {Phys. Lett. A}\ }\textbf {\bibinfo {volume}
  {153}},\ \bibinfo {pages} {321} (\bibinfo {year} {1991})}\BibitemShut
  {NoStop}%
\bibitem [{\citenamefont {Pozas-Kerstjens}\ and\ \citenamefont
  {Mart{\'{i}}n-Mart{\'{i}}nez}(2015)}]{Pozas-Kerstjens2015}%
  \BibitemOpen
  \bibfield  {author} {\bibinfo {author} {\bibfnamefont {A.}~\bibnamefont
  {Pozas-Kerstjens}}\ and\ \bibinfo {author} {\bibfnamefont {E.}~\bibnamefont
  {Mart{\'{i}}n-Mart{\'{i}}nez}},\ }\href {\doibase 10.1103/PhysRevD.92.064042}
  {\bibfield  {journal} {\bibinfo  {journal} {Phys. Rev. D}\ }\textbf {\bibinfo
  {volume} {92}},\ \bibinfo {pages} {064042} (\bibinfo {year}
  {2015})}\BibitemShut {NoStop}%
\bibitem [{\citenamefont {Mart{\'{i}}n-Mart{\'{i}}nez}\ \emph
  {et~al.}(2013{\natexlab{a}})\citenamefont {Mart{\'{i}}n-Mart{\'{i}}nez},
  \citenamefont {Brown}, \citenamefont {Donnelly},\ and\ \citenamefont
  {Kempf}}]{Martin-Martinez2013a}%
  \BibitemOpen
  \bibfield  {author} {\bibinfo {author} {\bibfnamefont {E.}~\bibnamefont
  {Mart{\'{i}}n-Mart{\'{i}}nez}}, \bibinfo {author} {\bibfnamefont {E.~G.}\
  \bibnamefont {Brown}}, \bibinfo {author} {\bibfnamefont {W.}~\bibnamefont
  {Donnelly}}, \ and\ \bibinfo {author} {\bibfnamefont {A.}~\bibnamefont
  {Kempf}},\ }\href {\doibase 10.1103/PhysRevA.88.052310} {\bibfield  {journal}
  {\bibinfo  {journal} {Phys. Rev. A}\ }\textbf {\bibinfo {volume} {88}},\
  \bibinfo {pages} {052310} (\bibinfo {year} {2013}{\natexlab{a}})}\BibitemShut
  {NoStop}%
\bibitem [{\citenamefont {Salton}\ \emph {et~al.}(2015)\citenamefont {Salton},
  \citenamefont {Mann},\ and\ \citenamefont {Menicucci}}]{Salton2015}%
  \BibitemOpen
  \bibfield  {author} {\bibinfo {author} {\bibfnamefont {G.}~\bibnamefont
  {Salton}}, \bibinfo {author} {\bibfnamefont {R.~B.}\ \bibnamefont {Mann}}, \
  and\ \bibinfo {author} {\bibfnamefont {N.~C.}\ \bibnamefont {Menicucci}},\
  }\href {\doibase 10.1088/1367-2630/17/3/035001} {\bibfield  {journal}
  {\bibinfo  {journal} {New J. Phys.}\ }\textbf {\bibinfo {volume} {17}},\
  \bibinfo {pages} {035001} (\bibinfo {year} {2015})}\BibitemShut {NoStop}%
\bibitem [{\citenamefont {Steeg}\ and\ \citenamefont
  {Menicucci}(2009)}]{Steeg2009}%
  \BibitemOpen
  \bibfield  {author} {\bibinfo {author} {\bibfnamefont {G.~V.}\ \bibnamefont
  {Steeg}}\ and\ \bibinfo {author} {\bibfnamefont {N.~C.}\ \bibnamefont
  {Menicucci}},\ }\href {\doibase 10.1103/PhysRevD.79.044027} {\bibfield
  {journal} {\bibinfo  {journal} {Phys. Rev. D}\ }\textbf {\bibinfo {volume}
  {79}},\ \bibinfo {pages} {044027} (\bibinfo {year} {2009})}\BibitemShut
  {NoStop}%
\bibitem [{\citenamefont {Mart{\'{i}}n-Mart{\'{i}}nez}\ and\ \citenamefont
  {Menicucci}(2012)}]{Martin-Martinez2012}%
  \BibitemOpen
  \bibfield  {author} {\bibinfo {author} {\bibfnamefont {E.}~\bibnamefont
  {Mart{\'{i}}n-Mart{\'{i}}nez}}\ and\ \bibinfo {author} {\bibfnamefont
  {N.~C.}\ \bibnamefont {Menicucci}},\ }\href {\doibase
  10.1088/0264-9381/29/22/224003} {\bibfield  {journal} {\bibinfo  {journal}
  {Class. Quantum Gravity}\ }\textbf {\bibinfo {volume} {29}},\ \bibinfo
  {pages} {224003} (\bibinfo {year} {2012})}\BibitemShut {NoStop}%
\bibitem [{\citenamefont {Mart{\'{i}}n-Mart{\'{i}}nez}\ \emph
  {et~al.}(2016)\citenamefont {Mart{\'{i}}n-Mart{\'{i}}nez}, \citenamefont
  {Smith},\ and\ \citenamefont {Terno}}]{Martin-Martinez2016a}%
  \BibitemOpen
  \bibfield  {author} {\bibinfo {author} {\bibfnamefont {E.}~\bibnamefont
  {Mart{\'{i}}n-Mart{\'{i}}nez}}, \bibinfo {author} {\bibfnamefont {A.~R.~H.}\
  \bibnamefont {Smith}}, \ and\ \bibinfo {author} {\bibfnamefont {D.~R.}\
  \bibnamefont {Terno}},\ }\href {\doibase 10.1103/PhysRevD.93.044001}
  {\bibfield  {journal} {\bibinfo  {journal} {Phys. Rev. D}\ }\textbf {\bibinfo
  {volume} {93}},\ \bibinfo {pages} {044001} (\bibinfo {year}
  {2016})}\BibitemShut {NoStop}%
\bibitem [{\citenamefont {DeWitt}\ \emph {et~al.}(1979)\citenamefont {DeWitt},
  \citenamefont {Hawking},\ and\ \citenamefont {Israel}}]{DeWittBook}%
  \BibitemOpen
  \bibfield  {author} {\bibinfo {author} {\bibfnamefont {B.~S.}\ \bibnamefont
  {DeWitt}}, \bibinfo {author} {\bibfnamefont {S.~W.}\ \bibnamefont {Hawking}},
  \ and\ \bibinfo {author} {\bibfnamefont {W.}~\bibnamefont {Israel}},\
  }\href@noop {} {\emph {\bibinfo {title} {{General Relativity: An Einstein
  Centenary Survey}}}}\ (\bibinfo  {publisher} {Cambridge University Press},\
  \bibinfo {year} {1979})\BibitemShut {NoStop}%
\bibitem [{\citenamefont {Braun}(2002)}]{Braun2002}%
  \BibitemOpen
  \bibfield  {author} {\bibinfo {author} {\bibfnamefont {D.}~\bibnamefont
  {Braun}},\ }\href {\doibase 10.1103/PhysRevLett.89.277901} {\bibfield
  {journal} {\bibinfo  {journal} {Phys. Rev. Lett.}\ }\textbf {\bibinfo
  {volume} {89}},\ \bibinfo {pages} {277901} (\bibinfo {year}
  {2002})}\BibitemShut {NoStop}%
\bibitem [{\citenamefont {Braun}(2005)}]{Braun2005}%
  \BibitemOpen
  \bibfield  {author} {\bibinfo {author} {\bibfnamefont {D.}~\bibnamefont
  {Braun}},\ }\href {\doibase 10.1103/PhysRevA.72.062324} {\bibfield  {journal}
  {\bibinfo  {journal} {Phys. Rev. A}\ }\textbf {\bibinfo {volume} {72}},\
  \bibinfo {pages} {062324} (\bibinfo {year} {2005})}\BibitemShut {NoStop}%
\bibitem [{\citenamefont {Mart{\'{i}}n-Mart{\'{i}}nez}\ \emph
  {et~al.}(2013{\natexlab{b}})\citenamefont {Mart{\'{i}}n-Mart{\'{i}}nez},
  \citenamefont {Montero},\ and\ \citenamefont {del
  Rey}}]{Martin-Martinez2013}%
  \BibitemOpen
  \bibfield  {author} {\bibinfo {author} {\bibfnamefont {E.}~\bibnamefont
  {Mart{\'{i}}n-Mart{\'{i}}nez}}, \bibinfo {author} {\bibfnamefont
  {M.}~\bibnamefont {Montero}}, \ and\ \bibinfo {author} {\bibfnamefont
  {M.}~\bibnamefont {del Rey}},\ }\href {\doibase 10.1103/PhysRevD.87.064038}
  {\bibfield  {journal} {\bibinfo  {journal} {Phys. Rev. D}\ }\textbf {\bibinfo
  {volume} {87}},\ \bibinfo {pages} {064038} (\bibinfo {year}
  {2013}{\natexlab{b}})}\BibitemShut {NoStop}%
\bibitem [{\citenamefont {Alhambra}\ \emph {et~al.}(2014)\citenamefont
  {Alhambra}, \citenamefont {Kempf},\ and\ \citenamefont
  {Mart{\'{i}}n-Mart{\'{i}}nez}}]{Alhambra2014}%
  \BibitemOpen
  \bibfield  {author} {\bibinfo {author} {\bibfnamefont {{\'{A}}.~M.}\
  \bibnamefont {Alhambra}}, \bibinfo {author} {\bibfnamefont {A.}~\bibnamefont
  {Kempf}}, \ and\ \bibinfo {author} {\bibfnamefont {E.}~\bibnamefont
  {Mart{\'{i}}n-Mart{\'{i}}nez}},\ }\href {\doibase 10.1103/PhysRevA.89.033835}
  {\bibfield  {journal} {\bibinfo  {journal} {Phys. Rev. A}\ }\textbf {\bibinfo
  {volume} {89}},\ \bibinfo {pages} {033835} (\bibinfo {year}
  {2014})}\BibitemShut {NoStop}%
\bibitem [{\citenamefont {Olson}\ and\ \citenamefont
  {Ralph}(2011)}]{Olson2011}%
  \BibitemOpen
  \bibfield  {author} {\bibinfo {author} {\bibfnamefont {S.~J.}\ \bibnamefont
  {Olson}}\ and\ \bibinfo {author} {\bibfnamefont {T.~C.}\ \bibnamefont
  {Ralph}},\ }\href {\doibase 10.1103/PhysRevLett.106.110404} {\bibfield
  {journal} {\bibinfo  {journal} {Phys. Rev. Lett.}\ }\textbf {\bibinfo
  {volume} {106}},\ \bibinfo {pages} {110404} (\bibinfo {year}
  {2011})}\BibitemShut {NoStop}%
\bibitem [{\citenamefont {Olson}\ and\ \citenamefont
  {Ralph}(2012)}]{Olson2012}%
  \BibitemOpen
  \bibfield  {author} {\bibinfo {author} {\bibfnamefont {S.~J.}\ \bibnamefont
  {Olson}}\ and\ \bibinfo {author} {\bibfnamefont {T.~C.}\ \bibnamefont
  {Ralph}},\ }\href {\doibase 10.1103/PhysRevA.85.012306} {\bibfield  {journal}
  {\bibinfo  {journal} {Phys. Rev. A}\ }\textbf {\bibinfo {volume} {85}},\
  \bibinfo {pages} {012306} (\bibinfo {year} {2012})}\BibitemShut {NoStop}%
\bibitem [{\citenamefont {Sab{\'{i}}n}\ \emph {et~al.}(2012)\citenamefont
  {Sab{\'{i}}n}, \citenamefont {Peropadre}, \citenamefont {del Rey},\ and\
  \citenamefont {Mart{\'{i}}n-Mart{\'{i}}nez}}]{Sabin2012}%
  \BibitemOpen
  \bibfield  {author} {\bibinfo {author} {\bibfnamefont {C.}~\bibnamefont
  {Sab{\'{i}}n}}, \bibinfo {author} {\bibfnamefont {B.}~\bibnamefont
  {Peropadre}}, \bibinfo {author} {\bibfnamefont {M.}~\bibnamefont {del Rey}},
  \ and\ \bibinfo {author} {\bibfnamefont {E.}~\bibnamefont
  {Mart{\'{i}}n-Mart{\'{i}}nez}},\ }\href {\doibase
  10.1103/PhysRevLett.109.033602} {\bibfield  {journal} {\bibinfo  {journal}
  {Phys. Rev. Lett.}\ }\textbf {\bibinfo {volume} {109}},\ \bibinfo {pages}
  {033602} (\bibinfo {year} {2012})}\BibitemShut {NoStop}%
\bibitem [{\citenamefont {Mart{\'{i}}n-Mart{\'{i}}nez}\ and\ \citenamefont
  {Sanders}(2016)}]{Martin-Martinez2016}%
  \BibitemOpen
  \bibfield  {author} {\bibinfo {author} {\bibfnamefont {E.}~\bibnamefont
  {Mart{\'{i}}n-Mart{\'{i}}nez}}\ and\ \bibinfo {author} {\bibfnamefont
  {B.~C.}\ \bibnamefont {Sanders}},\ }\href {\doibase
  10.1088/1367-2630/18/4/043031} {\bibfield  {journal} {\bibinfo  {journal}
  {New J. Phys.}\ }\textbf {\bibinfo {volume} {18}},\ \bibinfo {pages} {043031}
  (\bibinfo {year} {2016})}\BibitemShut {NoStop}%
\bibitem [{\citenamefont {Schlicht}(2004)}]{Schlicht2004}%
  \BibitemOpen
  \bibfield  {author} {\bibinfo {author} {\bibfnamefont {S.}~\bibnamefont
  {Schlicht}},\ }\href {\doibase 10.1088/0264-9381/21/19/011} {\bibfield
  {journal} {\bibinfo  {journal} {Class. Quantum Gravity}\ }\textbf {\bibinfo
  {volume} {21}},\ \bibinfo {pages} {4647} (\bibinfo {year}
  {2004})}\BibitemShut {NoStop}%
\bibitem [{\citenamefont {Ju{\'{a}}rez-Aubry}\ and\ \citenamefont
  {Louko}(2014)}]{Juarez-Aubry2014}%
  \BibitemOpen
  \bibfield  {author} {\bibinfo {author} {\bibfnamefont {B.~A.}\ \bibnamefont
  {Ju{\'{a}}rez-Aubry}}\ and\ \bibinfo {author} {\bibfnamefont
  {J.}~\bibnamefont {Louko}},\ }\href {\doibase 10.1088/0264-9381/31/24/245007}
  {\bibfield  {journal} {\bibinfo  {journal} {Class. Quantum Gravity}\ }\textbf
  {\bibinfo {volume} {31}},\ \bibinfo {pages} {245007} (\bibinfo {year}
  {2014})}\BibitemShut {NoStop}%
\bibitem [{\citenamefont {Birrell}\ and\ \citenamefont
  {Davies}(1984)}]{Birrell1984}%
  \BibitemOpen
  \bibfield  {author} {\bibinfo {author} {\bibfnamefont {N.~D.}\ \bibnamefont
  {Birrell}}\ and\ \bibinfo {author} {\bibfnamefont {P.~C.~W.}\ \bibnamefont
  {Davies}},\ }\href@noop {} {\emph {\bibinfo {title} {{Quantum Fields in
  Curved Space}}}}\ (\bibinfo  {publisher} {Cambridge University Press},\
  \bibinfo {year} {1984})\BibitemShut {NoStop}%
\bibitem [{\citenamefont {Takagi}(1986)}]{Takagi1986}%
  \BibitemOpen
  \bibfield  {author} {\bibinfo {author} {\bibfnamefont {S.}~\bibnamefont
  {Takagi}},\ }\href {\doibase 10.1143/PTPS.88.1} {\bibfield  {journal}
  {\bibinfo  {journal} {Prog. Theor. Phys. Suppl.}\ }\textbf {\bibinfo {volume}
  {88}},\ \bibinfo {pages} {1} (\bibinfo {year} {1986})}\BibitemShut {NoStop}%
\bibitem [{\citenamefont {Lin}\ \emph {et~al.}(2016)\citenamefont {Lin},
  \citenamefont {Chou},\ and\ \citenamefont {Hu}}]{Lin2016}%
  \BibitemOpen
  \bibfield  {author} {\bibinfo {author} {\bibfnamefont {S.-Y.}\ \bibnamefont
  {Lin}}, \bibinfo {author} {\bibfnamefont {C.-H.}\ \bibnamefont {Chou}}, \
  and\ \bibinfo {author} {\bibfnamefont {B.~L.}\ \bibnamefont {Hu}},\ }\href
  {\doibase 10.1007/JHEP03(2016)047} {\bibfield  {journal} {\bibinfo  {journal}
  {JHEP}\ }\textbf {\bibinfo {volume} {2016}},\ \bibinfo {pages} {47} (\bibinfo
  {year} {2016})}\BibitemShut {NoStop}%
\bibitem [{\citenamefont {Mart{\'{i}}n-Mart{\'{i}}nez}\ and\ \citenamefont
  {Louko}(2014)}]{Martin-Martinez2014}%
  \BibitemOpen
  \bibfield  {author} {\bibinfo {author} {\bibfnamefont {E.}~\bibnamefont
  {Mart{\'{i}}n-Mart{\'{i}}nez}}\ and\ \bibinfo {author} {\bibfnamefont
  {J.}~\bibnamefont {Louko}},\ }\href {\doibase 10.1103/PhysRevD.90.024015}
  {\bibfield  {journal} {\bibinfo  {journal} {Phys. Rev. D}\ }\textbf {\bibinfo
  {volume} {90}},\ \bibinfo {pages} {024015} (\bibinfo {year}
  {2014})}\BibitemShut {NoStop}%
\bibitem [{\citenamefont {Thinh}\ \emph {et~al.}(2016)\citenamefont {Thinh},
  \citenamefont {Bancal},\ and\ \citenamefont
  {Mart\'{\i}n-Mart\'{\i}nez}}]{Thinh2016}%
  \BibitemOpen
  \bibfield  {author} {\bibinfo {author} {\bibfnamefont {L.~P.}\ \bibnamefont
  {Thinh}}, \bibinfo {author} {\bibfnamefont {J.-D.}\ \bibnamefont {Bancal}}, \
  and\ \bibinfo {author} {\bibfnamefont {E.}~\bibnamefont
  {Mart\'{\i}n-Mart\'{\i}nez}},\ }\href {\doibase 10.1103/PhysRevA.94.022321}
  {\bibfield  {journal} {\bibinfo  {journal} {Phys. Rev. A}\ }\textbf {\bibinfo
  {volume} {94}},\ \bibinfo {pages} {022321} (\bibinfo {year}
  {2016})}\BibitemShut {NoStop}%
\bibitem [{\citenamefont {Scully}\ and\ \citenamefont
  {Zubairy}(1997)}]{ScullyBook}%
  \BibitemOpen
  \bibfield  {author} {\bibinfo {author} {\bibfnamefont {M.~O.}\ \bibnamefont
  {Scully}}\ and\ \bibinfo {author} {\bibfnamefont {M.~S.}\ \bibnamefont
  {Zubairy}},\ }\href@noop {} {\emph {\bibinfo {title} {{Quantum Optics}}}}\
  (\bibinfo  {publisher} {Cambridge University Press},\ \bibinfo {year}
  {1997})\BibitemShut {NoStop}%
\bibitem [{\citenamefont {Bransden}\ and\ \citenamefont
  {Joachain}(2003)}]{BransdenBook}%
  \BibitemOpen
  \bibfield  {author} {\bibinfo {author} {\bibfnamefont {B.~H.}\ \bibnamefont
  {Bransden}}\ and\ \bibinfo {author} {\bibfnamefont {C.~J.}\ \bibnamefont
  {Joachain}},\ }\href@noop {} {\emph {\bibinfo {title} {{Physics of Atoms and
  Molecules}}}}\ (\bibinfo  {publisher} {Prentice Hall},\ \bibinfo {year}
  {2003})\BibitemShut {NoStop}%
\bibitem [{\citenamefont {Lamb}\ \emph {et~al.}(1987)\citenamefont {Lamb},
  \citenamefont {Schlicher},\ and\ \citenamefont {Scully}}]{Lamb1987}%
  \BibitemOpen
  \bibfield  {author} {\bibinfo {author} {\bibfnamefont {W.~E.}\ \bibnamefont
  {Lamb}}, \bibinfo {author} {\bibfnamefont {R.~R.}\ \bibnamefont {Schlicher}},
  \ and\ \bibinfo {author} {\bibfnamefont {M.~O.}\ \bibnamefont {Scully}},\
  }\href {\doibase 10.1103/PhysRevA.36.2763} {\bibfield  {journal} {\bibinfo
  {journal} {Phys. Rev. A}\ }\textbf {\bibinfo {volume} {36}},\ \bibinfo
  {pages} {2763} (\bibinfo {year} {1987})}\BibitemShut {NoStop}%
\bibitem [{\citenamefont {Louko}\ and\ \citenamefont
  {Mart{\'{i}}n-Mart{\'{i}}nez}()}]{Loukoprep}%
  \BibitemOpen
  \bibfield  {author} {\bibinfo {author} {\bibfnamefont {J.}~\bibnamefont
  {Louko}}\ and\ \bibinfo {author} {\bibfnamefont {E.}~\bibnamefont
  {Mart{\'{i}}n-Mart{\'{i}}nez}},\ }\href@noop {} {\bibinfo  {journal} {in
  preparation}\ }\BibitemShut {NoStop}%
\bibitem [{\citenamefont {Sagu{\'{e}}}\ \emph {et~al.}(2007)\citenamefont
  {Sagu{\'{e}}}, \citenamefont {Vetsch}, \citenamefont {Alt}, \citenamefont
  {Meschede},\ and\ \citenamefont {Rauschenbeutel}}]{Sague2007}%
  \BibitemOpen
\bibfield  {journal} {  }\bibfield  {author} {\bibinfo {author} {\bibfnamefont
  {G.}~\bibnamefont {Sagu{\'{e}}}}, \bibinfo {author} {\bibfnamefont
  {E.}~\bibnamefont {Vetsch}}, \bibinfo {author} {\bibfnamefont
  {W.}~\bibnamefont {Alt}}, \bibinfo {author} {\bibfnamefont {D.}~\bibnamefont
  {Meschede}}, \ and\ \bibinfo {author} {\bibfnamefont {A.}~\bibnamefont
  {Rauschenbeutel}},\ }\href {\doibase 10.1103/PhysRevLett.99.163602}
  {\bibfield  {journal} {\bibinfo  {journal} {Phys. Rev. Lett.}\ }\textbf
  {\bibinfo {volume} {99}},\ \bibinfo {pages} {163602} (\bibinfo {year}
  {2007})}\BibitemShut {NoStop}%
\bibitem [{\citenamefont {Kramida}(2010)}]{Kramida2010}%
  \BibitemOpen
  \bibfield  {author} {\bibinfo {author} {\bibfnamefont {A.}~\bibnamefont
  {Kramida}},\ }\href {\doibase 10.1016/j.adt.2010.05.001} {\bibfield
  {journal} {\bibinfo  {journal} {At. Data. Nucl. Data Tables}\ }\textbf
  {\bibinfo {volume} {96}},\ \bibinfo {pages} {586} (\bibinfo {year}
  {2010})}\BibitemShut {NoStop}%
\bibitem [{\citenamefont {Le{\'{o}}n}\ and\ \citenamefont
  {Sab{\'{i}}n}(2009)}]{Leon2009}%
  \BibitemOpen
  \bibfield  {author} {\bibinfo {author} {\bibfnamefont {J.}~\bibnamefont
  {Le{\'{o}}n}}\ and\ \bibinfo {author} {\bibfnamefont {C.}~\bibnamefont
  {Sab{\'{i}}n}},\ }\href {\doibase 10.1103/PhysRevA.79.012304} {\bibfield
  {journal} {\bibinfo  {journal} {Phys. Rev. A}\ }\textbf {\bibinfo {volume}
  {79}},\ \bibinfo {pages} {012304} (\bibinfo {year} {2009})}\BibitemShut
  {NoStop}%
\bibitem [{\citenamefont {Intravaia}\ \emph {et~al.}(2015)\citenamefont
  {Intravaia}, \citenamefont {Mkrtchian}, \citenamefont {Buhmann},
  \citenamefont {Scheel}, \citenamefont {Dalvit},\ and\ \citenamefont
  {Henkel}}]{Intravaia2015}%
  \BibitemOpen
  \bibfield  {author} {\bibinfo {author} {\bibfnamefont {F.}~\bibnamefont
  {Intravaia}}, \bibinfo {author} {\bibfnamefont {V.~E.}\ \bibnamefont
  {Mkrtchian}}, \bibinfo {author} {\bibfnamefont {S.~Y.}\ \bibnamefont
  {Buhmann}}, \bibinfo {author} {\bibfnamefont {S.}~\bibnamefont {Scheel}},
  \bibinfo {author} {\bibfnamefont {D.~A.~R.}\ \bibnamefont {Dalvit}}, \ and\
  \bibinfo {author} {\bibfnamefont {C.}~\bibnamefont {Henkel}},\ }\href
  {\doibase 10.1088/0953-8984/27/21/214020} {\bibfield  {journal} {\bibinfo
  {journal} {J. Phys. Condens. Matter}\ }\textbf {\bibinfo {volume} {27}},\
  \bibinfo {pages} {214020} (\bibinfo {year} {2015})}\BibitemShut {NoStop}%
\bibitem [{\citenamefont {Jackson}(2002)}]{Jackson2002}%
  \BibitemOpen
  \bibfield  {author} {\bibinfo {author} {\bibfnamefont {J.~D.}\ \bibnamefont
  {Jackson}},\ }\href {\doibase 10.1119/1.1491265} {\bibfield  {journal}
  {\bibinfo  {journal} {Am. J. Phys.}\ }\textbf {\bibinfo {volume} {70}},\
  \bibinfo {pages} {917} (\bibinfo {year} {2002})},\ \Eprint
  {http://arxiv.org/abs/0204034} {0204034} \BibitemShut {NoStop}%
\bibitem [{\citenamefont
  {Mart{\'{i}}n-Mart{\'{i}}nez}(2015)}]{Martin-Martinez2015}%
  \BibitemOpen
  \bibfield  {author} {\bibinfo {author} {\bibfnamefont {E.}~\bibnamefont
  {Mart{\'{i}}n-Mart{\'{i}}nez}},\ }\href {\doibase 10.1103/PhysRevD.92.104019}
  {\bibfield  {journal} {\bibinfo  {journal} {Phys. Rev. D}\ }\textbf {\bibinfo
  {volume} {92}},\ \bibinfo {pages} {104019} (\bibinfo {year}
  {2015})}\BibitemShut {NoStop}%
\bibitem [{\citenamefont {Vidal}\ and\ \citenamefont
  {Werner}(2002)}]{Vidal2002}%
  \BibitemOpen
  \bibfield  {author} {\bibinfo {author} {\bibfnamefont {G.}~\bibnamefont
  {Vidal}}\ and\ \bibinfo {author} {\bibfnamefont {R.~F.}\ \bibnamefont
  {Werner}},\ }\href {\doibase 10.1103/PhysRevA.65.032314} {\bibfield
  {journal} {\bibinfo  {journal} {Phys. Rev. A}\ }\textbf {\bibinfo {volume}
  {65}},\ \bibinfo {pages} {032314} (\bibinfo {year} {2002})}\BibitemShut
  {NoStop}%
\bibitem [{\citenamefont {Wootters}(1998)}]{Wooters1998}%
  \BibitemOpen
  \bibfield  {author} {\bibinfo {author} {\bibfnamefont {W.~K.}\ \bibnamefont
  {Wootters}},\ }\href {\doibase 10.1103/PhysRevLett.80.2245} {\bibfield
  {journal} {\bibinfo  {journal} {Phys. Rev. Lett.}\ }\textbf {\bibinfo
  {volume} {80}},\ \bibinfo {pages} {2245} (\bibinfo {year}
  {1998})}\BibitemShut {NoStop}%
\bibitem [{\citenamefont {Svelto}(2010)}]{SveltoBook}%
  \BibitemOpen
  \bibfield  {author} {\bibinfo {author} {\bibfnamefont {O.}~\bibnamefont
  {Svelto}},\ }\href@noop {} {\emph {\bibinfo {title} {{Principles of
  Lasers}}}}\ (\bibinfo  {publisher} {Springer Science {\&} Business Media},\
  \bibinfo {year} {2010})\ p.\ \bibinfo {pages} {150}\BibitemShut {NoStop}%
\bibitem [{\citenamefont {Galindo}\ and\ \citenamefont
  {Pascual}(1992)}]{GalindoBook}%
  \BibitemOpen
  \bibfield  {author} {\bibinfo {author} {\bibfnamefont {A.}~\bibnamefont
  {Galindo}}\ and\ \bibinfo {author} {\bibfnamefont {P.}~\bibnamefont
  {Pascual}},\ }\href@noop {} {\emph {\bibinfo {title} {{Quantum Mechanics
  I}}}}\ (\bibinfo  {publisher} {Springer-Verlag},\ \bibinfo {year} {1992})\
  p.\ \bibinfo {pages} {243}\BibitemShut {NoStop}%
\bibitem [{\citenamefont {Retzker}\ \emph {et~al.}(2005)\citenamefont
  {Retzker}, \citenamefont {Cirac},\ and\ \citenamefont
  {Reznik}}]{Retzker2005}%
  \BibitemOpen
  \bibfield  {author} {\bibinfo {author} {\bibfnamefont {A.}~\bibnamefont
  {Retzker}}, \bibinfo {author} {\bibfnamefont {J.~I.}\ \bibnamefont {Cirac}},
  \ and\ \bibinfo {author} {\bibfnamefont {B.}~\bibnamefont {Reznik}},\ }\href
  {\doibase 10.1103/PhysRevLett.94.050504} {\bibfield  {journal} {\bibinfo
  {journal} {Phys. Rev. Lett.}\ }\textbf {\bibinfo {volume} {94}},\ \bibinfo
  {pages} {050504} (\bibinfo {year} {2005})}\BibitemShut {NoStop}%
\bibitem [{\citenamefont {Cohen-Tannoudji}\ \emph {et~al.}(1987)\citenamefont
  {Cohen-Tannoudji}, \citenamefont {Dupont-Roc},\ and\ \citenamefont
  {Grynberg}}]{Cohen-TannoudjiQEDBook}%
  \BibitemOpen
  \bibfield  {author} {\bibinfo {author} {\bibfnamefont {C.}~\bibnamefont
  {Cohen-Tannoudji}}, \bibinfo {author} {\bibfnamefont {J.}~\bibnamefont
  {Dupont-Roc}}, \ and\ \bibinfo {author} {\bibfnamefont {G.}~\bibnamefont
  {Grynberg}},\ }\href@noop {} {\emph {\bibinfo {title} {{Photons and Atoms:
  Introduction to Quantum Electrodynamics}}}}\ (\bibinfo  {publisher}
  {Wiley-VCH},\ \bibinfo {year} {1987})\ p.~\bibinfo {pages} {36}\BibitemShut
  {NoStop}%
\bibitem [{\citenamefont {Morrison}\ and\ \citenamefont
  {Parker}(1987)}]{Morrison1987}%
  \BibitemOpen
  \bibfield  {author} {\bibinfo {author} {\bibfnamefont {M.~A.}\ \bibnamefont
  {Morrison}}\ and\ \bibinfo {author} {\bibfnamefont {G.~A.}\ \bibnamefont
  {Parker}},\ }\href {\doibase 10.1071/PH870465} {\bibfield  {journal}
  {\bibinfo  {journal} {Austral. J. Phys.}\ }\textbf {\bibinfo {volume} {40}},\
  \bibinfo {pages} {465} (\bibinfo {year} {1987})}\BibitemShut {NoStop}%
\end{thebibliography}%
\end{document}